\documentclass[appendixfloats]{emulateapj}
\usepackage{natbib}
\usepackage{graphicx}
\usepackage{epsfig}
\usepackage{subfigure}
\usepackage{rotating}
\usepackage{lmodern}
\usepackage{slantsc}
\usepackage{color}
\usepackage[normalem]{ulem}
\usepackage{booktabs}

\slugcomment{ApJ Accepted 06/02/2017.}

\shorttitle{Four component fits to the ``7.7"~\micron\ PAH band}
\shortauthors{D. J. Stock \& E. Peeters}

\begin{document}
\newcommand{\HII}{H~{\sc ii}}
\newcommand{\Gnaught}{G$_0$}
\newcommand{\cii}{[C~{\sc ii}]}
\newcommand{\oi}{[O~{\sc i}]}
\newcommand{\ci}{[C~{\sc i}]}
\newcommand{\pasa}{Publications of the Astronomical Society of Australia}

\title{Polycyclic Aromatic Hydrocarbon emission in \textit{Spitzer}/IRS maps II:\\ A direct link between band profiles and the radiation field strength}

\author{D.~J.~Stock\altaffilmark{1}, E.~Peeters\altaffilmark{1,2}}

\altaffiltext{1}{Department of Physics and Astronomy, University of Western Ontario, London, ON, N6A 3K7, Canada}
\altaffiltext{2}{Carl Sagan Center, SETI Institute, 189 Bernardo Avenue, Suite 100, Mountain View, CA 94043, USA}
\email{dstock84@gmail.com}

\begin{abstract}
We decompose the observed 7.7~\micron\ polycyclic aromatic hydrocarbon (PAH) emission complexes in a large sample of over 7000 mid-infrared spectra of the interstellar medium (ISM) using spectral cubes observed with the \textit{Spitzer}/IRS-SL instrument. In order to fit the 7.7~\micron\ PAH emission complex we invoke four Gaussian components which are found to be very stable in terms of their peak positions and widths across all of our spectra, and subsequently define a decomposition with fixed parameters which gives an acceptable fit for all the spectra. We see a strong environmental dependence on the inter-relationships between our band fluxes -- in the \HII\ regions all four components are inter-correlated, while in the reflection nebulae (RNe) the inner and outer pairs of bands correlate in the same manner as previously seen for NGC~2023. We show that this effect arises because the RNe maps are dominated by strongly irradiated PDR emission, while the much larger \HII\ region maps are dominated by emission from regions much more distant from the exciting stars, leading to subtly different spectral behavior. Further investigation of this dichotomy reveals that the ratio of two of these components (centered at 7.6 and 7.8 \micron) is linearly related to the UV field intensity (log G$_0$). We find that this relationship does not hold for sources consisting of circumstellar material, which are known to have variable 7.7~\micron\ spectral profiles. 
\end{abstract}


\section{Introduction}

It is now generally accepted that the `unidentified' mid-infrared (MIR) emission bands between 3 and 20~\micron\ are emitted by polycyclic aromatic hydrocarbon molecules (PAHs;  see e.g., \citealt{1984A&A...137L...5L,1985ApJ...290L..25A,1989ApJS...71..733A,1989ARA&A..27..161P}) as well as closely related carbonaceous species; e.g., PAH clusters, VSGs \citep{1984A&A...137L...5L,1989ApJS...71..733A,2005A&A...429..193R}.
A key facet of the PAH model is that the emission is generated by an ensemble of emitting PAH molecules, all, in principle, with different structures but generally comprised of fused, benzene-like, carbon rings, with a peripheral set of hydrogen atoms. The emission of this ensemble of PAH molecules takes the form of the aforementioned `unidentified' MIR bands, a collection of broad spectral features at around 3.3, 6.2, 7.7, 8.6, 11.2 and 12.7~\micron\ which dominates the infrared emission of many astronomical sources (as well as weaker bands at 5.2, 5.7, 6.0, 11.0, 12.0, 13.5 and 14.2 \micron). The bands can be classified along several lines to understand their emission, with the primary division being between neutral species, which emit primarily in the 3.3 and 11.2~\micron\ bands, and ionic species, which emit primarily in the 6 -- 9~\micron\ region (e.g. \citealt{1999ApJ...511L.115A}). However, it has been shown (e.g. \citealt{2002A&A...390.1089P, 2016ApJ...824..111S}) that there is additional substructure within the bands in the form of spectral profile variations, which suggests the existence of separate, but blended, emission components within each of the bands.

Observationally, these separate components can be examined by looking for changes in the PAH spectra between objects. The class `A', `B' and `C' profiles for the `7.7' PAH complex (i.e. the 7.4-8.0~\micron\ PAH emission) are defined by examining the spectral shape of the band, principally in terms of its peak position (from `A' being the shortest wavelength to `C' the longest; \citealt{2002A&A...390.1089P}). Given the fact that the `7.7'~\micron\ PAH complex emits roughly half of the PAH flux (e.g. \citealt{2002A&A...390.1089P, 2007ApJ...656..770S}), there have been many studies into its spectral and spatial behavior, including the creation of a class `D' profile (\citealt{2014MNRAS.439.1472M, 2014ApJ...791...28S}).  The initial studies of decomposing the `7.7'~\micron\ PAH complex using either Gaussians or Lorentzians found that there are several sub-components, with most the flux being present in those centered at around 7.6 and 7.8~\micron\ (\citealt{2001A&A...372..981V, 2002PhDT........63V, 2002A&A...390.1089P}). Subsequent studies by other groups used other approaches, notably the \citet{2005ApJ...632..956S, 2007ApJ...664.1144S} technique of determining the central wavelength of the feature (defined as the wavelength which exactly divides the flux in half). Motivated by these findings, \citet{P15} fit the 7--9~\micron\ PAH emission observed in each pixel of a \textit{Spitzer}/IRS spectral map of the reflection nebula\footnote{Abbreviated as RN (singular) or RNe (plural).} NGC~2023 with four Gaussian components, and found remarkable consistency in the peak positions and widths of the individual components between the various spectra. In addition it was found that the subcomponents correlated with each other slightly better than the equivalent measurements using the spline method, indicating that the Gaussian decomposition method better isolates the underlying emission components within the 7.7~\micron\ PAH complex. The four Gaussian components used by \citet{P15} were centered on approximately 7.6, 7.8, 8.2 and 8.6~\micron\ (which we refer to as `G7.6', `G7.8', `G8.2' and `G8.6' throughout). \citet{P15} found that the outer pair of bands, peaking at 7.6 and 8.6~\micron\ respectively, had the best correlation coefficient of their complete sample of bands, while the inner components also correlated with each other, but to a lesser degree. Upon further inspection, it was found that maps of the inner emission components seemed to correlate better with the continuum emission, rather than the other PAH bands.

In this paper we apply the \citet{P15} decomposition to a sample of \textit{Spitzer}/IRS SL spectral cubes of \HII\ regions and RNe in order to determine whether the Gaussian decomposition parameters found for NGC~2023 are applicable to other objects, specifically our sample of \HII\ regions. Furthermore we aim to discover whether the behavior of the various bands has any environmental dependence by comparing the results from the harsher environments around \HII\ region PDRs with those found for the RNe. The paper is organized as follows: Section~2 briefly describes the data sample and reduction process; Section~3 gives details of the fitting process; in Section~4 we describe the results; and in Section~5 we discuss our results in terms of the PAH emission from different environments. Finally, Section~6 summarizes our conclusions.

\section{Data} \label{sec:data}

Our data sample is taken from \citet[][hereafter Paper I]{S15} and consists of a sample of five \HII\ regions and star-forming complexes : W49A, IRAS 12063-6259, G37.55-0.11, M 17 and the Horsehead nebula as well as three RNe: NGC~1333 SVS3, NGC~2023 (south) and NGC~7023 NW\footnote{We will henceforth refer to these fields as simply NGC~1333, NGC~2023 and NGC~7023 for brevity.}, each of which was observed using the Spitzer-IRS instrument's low resolution SL module in the wavelength range 5 $< \lambda < 15$ \micron. We include NGC~2023 for the purposes of comparison to the results of \citet{P15} which exclusively considered NGC~2023. For discussion of the details of each source, in terms of its exciting stars, densities, radiation fields etc, please refer to Table~1 of Paper~I.

Paper~I also included the star forming region G11.94-0.62, however we have excluded it from this work as the presence of the methanol absorption in G11.94-0.62 leads to additional uncertainties in the dereddening process. We show a clear example of this absorption in Paper~I, Figure 3. As G11.94-0.62 presents the same spectral behavior as the other UC-\HII\ regions (i.e., similar PAH band correlation parameters etc; see paper~I), we chose to exclude it from the sample. 

Each spectral cube was constructed, extinction-corrected as described in Paper~I. Briefly, this involved:

\begin{enumerate}
\item cleaning the individual cubes of bad pixels and otherwise faulty data (e.g., hot pixels etc), 
\item scaling and stitching the cubes observed for different orders (i.e. SL1 and SL2) together,
\item resampling the cube using a 2$\times$2 aperture to roughly match the PSF\footnote{I.e., when we quote pixel coordinates thoughout the paper they must be doubled to match the same location in the original cube.},
\item correcting for extinction by quantifying the influence of the 9.8~\micron\ silicate feature via the modified Spoon method \citep{2013ApJ...771...72S}, 
\item defining and subtracting spline continua for each pixel,
\item measuring the strengths of the PAH emission features via either direct integration or Gaussian fits.
\end{enumerate}

This process resulted in a consistent sample of spectral maps and measurements for each object in our sample. 

In this work we disregard the $\sim$ 30 pixels in the W49A, G37.55-0.11 and IRAS~12063-6259 cubes associated with the peaks in the dust continuum emission (i.e. near the cores of the UC-\HII\ regions) as these pixels display strong rising dust emission continuum and aberrant 7.7~\micron\ PAH complex emission (very weak plateau and 8.2~\micron\ emission; see Figure~\ref{fig:spectra_w49a} for an example). These spectral features lead to poor continuum subtraction using our standard methodology described above and in detail in Paper~I. Instead, we discuss their unique spectral appearance in Section~\ref{sec:disc}. This simple strategy was slightly altered for IRAS~12063-6259, a dual lobed UC-\HII\ region (e.g., \citealt{2003A&A...407..957M}). It was found that the most intense continuum source in IRAS~12063-6259 (the northern lobe) did not display discrepant 7.7~\micron\ PAH complex profiles -- and hence our measurement approach was able to provide good continuum subtractions, while the southern lobe displayed discrepant PAH emission profiles leading to poor continuum subtraction. As such, we masked the southern pixels.

\section{Analyses}

\subsection{Decompositions, fits and models}

Before proceeding to a discussion of the various decomposition strategies we have used to ensure that the four component model is reliable, we will outline some of the underlying assumptions of the \citet{P15} four Gaussian decomposition method. Firstly, we assume that PAH emission at specific wavelengths is generated by the comparable populations of PAH molecules in different objects, i.e. that our Gaussian components have to be stable (i.e. low variability in peak position and width) in order to be meaningful, allowing comparison with the other PAH bands and their identifications. Secondly, whilst we use the mechanics and apparatus of model fitting, our four fixed parameter Gaussian components vary only in intensity and hence is really a decomposition, rather than a traditional `fit'. Indeed, previous higher spectral resolution studies, including \citet{Moutou:99} and \citet{2002A&A...390.1089P}, have shown that there are more than four components (e.g. the main peak at around 7.6~\micron\ is often sharper than can be fit using the four Gaussian approach). 

\subsection{Unconstrained Four Gaussian Fits} 

\begin{figure*}
	\begin{center}
	\includegraphics[width=18cm]{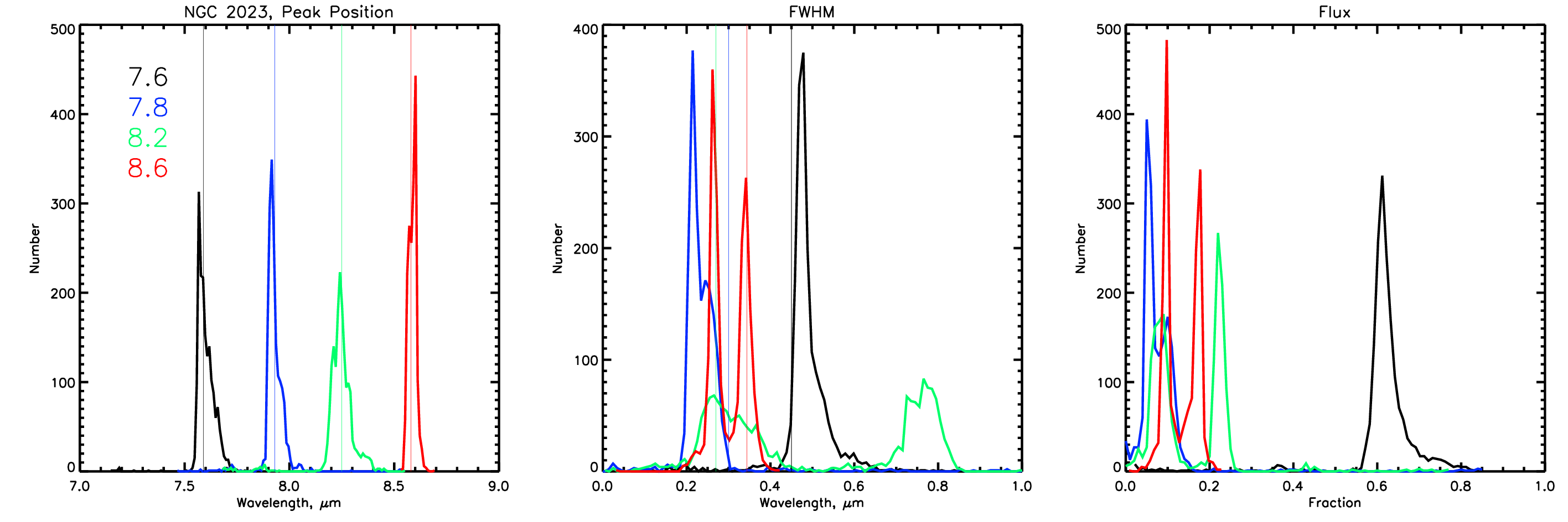}
	\includegraphics[width=18cm]{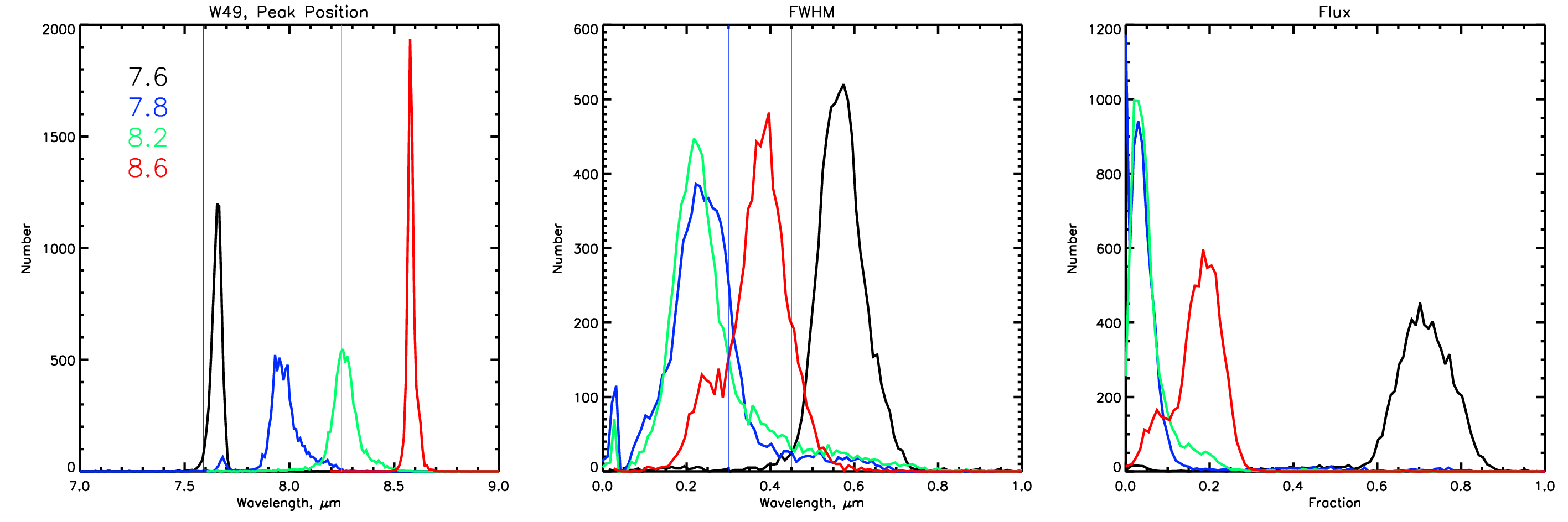}
	\end{center}
	\caption{Histograms (with boxes suppressed for clarity) showing the distribution of peak position, FWHM, and flux fraction for every pixel of the NGC~2023 and W49A spectral maps for an unconstrained fit. In each case the bin size is set to 0.01. Thin vertical lines show the parameters adopted by \citet{P15}.}
	\bigskip
	\label{fig:hists}
\end{figure*}

\begin{figure}
	\begin{center}
	\includegraphics[width=8cm]{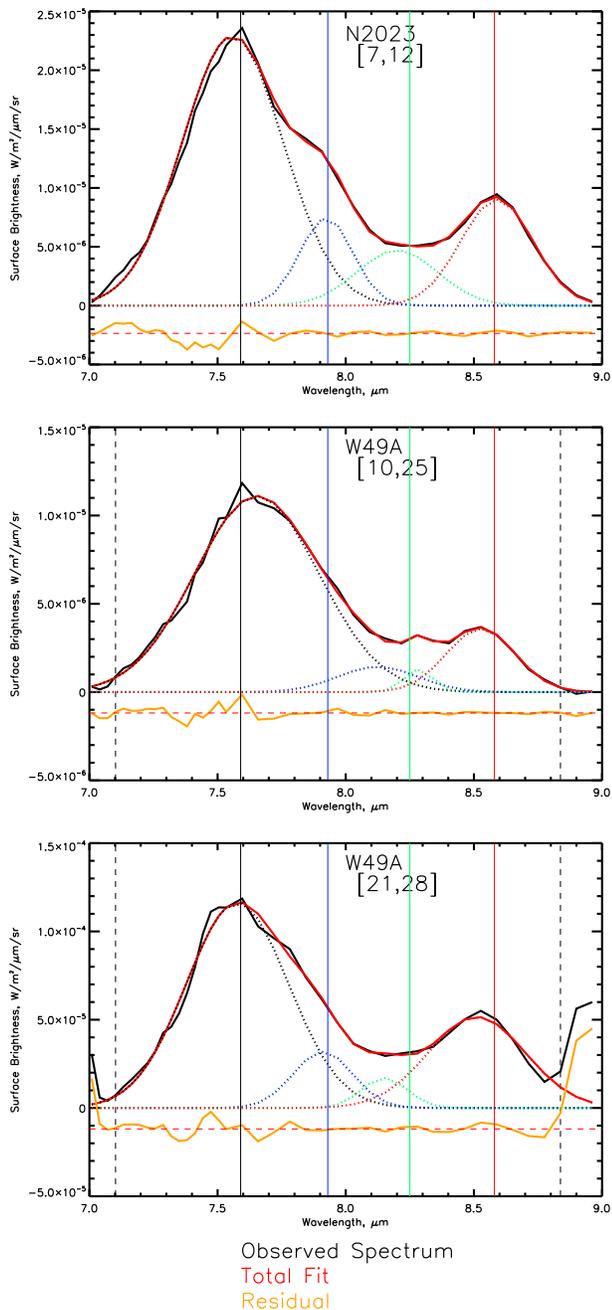}
	\end{center}
	\caption{Examples of fits to the 7.7~\micron\ PAH complex for NGC~2023 and W49A showing the two different `modes' of unconstrained fits. Central wavelengths adopted by \citet{P15} shown as vertical lines in the same colors as the four components: black, blue, green and red for the 7.6, 7.8, 8.2 and 8.6 Gaussians respectively. The top two panels display the `typical' modes for each source, while the bottom panel shows a W49A spectrum similar to the NGC~2023 spectrum.}
	\bigskip
	\label{fig:freefits}
\end{figure}

Firstly we investigate whether the relatively stable four-gaussian fit described by \citet{P15} is replicated in our sample (i.e. peaks at 7.6, 7.85, 8.2 and 8.6 \micron), or whether there are 7.7~\micron\ PAH complex spectra which give rise to a different set of components. We have tested this idea by performing unconstrained four-Gaussian fits on each spectral pixel in our maps. These fits use the IDL least-squares minimization routines prodvided by \citet{2009ASPC..411..251M} and take into account the flux uncertainties in the measured spectra. We will primarily discuss in details the results for NGC~2023 and W49A as they are representative of the RNe and \HII\ regions respectively. In Figure~\ref{fig:hists} we show histograms of the peak positions, widths and relative fluxes for the four components in each pixel of NGC~2023 and W49A.

At this stage it is clear that the central wavelengths adopted by \citet{P15} are consistent with those found in the top left panel of Figure~\ref{fig:hists}, however the feature widths differ slightly. In particular there is a degeneracy between wide and narrow 8.2 and 8.6~\micron\ components respectively. \citet{P15} chose to break this degeneracy by requiring that the 8.2~\micron\ component have the smaller FWHM such that the 8.2~\micron\ component is now allowed to become a pseudo-plateau. This choice breaks the degeneracy between the 8.2 and 8.6 \micron\ Gaussian widths and recovers the parameters given by \citet{P15}.

It is clear from the bottom panels of Figure~\ref{fig:hists} that the average behaviour for W49A is different from NGC~2023. Each of the components adopts a similar average wavelength to that of NGC~2023 except the 7.6~\micron\ Gaussian, which is significantly redshifted. The strongest difference is with the feature widths, with there being a much higher degree of variability in the Gaussian FWHMs leading to much wider histograms. This reflects the fact that the fits for W49A follow a different pattern from those in NGC~2023, with the 7.6~\micron\ Gaussian moving to dominate the peak of the 7.7~\micron\ PAH complex, and the 7.8 and 8.2~\micron\ components moving to the red to better fit the trough between the 7.7~\micron\ peak and the 8.6~\micron\ peak. We show comparisons of the typical NGC~2023 and W49A fits in Figure~\ref{fig:freefits}. 

Inspection of Figure~\ref{fig:freefits} explains the differences seen in Figure~\ref{fig:hists}. The W49A spectrum shown in the middle panel of Figure~\ref{fig:freefits} shows little trace of the red shoulder to the main 7.7~\micron\ PAH complex peak which is so clear in the NGC~2023 spectrum. Tellingly though, the 7.6~\micron\ Gaussian from the W49A fit is much wider than that of the NGC~2023 fit (which is true on average, see Figure~\ref{fig:hists}) which can be understood in two distinct ways. Firstly, it could be that the \HII\ region emission is fundamentally different from that of the RNe, such that the 7.6~\micron\ component we see is a different shape. Alternatively it could be that the 7.6 and 7.8~\micron\ Gaussian components combine to produce a wider yet still roughly Gaussian shaped feature. The first explanation is difficult to reconcile with the existence of spectra in the W49A cube which are similar to that seen in NGC~2023, i.e. that they show a pronounced `knee' on the red side of the 7.7~\micron\ PAH band (e.g. Figure~\ref{fig:freefits}, bottom panel). As such, we conclude that there is sufficient evidence to justify assuming that the main body of the 7.7~\micron\ PAH complex is comprised of two separate components in both environments. However, as the best `free' fit for some spectra will fit the entire 7.7~\micron\ peak with a single Gaussian component, it is necesary to constrain the peak positions and widths of the features.

\subsection{Constrained Fits}

\begin{figure*}
	\begin{center}
	\includegraphics[width=18cm]{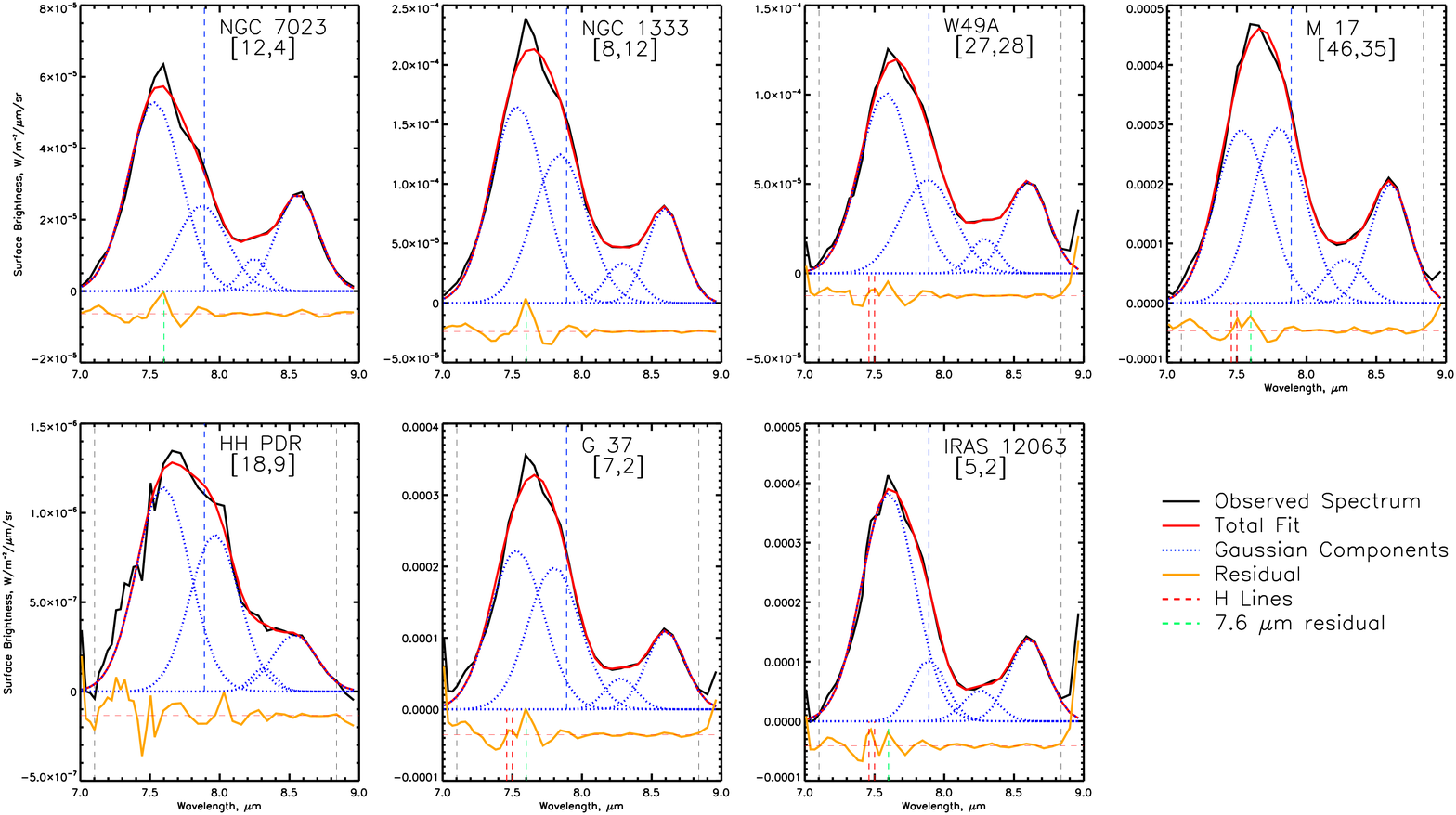}
	\end{center}
	\caption{Example decompositions of the 7.7 PAH complex using four constrained Gaussian components for each object except NGC~2023 which is shown in Figure~\ref{fig:comp_comp}. In each case the black line represents the data, the red line the overall fit (within the boundaries of the vertical black dashed lines for the \HII\ regions so as to exclude atomic lines); the blue dashed lines the individual Gaussian components; the yellow line the residual offset for clarity to the level of the dashed red horizontal line; the red dashed vertical lines showing residual features common to multiple objects; and the blue vertical line showing the average central wavelength of the 7.8~\micron\ feature for the RNe to illustrate the variance amongst the \HII\ region spectra. }
	\bigskip
	\label{fig:decomp}
\end{figure*}

Taking the \citet{P15} fit parameters as a starting point, we fitted the 7.7~\micron\ PAH emission complex in each spectrum of each cube for a total of 7052 spectra. In each spectrum we allowed the central wavelengths and widths to vary slightly for all four parameters (typically in a 0.2~\micron\ window for the peak position and around a 0.25~\micron\ window for the FWHM). The fluxes of each component were required to be positive. Hereafter we refer to this process as the `constrained' fits. Examples of typical fits for each object are shown in Figure~\ref{fig:decomp}. There appear to be two features in the residuals for our fits that are relatively prominent and repeated in several objects. The first of these, appearing at a wavelength of around 7.46~\micron\ for the \HII\ regions exclusively, appears to be the blended Pf$\alpha$ (6-5; 7.46 \micron) and Hu$\beta$ (8-6; 7.50 \micron) hydrogen recombination lines, with Pf$\alpha$ dominating in flux. The second prominent residual, at around 7.6 \micron, does not correspond to any known emission line, rather, it is likely to represent the sharp peak of the 7.6~\micron\ component which is seen in some sources (see e.g., \citet{2002A&A...390.1089P} for higher resolution ISO-SWS spectra of typical ISM sources). The flux and width of the 7.6~\micron\  component is dominated by the broader underlying emission and the steepness of the rising flux on the blue side, as such our Gaussian fits cannot capture the additional sharp peak seen at 7.6~\micron\ without severely compromising the rest of the fit.

The results of this process, along with uncertainties, are presented in Table~\ref{tables:params}. The uncertainties were calculated for each object by calculating the standard deviation of the set of values, i.e. the actual scatter of the data points rather than the fitting uncertainties which were generally very small. At the foot of the table, we have included the averages for all of the objects along with their uncertainties, again, calculated using the standard deviation of the set of best fits rather than combining all of the individual uncertainties. For NGC~2023, we recover the same parameters as \citet{P15} within our uncertainties, however when compared to the other objects, we see that the 7.6~\micron\ component of the NGC~2023 data is shifted to the red slightly. We have also included the averages for the \HII\ regions and RNe samples.

\subsection{Final Decomposition}\label{sec:genfits}

\begin{figure}
	\begin{center}
	\includegraphics[width=8cm]{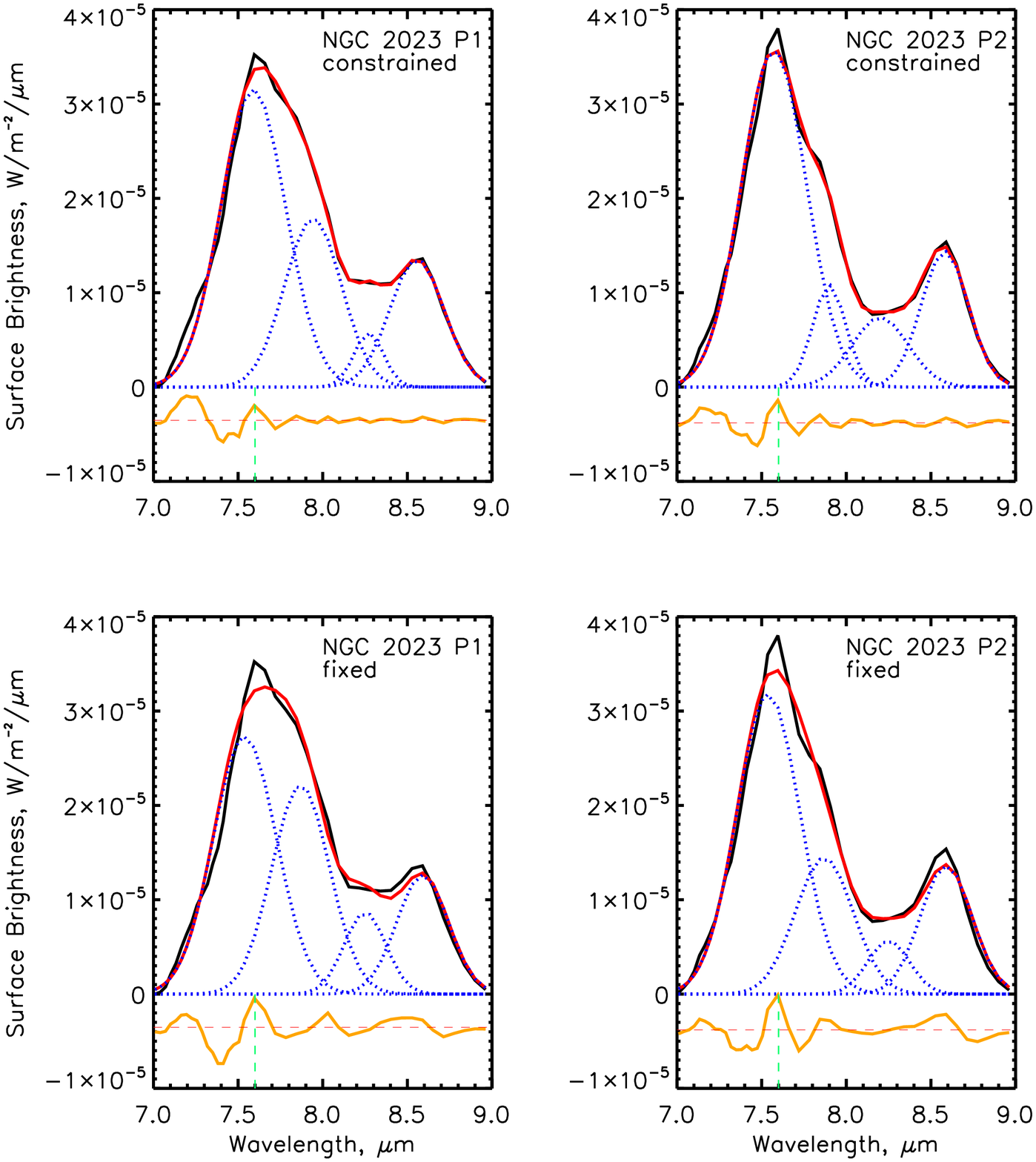}
	\end{center}
	\caption{Comparison of the fits produced by the constrained method and those produced using the average fit parameters using two spectra of NGC~2023 which showed the largest (P1; [17, 8]) and smallest (P2; [12, 4]) widths of the 7.8 feature in the constrained fits. In each case the black line represents the data, the red line the overall fit, the blue dashed lines the individual Gaussian components and the yellow line the residual (offset to the zero level represented by the red dashed line for clarity). }
	\bigskip
	\label{fig:comp_comp}
\end{figure}

We then re-ran the fitting procedure for each object with the peak positions and full width half maxima fixed at the average value presented in Table~\ref{tables:params} for every pixel of each cube, i.e., only allowing the flux in each component to vary. Figure~\ref{fig:comp_comp} shows the differences between the constrained fits and the final decomposition (i.e. with fixed parameters) for two NGC~2023 spectra with the highest (P1) and lowest (P2) widths for the 7.8 feature. The component fluxes used throughout the rest of this paper refer to those measured this way. For NGC~2023, the process makes the quality of the fit to the peak of the 7.7~\micron\ PAH complex somewhat worse because the fixed fit central wavelengths for the 7.6 and 7.8 components are somewhat bluer than we found for unconstrained fits. However, each of the bottom two panels in Figure~\ref{fig:comp_comp} (those with fixed $\lambda$ and FWHM) still show reasonable fits even though the magnitude of the residuals has increased. For the other sources this effect is less strong as their spectra are generally much closer to the average fit parameters.

\begin{table*}
\caption{\label{tables:params} Decomposition Parameters}
\begin{center}
\begin{tabular}{c c c c c c}
\toprule
				&  &\multicolumn{4}{c}{Components} \\
Object	 & Property			& G7.6	& G7.8	& G8.2	& G8.6 \\
\midrule
\\
\multicolumn{2}{l}{\textit{\citet{P15}:}}\\ \midrule
\\
NGC~2023	& Central $\lambda$ [\micron]	& 7.59	& 7.93	& 8.25	& 8.58	\\
		& FWHM [\micron]		& 0.45	& 0.3	& 0.27	& 0.344	\\

\\
\multicolumn{2}{l}{\textit{This Work: RNe}}\\ \midrule
\\
NGC~2023	& Central $\lambda$ [\micron]	& 7.58 $\pm$ 0.02 & 7.92 $\pm$ 0.02 & 8.25 $\pm$ 0.04 & 8.58 $\pm$ 0.01\\
		& FWHM [\micron]		& 0.45 $\pm$ 0.01 & 0.32 $\pm$ 0.07 & 0.30 $\pm$ 0.07 & 0.34 $\pm$ 0.02\\
\\
NGC~7023	& Central $\lambda$ [\micron]	& 7.53 $\pm$ 0.02 & 7.89 $\pm$ 0.03 & 8.24 $\pm$ 0.03 & 8.57 $\pm$ 0.02\\
		& FWHM [\micron]		& 0.45 $\pm$ 0.01 & 0.40 $\pm$ 0.07 & 0.26 $\pm$ 0.07 & 0.36 $\pm$ 0.02\\
\\
NGC~1333	& Central $\lambda$ [\micron]	& 7.57 $\pm$ 0.06 & 7.89 $\pm$ 0.05 & 8.26 $\pm$ 0.04 & 8.60 $\pm$ 0.04\\
		& FWHM [\micron]		& 0.42 $\pm$ 0.06 & 0.37 $\pm$ 0.09 & 0.29 $\pm$ 0.09 & 0.37 $\pm$ 0.08\\
\\
\multicolumn{2}{l}{\textsl{This Work: H~\textsc{ii} regions}}\\ \midrule
\\
W49A		& Central $\lambda$ [\micron]	& 7.54 $\pm$ 0.03 & 7.84 $\pm$ 0.04 & 8.25 $\pm$ 0.04 & 8.59 $\pm$ 0.02\\
		& FWHM [\micron]		& 0.45 $\pm$ 0.01 & 0.43 $\pm$ 0.04 & 0.32 $\pm$ 0.09 & 0.37 $\pm$ 0.06\\
\\
M 17		& Central $\lambda$ [\micron]	& 7.54 $\pm$ 0.03 & 7.83 $\pm$ 0.04 & 8.25 $\pm$ 0.02 & 8.59 $\pm$ 0.02\\
		& FWHM [\micron]		& 0.45 $\pm$ 0.01 & 0.44 $\pm$ 0.03 & 0.29 $\pm$ 0.08 & 0.37 $\pm$ 0.03\\
\\
Horsehead	& Central $\lambda$ [\micron]	& 7.55 $\pm$ 0.06 & 7.89 $\pm$ 0.06 & 8.26 $\pm$ 0.05 & 8.61 $\pm$ 0.04\\
		& FWHM [\micron]		& 0.45 $\pm$ 0.03 & 0.37 $\pm$ 0.09 & 0.29 $\pm$ 0.09 & 0.34 $\pm$ 0.09\\
\\
G37.55-0.11 	& Central $\lambda$ [\micron]	& 7.54 $\pm$ 0.04 & 7.84 $\pm$ 0.05 & 8.25 $\pm$ 0.03 & 8.59 $\pm$ 0.03\\
		& FWHM [\micron]		& 0.44 $\pm$ 0.03 & 0.43 $\pm$ 0.06 & 0.28 $\pm$ 0.08 & 0.39 $\pm$ 0.05\\
\\
IRAS 12063-6359	& Central $\lambda$ [\micron]	& 7.56 $\pm$ 0.03 & 7.86 $\pm$ 0.03 & 8.24 $\pm$ 0.03 & 8.59 $\pm$ 0.03\\
		& FWHM [\micron]		& 0.44 $\pm$ 0.04 & 0.38 $\pm$ 0.07 & 0.26 $\pm$ 0.09 & 0.38 $\pm$ 0.03\\
\\
\multicolumn{2}{l}{\textit{Sample Averages:}}\\
\midrule
\\
\HII\ Average   & Central $\lambda$ [\micron]	& 7.55 $\pm$ 0.01 & 7.85 $\pm$ 0.02 & 8.25 $\pm$ 0.01 & 8.59 $\pm$ 0.01\\
                & FWHM [\micron]		& 0.45 $\pm$ 0.01 & 0.41 $\pm$ 0.03 & 0.29 $\pm$ 0.02 & 0.37 $\pm$ 0.02\\
\\
RNe Average     & Central $\lambda$ [\micron]	& 7.56 $\pm$ 0.03 & 7.90 $\pm$ 0.02 & 8.25 $\pm$ 0.01 & 8.58 $\pm$ 0.02\\
                & FWHM [\micron]		& 0.44 $\pm$ 0.02 & 0.36 $\pm$ 0.04 & 0.28 $\pm$ 0.02 & 0.36 $\pm$ 0.02\\
\\
Average		& Central $\lambda$ [\micron]	& 7.55 $\pm$ 0.02 & 7.87 $\pm$ 0.03 & 8.25 $\pm$ 0.01 & 8.59 $\pm$ 0.01\\
		& FWHM [\micron]		& 0.44 $\pm$ 0.01 & 0.40 $\pm$ 0.04 & 0.29 $\pm$ 0.02 & 0.36 $\pm$ 0.02\\
\bottomrule
\end{tabular}
\end{center}

\end{table*}

\section{Results}

\subsection{Constrained Fits}

The key result from Table~\ref{tables:params} is that there appears to be a general decomposition which works for each of our sources with only very small scatter which we will refer to as the `fixed' fits or `fixed' decomposition. Secondly, the most distant outlier from the sample properties is NGC~2023 itself, which has the 7.6 and 7.8~\micron\ bands shifted to the red slightly ($\sim$0.03~\micron\ relative to the average) yet has similar widths compared to the other sources (although the 7.8 is actually consistent with a particular subset of the sample as will be discussed shortly). Interestingly, the G8.2 and G8.6 features adopt the same central wavelengths for NGC~2023 as the rest of our sample ($\lambda$ = 8.25, 8.58 \micron), so it is unlikely that it is a calibration error with the wavelength scale.

Of our four components, the 7.6, 8.2 and 8.6~\micron\ bands are very stable, with uncertainties of the order 0.01-0.02~\micron\ for peak position and width. The parameters of the 7.8~\micron\ component on the other hand have much more scatter, which is reflected in the uncertainties being around twice that found for the other bands. In fact, the band peak position seems to roughly segregate by object type (this observation is further investigated in Section~\ref{sec:disc78}). The 7.8 bands in the RNe and the Horsehead adopt an average value of around 7.89 \micron, while in the \HII\ regions, it appears systematically lower, at around 7.84 \micron. This effect is also slightly apparent in the widths, with the RNe having slightly narrower 7.8~\micron\ components than the \HII\ regions, however this relationship is less clear than for the peak positions.

Therefore, only the 7.6 band of NGC~2023 is actually discrepant as the central wavelength of the 7.8~\micron\ band is consistent within errors with the other RNe measurements. We have investigated whether faulty stitching of the SL2 and SL1 orders (this join actually occurs within the blue wing of the 7.7 PAH complex) of the IRS-SL instrument could be to blame, i.e., that there could have been a one to two pixel offset between the wavelength and data arrays. However this appears not to be the case, as there would be a consistent wavelength offset for all of the blueward features of the 7.7~\micron\ complex, which is not observed. \citet{P15} comment on slight wavelength offsets within the same NGC~2023 IRS-SL cube, and find that while there are measurable wavelength offsets of order 0.01 \micron, these offsets affect the entire spectrum and not one small part of it. As such, we regard the slight difference between the blue wing present in the NGC~2023 set of 7.7~\micron\ PAH complexes and the rest of our sample as genuine. Using the fixed decomposition derived from the average of all of the bands results in a slight overfitting of the blue wing of the 7.7~\micron\ PAH feature (see e.g., Figure~\ref{fig:comp_comp} and the discussion in Section~\ref{sec:genfits}). 

On the basis of these results we will henceforth group the sources as follows: \HII\ regions (W49A, IRAS 12063, G37.55-0.11); RNe (NGC~1333, 2023). The remaining sources (NGC~7023, M 17 and the Horsehead) we present separately on the basis that they were observed to behave peculiarly in Paper I. NGC~7023 displays a diffuse component of primarily ionized PAH spectra (e.g., depressed 11.2~\micron\ band fluxes; Paper I, \citealt{2014ApJ...795..110B}); the Horsehead, despite being classified as an \HII\ region, seemed to behave more like the RNe in terms of its PAH characteristics, perhaps due to its large separation from the exciting star and hence lower G$_0$ value than the other \HII\ regions; and M 17 was observed to have different gradients in some PAH band ratio plots than the other \HII\ regions.

\subsection{Flux Correlations}

\begin{figure*}
  \centering
    \includegraphics[width=18cm]{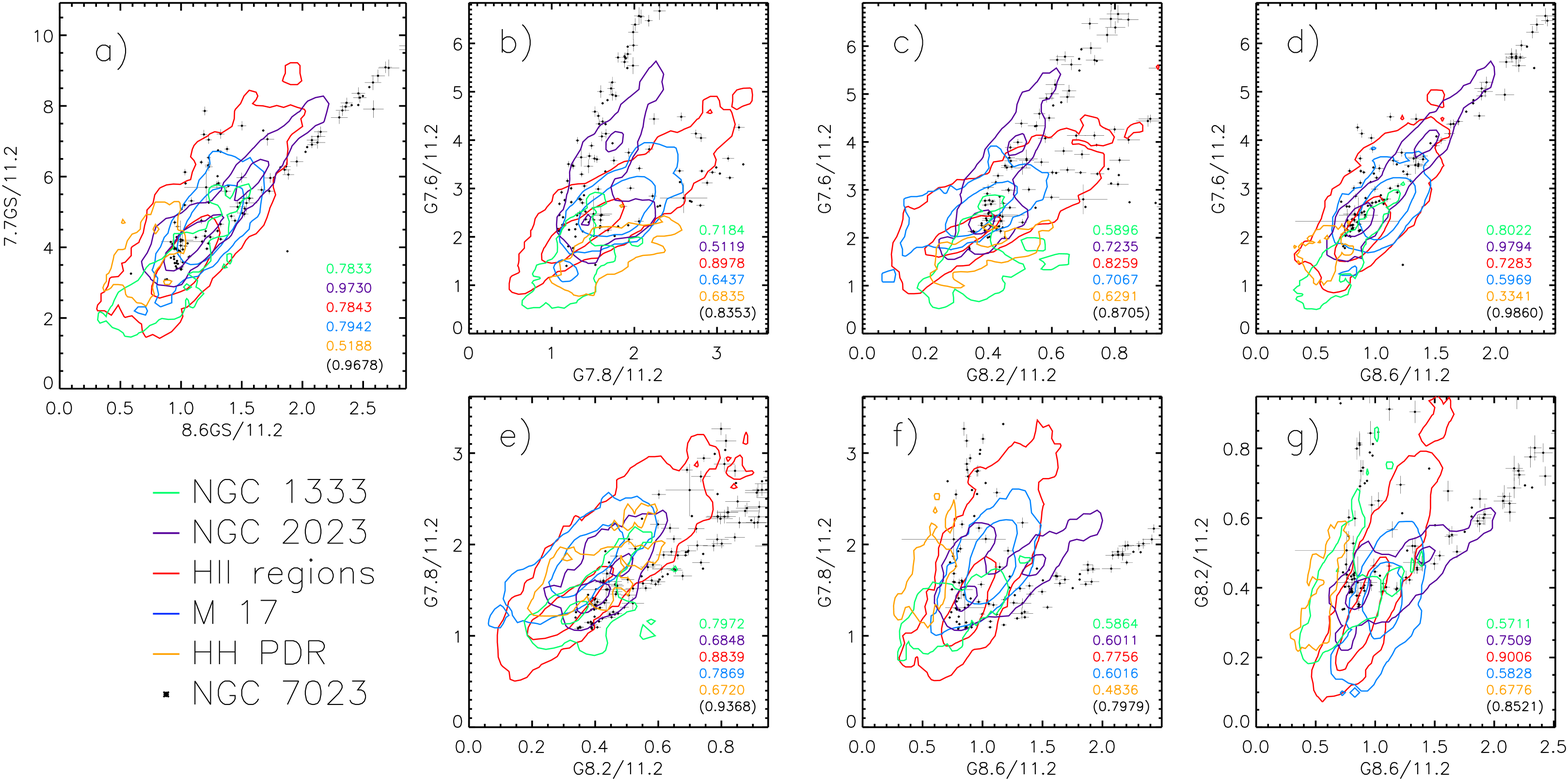}
    \caption{Inter-correlations between the four Gaussian components making up the 7.7~\micron\ PAH feature (b-g), with the equivalent plot using the traditional spline method band measurements  7.7/11.2 vs. 8.6/11.2 plot (see Paper I) included to give a sense of the expected scatter (a). Each dataset (except NGC~7023) has been presented as slightly smoothed point density contours (with levels at [0.5, 5], except for the \HII\ regions sample for which levels of [0.5, 10] were adopted) for clarity, for details please see the main text. Correlation coefficients are given in each plot color-coded to the relevant dataset. }
    \label{fig:plots}
\end{figure*}

\paragraph{General Comments} Using the flux measurements derived from the fixed decomposition, we can investigate how the integrated band strengths vary with respect to each other. The classical correlation plot is created by `normalizing' the band strengths of each spectrum via dividing by the flux of a band measured in that spectrum, typically the 11.2~\micron\ band. In our Figures, such measurements are denoted by a ratio, e.g., 6.2/11.2 meaning the 6.2~\micron\ PAH band strength divided by the 11.2~\micron\ band strength. Throughout this Section we have used the fluxes measured in Paper~I using direct integration for the 6.2 and 11.2~\micron\ PAH features. For comparison we also include the 7.7 and 8.6~\micron\ general spline, or `GS', PAH band measurements described in Paper~I to show the results of the `traditional' spline decomposition. Where we discuss correlation coefficients, these refer to Pearson weighted correlation coefficients taking into account both $x$ and $y$ uncertainties.

\paragraph{Inter-Correlations of the four components} Figure~\ref{fig:plots} shows an array of traditional correlation plots for the four components of the 7.7~\micron\ PAH band normalized to the 11.2~\micron\ band flux. The first panel presents one of the most well-known PAH band correlations, 7.7/11.2 vs. 8.6/11.2, to give the reader a sense of the quality of the correlations arising from the traditional analysis. In each case it has been necessary to reduce the points to contours of point density for the sake of clarity. This process is identical to that described in Paper I. Initially, the points were binned into a grid of 100 bins for each object in $x$ and $y$ spanning the dynamic range of the data and subsequently smoothing using a 3 pixel boxcar average routine. We show these data as contours, with levels of [0.5,5] points per bin in each case with the exception of the \HII\ regions sample, where we used levels of [0.5, 10] points per bin, reflecting the much larger number of points within the \HII\ region dataset. Throughout this process we only use data for which we have a SNR of at least 3 for each band measurements and 5 for the resulting band ratios.

Starting with NGC~2023, the behavior seen by \citet{P15} is recovered, in that the outer bands (7.6 and 8.6) correlate very well (correlation coefficient of 0.9794), and the inner bands correlate somewhat well (correlation coefficient of 0.6848). NGC~1333 seems to follow this pattern, although to a less extreme degree, with the inner and outer bands following each other to roughly the same degree (corr. coef. $\sim$ 0.8) but with obvious traces of the 7.6 and 7.8 bands being related (e.g., correlation coefficient of $\sim$0.7). The last of the RNe, NGC~7023, is somewhat difficult to compare to NGC~1333 and NGC~2023 as it possesses a large gradient in the flux of the 11.2~\micron\ PAH band across the map (see \citealt{2013ApJ...769..117B,2014ApJ...795..110B,2015ApJ...811..153S}, Paper I) which stretches the dynamic range of the correlation by a factor of 2-3. It does appear that NGC~7023 displays the same pattern as NGC~2023 and NGC~1333, however it is uncertain as to how meaningful the absolute correlation coefficients are in this case (as they are amplified by the larger range of 11.2 values) hence we include them in brackets in Figure~\ref{fig:plots}. Relative to each other, the NGC~7023 correlation coefficients display the same pattern as the other RNe, with the 7.6 and 8.6~\micron\ bands having the best correlation coefficient and the 7.8 and 8.2~\micron\ bands correlating, albeit to a lesser degree.

In stark contrast to the RNe, the \HII\ regions do not seem to follow the inner/outer pattern. Indeed, for the collected \HII\ sample, the outer correlation (7.6 vs. 8.6) is actually the \textit{worst} of all of the correlations examined in Figure~\ref{fig:plots}, albeit with a reasonable correlation coefficient of 0.72\footnote{The \HII\ region sample is dominated by the points from W49A, however the behavior described in the text is general and holds for each of the three regions collectively referred to as `\HII\ regions' when considered individually.}. However, the correlation coefficients for the \HII\ sample never drop below this value, and in fact seem to be highest for adjacent bands, i.e. 7.6 vs. 7.8, 7.8 vs. 8.2 and 8.2 vs. 8.6, with correlation coefficients around 0.9 in each case. The same pattern is seen (although more weakly) for the Horsehead and M~17 data. Further emphasis of this difference between the RNe and \HII\ regions can be added by inspecting the RNe contours in Figure~\ref{fig:plots} (panel C), which shows the correlation between the 7.6 and 8.2 components. In this figure the contours for each of the RNe are bifurcated, i.e. they show two distinct gradients, while the \HII\ region points (including M~17 and the Horsehead) do not show any such splitting.

\subsection{Emission Maps}

\begin{figure*}
  \centering
    \includegraphics[width=8cm]{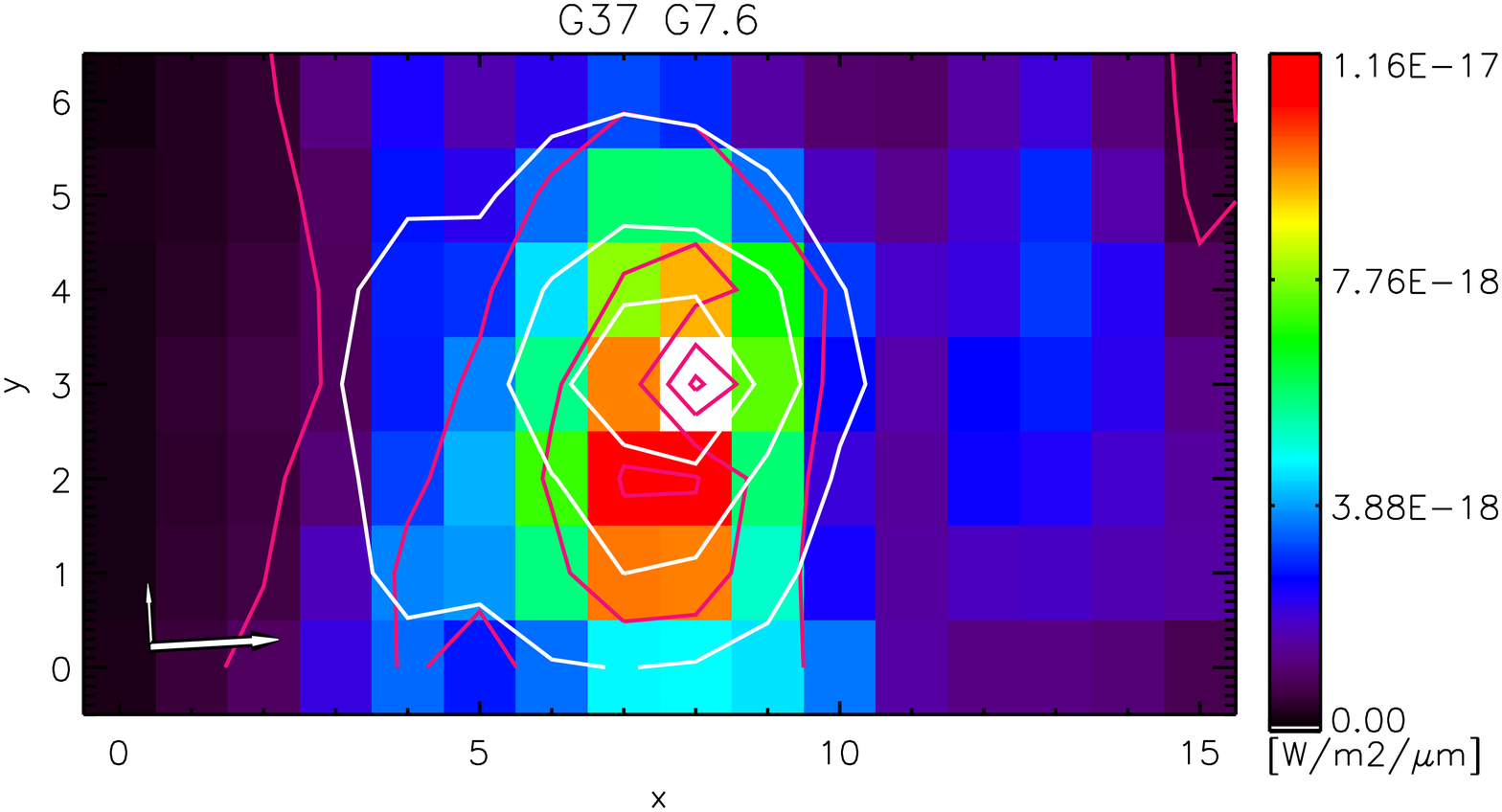}
    \includegraphics[width=8cm]{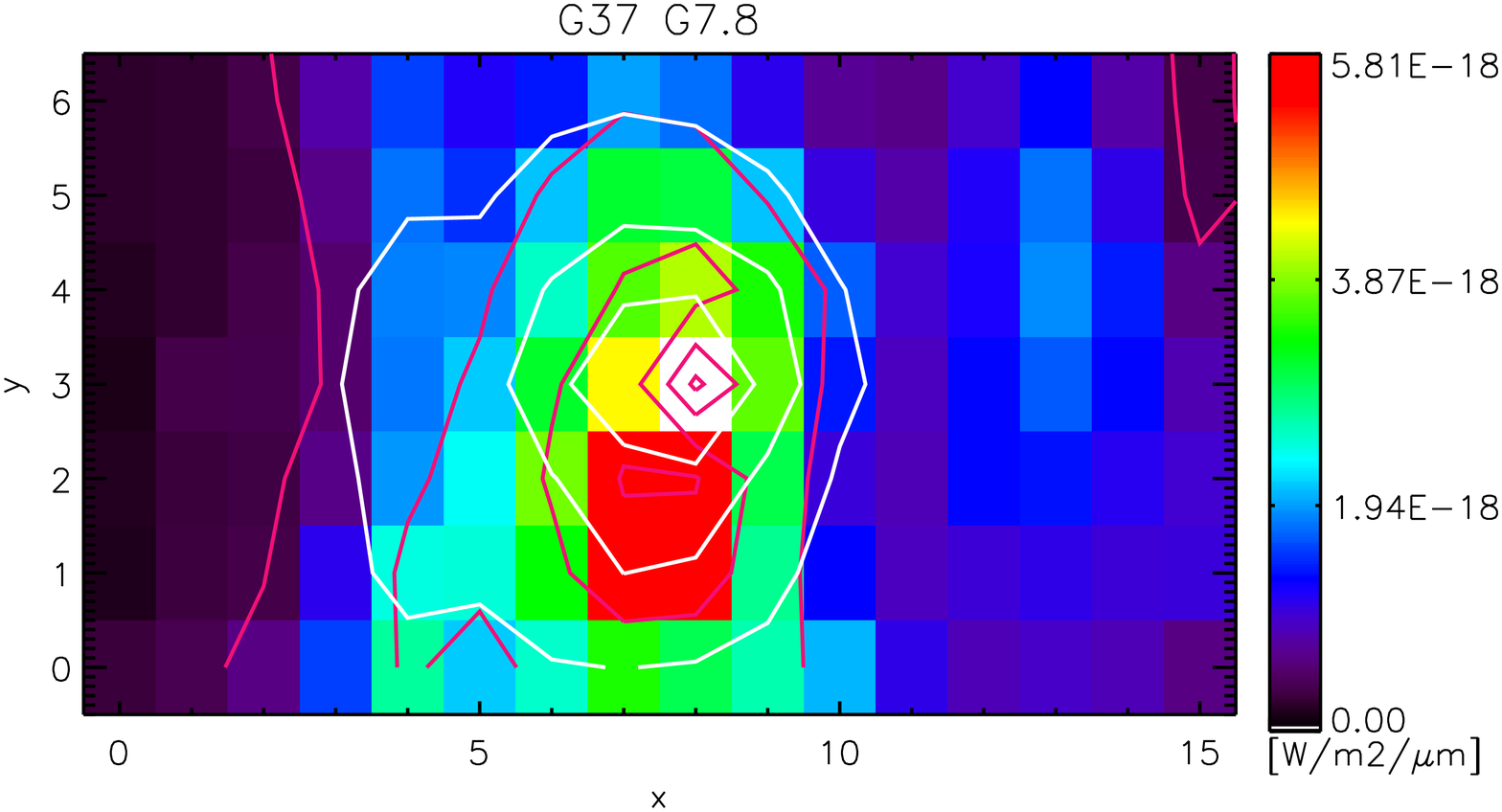}
    \includegraphics[width=8cm]{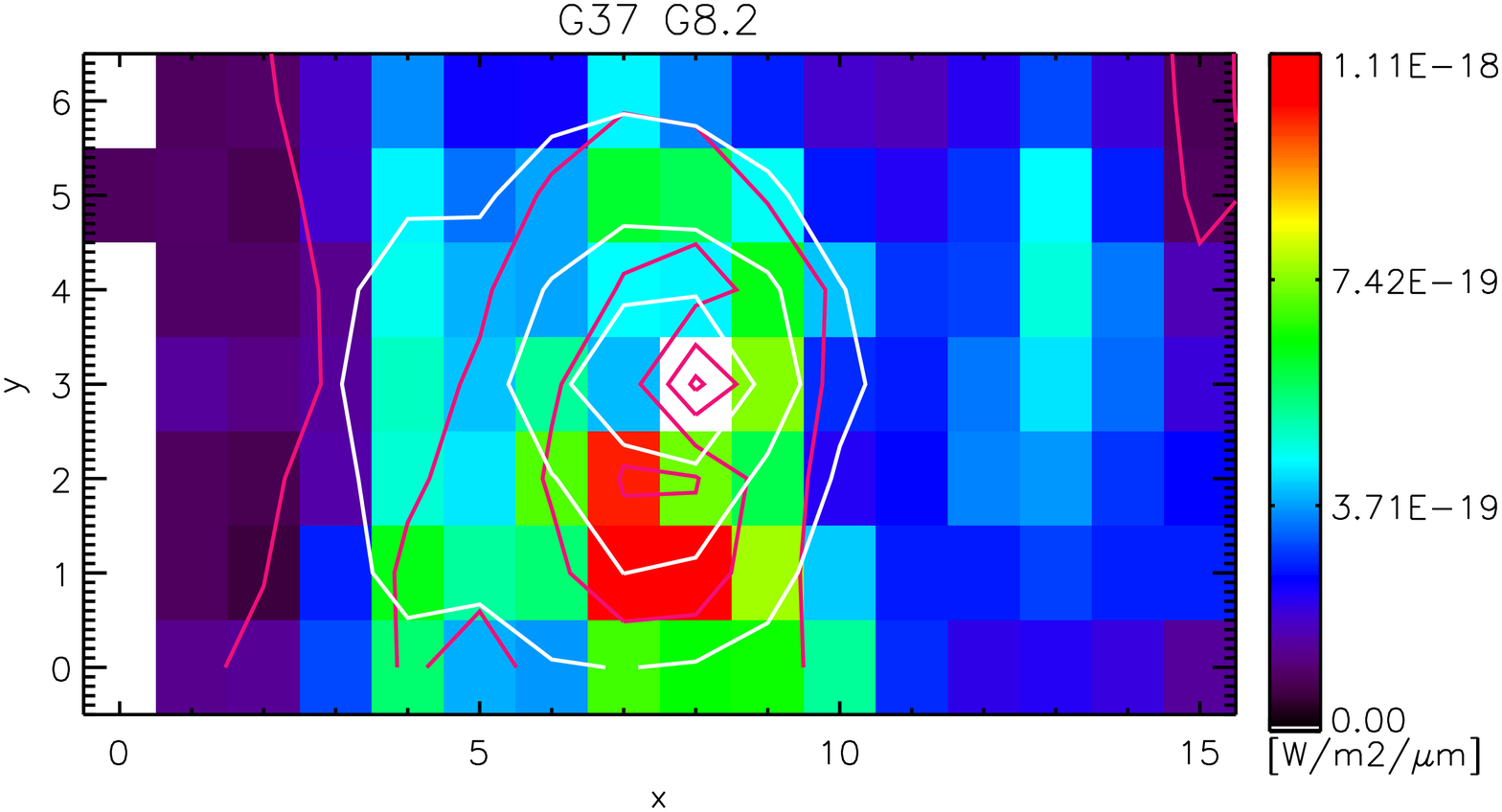}
    \includegraphics[width=8cm]{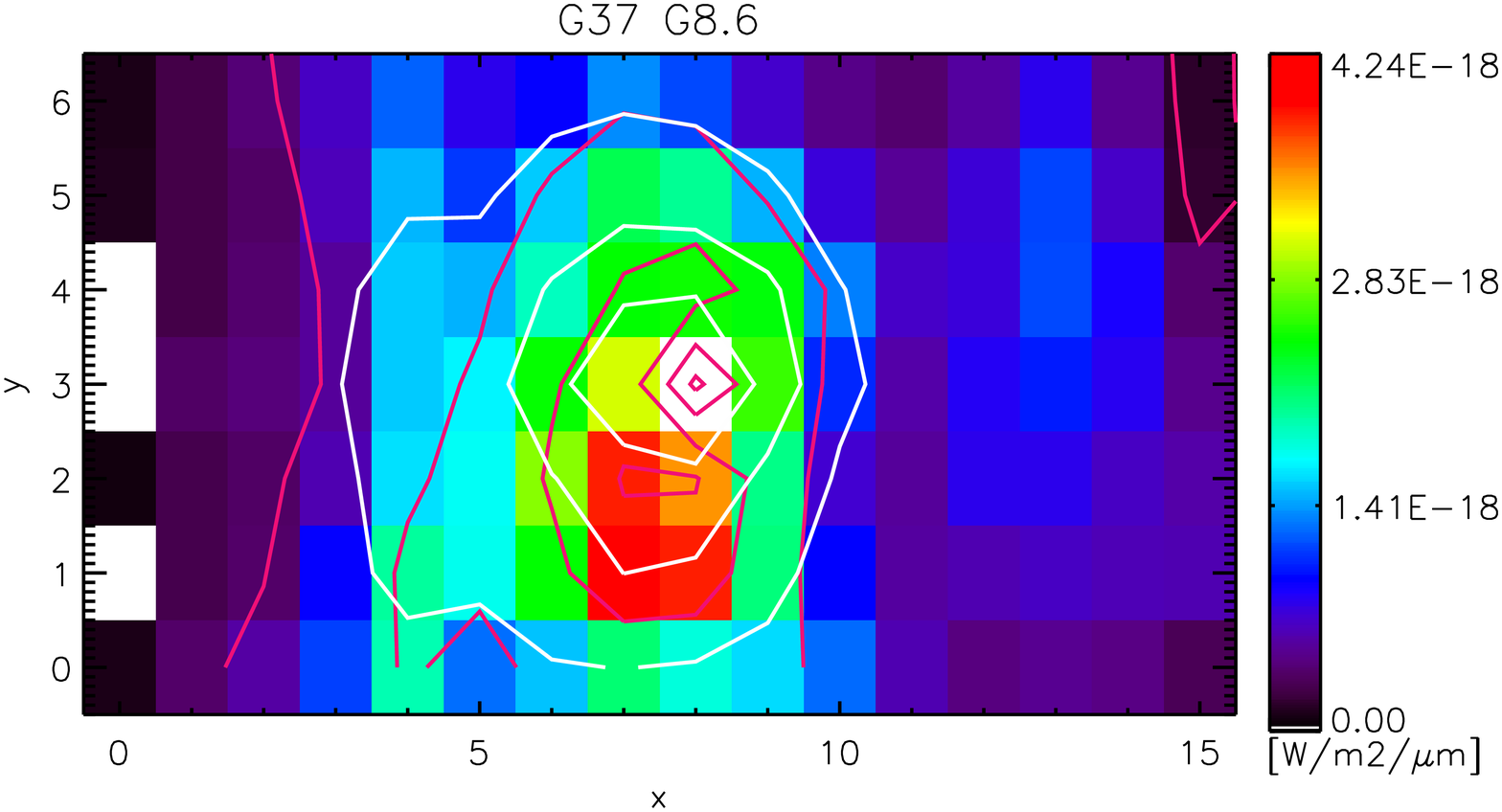}
    \includegraphics[width=8cm]{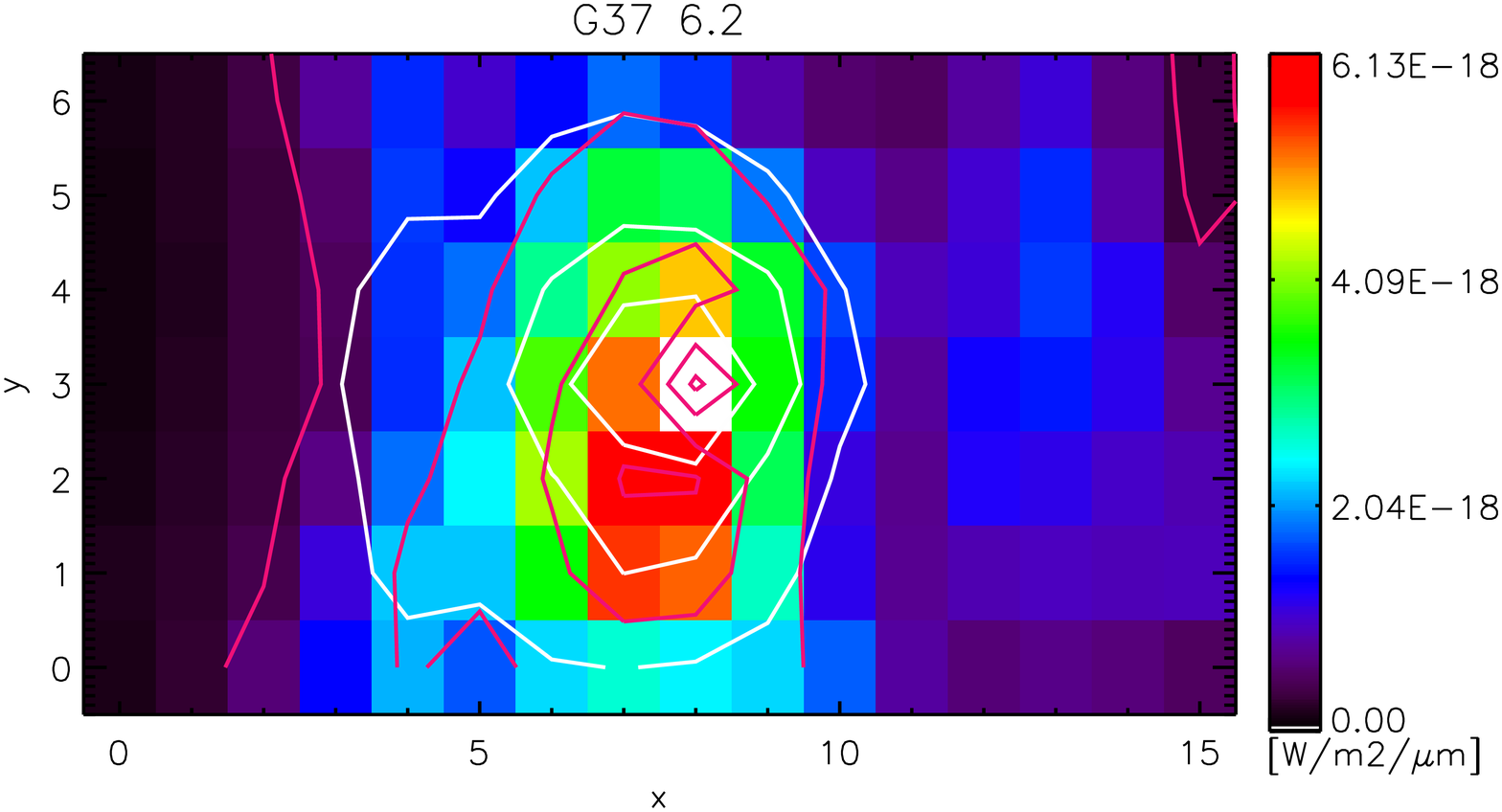}
    \includegraphics[width=8cm]{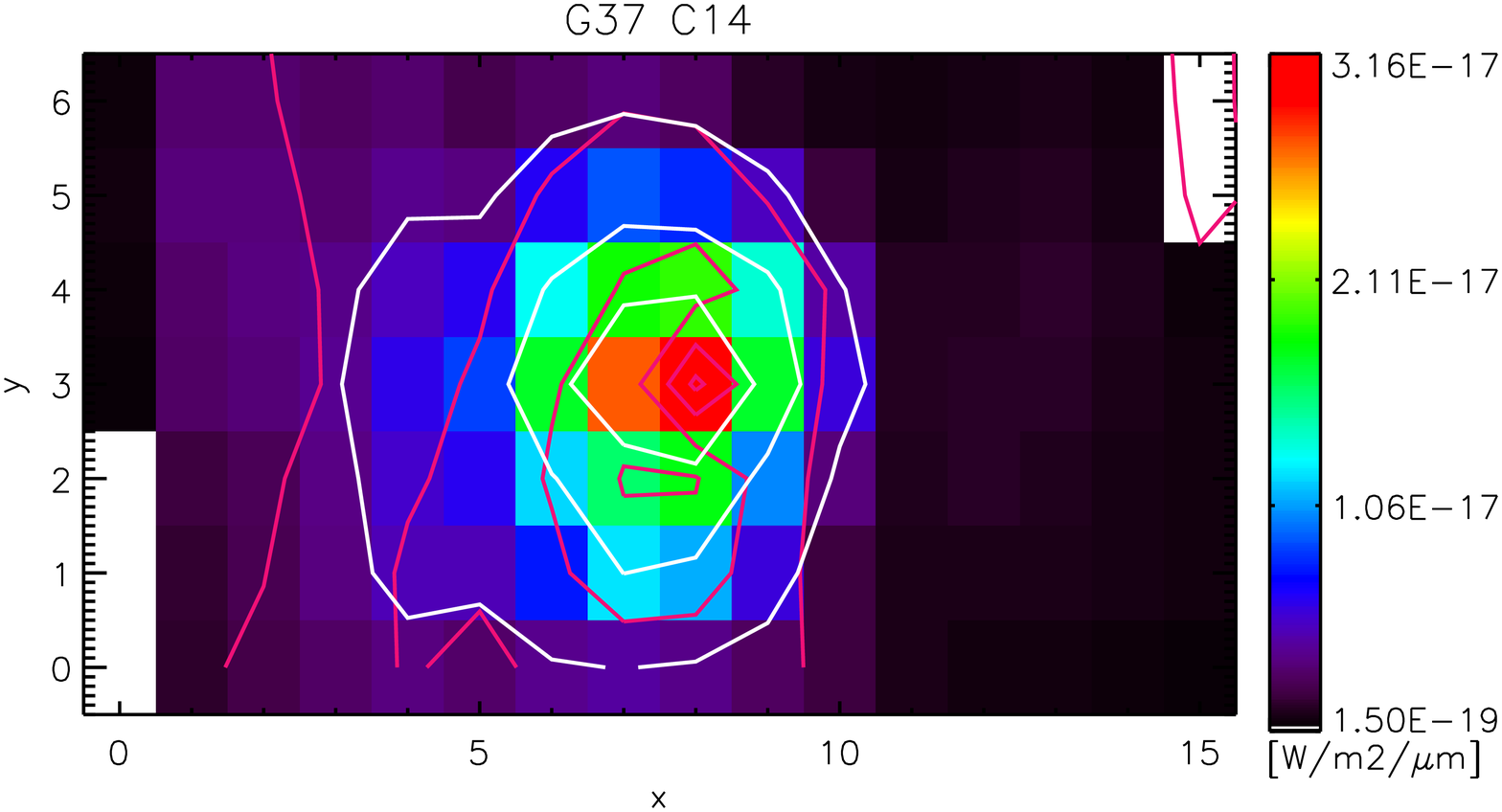}
    \caption{G37.55-0.11 spatial emission maps: (top two rows) the four components making up the 7.7~\micron\ PAH emission complex; (bottom row) maps of the `pure' PAH 6.2~\micron\ PAH band and the continuum strength at 14 \micron. The component maps and the 6.2~\micron\ PAH band emission map are in units of W m$^{-2}$ per pixel and the continuum is in units of W m$^{-2}$ \micron$^{-1}$ per pixel. We note that the central pixel [8,3] is white because of poor continuum subtraction in that region as opposed to the other white regions which represent low signal-to-noise ratio observations. The white and red contours represent the distribution of 6.2~\micron\ PAH emission and 14~\micron\ continuum emission respectively. North and east are indicated in the top left panel by the long and short arrows respectively.}   
    \label{fig:g37maps}
\end{figure*}

Spatial emission maps of the four components as well as the 6.2~\micron\ PAH band and 14~\micron\ continuum flux are given for each object in Figures~\ref{fig:g37maps},~\ref{fig:I12063maps},~\ref{fig:W49Amaps},~\ref{fig:M17maps},~\ref{fig:IC434maps},~\ref{fig:N1333maps},~\ref{fig:N2023maps} and~\ref{fig:N7023maps}. In each map the data were only included if the SNR of the feature in the corresponding spectrum was greater than three. Fluxes are quoted per 3.6\arcsec\ square pixel, 2 $\times$ 2 binned from the native 1.8 \arcsec\ square pixels to match the PSF. The spatial binning means that our maps appear rather coarser than those presented in \citet{P15} which did not suppress the non-independent pixels.

\paragraph{H\textsl{\textsc{ii}} Regions} In general the \HII\ region maps do not have sufficient spatial resolution to distinguish complex spatial structure as our sources are very distant (see Paper I, Table 1), in contrast to what has been achieved by using comparatively nearby RNe. However, the main apparent trends is that the four components all seem to follow the 6.2~\micron\ PAH band emission rather than the continuum emission. This finding is consistent with the assumption that the entire 7.7~\micron\ PAH complex is emitted by aromatic molecules similar to the carriers of the 6.2~\micron\ PAH band, consistent with the fact that these two bands are correlated (e.g., \citealt{2008ApJ...679..310G}).

\paragraph{Reflection Nebulae} The RNe maps generally follow the result found by \citet{P15}, with NGC~2023 and NGC~7023 showing that the 7.8 and 8.2~\micron\ bands follow the continuum, with the NGC~1333 maps  being less clear. The NGC~2023 maps remain the only clear example of this split between the inner and outer components of the 7.7~\micron\ PAH complex. An interesting note emerges from the NGC~7023 maps though, in that the 7.8 and 8.2 components, while largely matching the continuum emission morphology, also show emission similar to that of 7.6 and 8.6, albeit less intensely.

\section{Discussion}\label{sec:disc}

We have presented the results of the application of the four-Gaussian 7.7~\micron\ PAH band decomposition method to a large sample of Spitzer-IRS spectral maps. As the wider implications and context of this method are discussed by \citet{P15}, we will focus on the issues arising from our analysis, specifically the apparent differences between the \HII\ region and RNe spectra. The strongest evidence of this being the correlation plots themselves, where we see that our four Gaussian components inter-correlate reasonably well for the \HII\ regions, while for the RNe maps only the inner (G7.8 and G8.2) and outer (G7.6 and G8.6) sets of bands seem to correlate. In addition to this, we see spectral changes in the form of greatly reduced emission around 8~\micron\ in the spectra of the cores of some of the UC-\HII\ regions and the suggestion of a shift in the average peak position of the 7.8~\micron\ PAH band during our fitting procedures. In the remainder of this Section we first discuss the possible explanations for our observational findings in terms of the band assignments and subsequently the relationship between the PAH emission and the physical conditions. 

We note that while the exact parameters for our fixed fit are slightly altered from those found by \citet{P15} for NGC~2023 alone, the changes are small enough that the bands still almost entirely overlap. The key difference compared to the \citet{P15} parameters is that our decomposition puts slightly more flux into the G7.8 component by moving to the blue slightly and therefore reducing the G7.6 component, while the G8.2 and G8.6 components are almost identical. The upshot of this is that the discussion of \citet{P15} regarding the PAH populations contributing to each band and comparing the decomposition to more `bottom-up' approaches such as PAHFIT, PAHTAT and database fitting techniques is still applicable to this work.

\subsection{The weakening emission at 8 \micron} \label{sec:disc8}

\begin{figure}
  \centering
    \includegraphics[width=8cm]{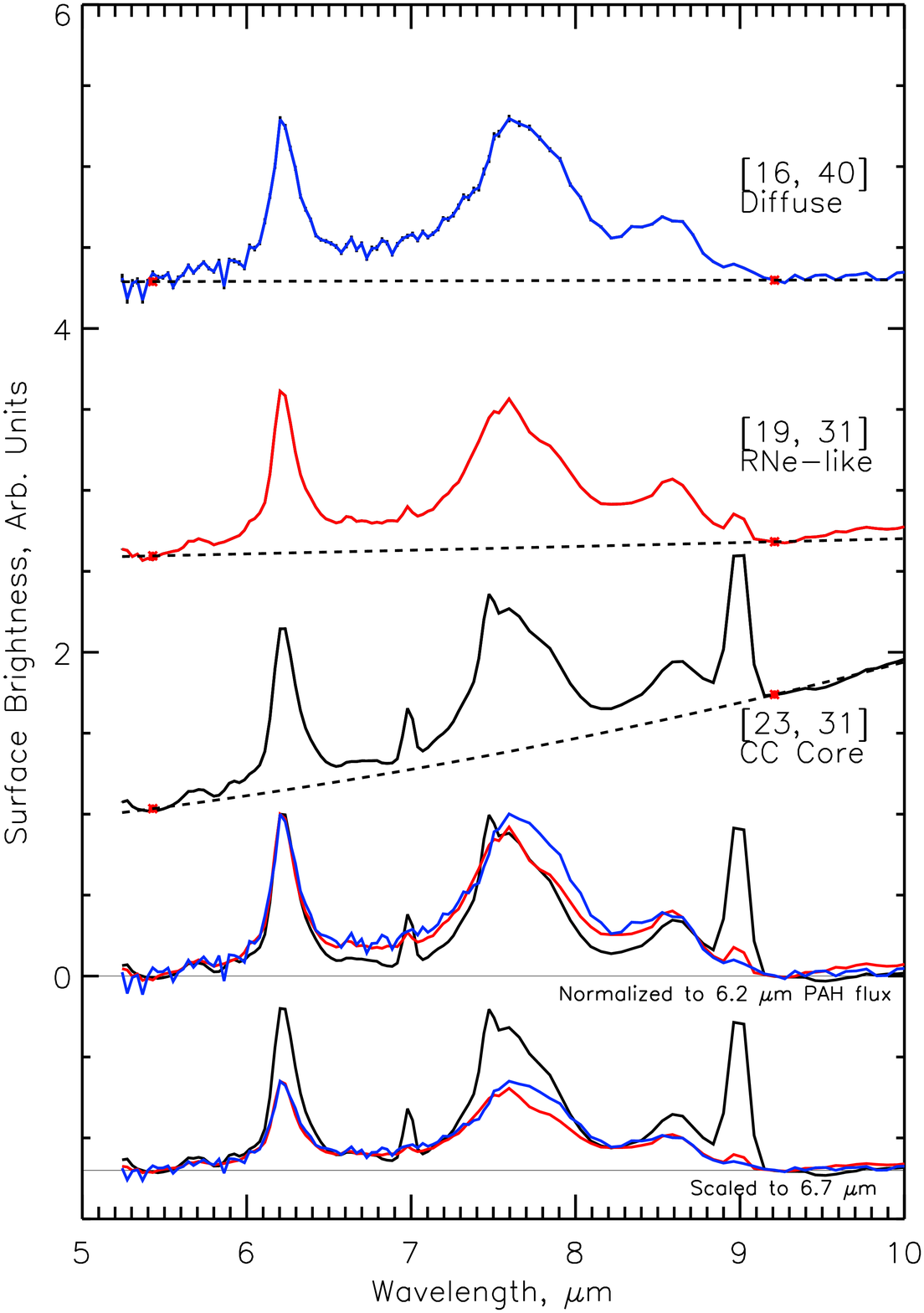}
    \caption{Changes to the 7.7~\micron\ PAH complex profile between the core of UC-\HII\ region W49A-CC (black), `\HII\ region'/RNe-like material 10\arcsec\ outside the core (red), and a more `diffuse' spectrum from $\sim$30\arcsec\ further out (blue) as well as continuum subtracted versions of these spectra at the bottom. Each of the individal spectra is normalized to the peak flux above the power-law continuum at 6.2 \micron. We show the continuum subtracted spectra both with this normalization (i.e. I(6.2 \micron) = 1, as well as scaled such that the plateau level at 6.7 \micron\ are equal. For each spectrum we have included a power-law continuum shown with black dashes using the red boxes as anchor points. The numbers in square brackets denote pixel positions, again relative to our 2$\times$2 binned cube. }
    \label{fig:spectra_w49a}
\end{figure}

In Section~\ref{sec:data} we mentioned that in regions of strong continuum emission we observed some changes to the 7.7~\micron\ PAH complex in that a) the plateau emission becomes much weaker; and b) the emission of the 7.7~\micron\ PAH complex at around 8~\micron\ becomes much weaker, i.e., the minimum between the 7.7 and 8.6~\micron\ peaks becomes more pronounced. We show a comparison of the core region of W49A-\HII\ region CC with `RNe-like'\footnote{We have adopted this nomenclature as it appears that the PAH spectra from highly-irradiated PDR regions are very similar in appearance to that of the spectra of RNe.} ~and `diffuse' spectra in Figure~\ref{fig:spectra_w49a} in which the spectra are shown with a power law fit to points at 5.5 and 9.2~\micron\ representing the continuum. Using a power law continuum is necessary here as the normal spline continuum approach is not appropriate for the `CC Core' spectrum as discussed in Section~\ref{sec:data}. We can see in the continuum-subtracted spectra at the bottom of Figure~\ref{fig:spectra_w49a} that the `RNe-like' and `diffuse' spectra seem to have higher flux at 8.2~\micron\ by a factor of around 50\% while the main PAH complex at 7.7~\micron\ and the 8.6~\micron\ feature appear to be the same, with the exception that the `diffuse' spectrum has a stronger emission at 7.8 \micron\ resulting in a more symmetric 7.7~\micron\ peak.

The main change between the spectra presented in Figure~\ref{fig:spectra_w49a} is the physical environment dominating the line of sight. The `CC-Core' spectrum is associated with the peak of the continuum emission from an UC-\HII\ region, while the other spectra are from significantly less energetic environments. The `RNe-like' spectrum is drawn from the bright halo of emission surrounding W49A/CC and the `Diffuse' spectrum is representative of regions distant from any exciting source. We have shown in previous work \citep{2014ApJ...791...99S} that emission from the outskirts of the W49A map is very similar to the spectra presented by \citet{boul_ism} for so-called `Diffuse'\footnote{This emission is very different from the `diffuse' emission discussed in the context of NGC~7023 by \citet{2013ApJ...769..117B,2014ApJ...795..110B,2015ApJ...806..121B}, where `diffuse' means the very highly irradiated regions next to the exciting star.} ~sightlines near quiescent molecular clouds. We note that the spectra presented by \citet{boul_ism} possesses the same characteristic wide 7.7 \micron\ peak (i.e., strong 7.8~\micron\ emission) as we see in our `diffuse' spectra. 

\bigskip
\citet[][Section 5.1.1]{P15} discussed the molecular origins of each of the 7--9~\micron\ subfeatures and came to the following conclusions relevant to the discussion at hand:
\begin{enumerate}
\item The 8.2~\micron\ emission originates in C-H in-plane bending modes at bay sites from molecules with multiple bay regions. This finding was arrived at by using density functional theory (DFT) calculations for a large compact molecule (C$_{150}$H$_{30}$) from which bay regions were created by removing carbon atoms from corners successively until each `corner' consisted of two duo hydrogen groups separated by a `bay'. The spectra of these molecules revealed emission peaking at 7.8 and 8.6~\micron\ for molecules with zero bay regions, which merge into a peak at around 8.1~\micron\ by the time each of the six corners had been converted into a `bay' (see \citealt{P15}, Figure~22). 
\item The 7.8~\micron\ emission originates in C-C stretch and CH in-plane bending modes of large PAHs and again, PAHs with bay regions as discussed above. The assignment to large PAH molecules is largely based on the work of \citet{2008ApJ...678..316B,2009ApJ...697..311B} who concluded that species with $N_C > 100$ contributed mainly to the 7.8~\micron\ side of the 7.7~\micron\ PAH complex. In addition it was found by inspection of the NASA Ames PAH database (PAHdb, \citealt{2010ApJS..189..341B,2014ApJS..211....8B}) that irregular edge structures (other than the aforementioned bay regions) can result in emission at around 7.8 \micron. Specifically nitrogen substitutions and addition or removal of hydrogen atoms to duo C-H groups.
\item It was confirmed, on the basis of the spectral maps, that the plateau features\footnote{In this work we define the underlying plateaus in the same manner as \citet{P15}. Note that the `8 \micron\ bump' in incorporated in the four Gaussian components. In Paper~I we showed that the 8 \micron\ bump correlated with the other PAH bands while the underlying plateau did not. }\ behave independently of the PAH features, in agreement with \citet{1989ApJ...344..791B,1989MNRAS.236..485R, 2012ApJ...747...44P}. 
\item Based on the spatial morphology of the 7-9~\micron\ features they argue that the emission from the plateau features as well as the G7.8 and G8.2 features is likely to arise from very large PAH or aromatic species. The clear candidates for the carriers of this emission are: a) very large PAH molecules ($100 < N_C < 150$); b) PAH clusters; or c) very small grains.
\end{enumerate}

These findings present a clear interpretation for the weaker 8.2~\micron\ emission in the high continuum flux regions (i.e. the cores of the UC-\HII\ regions): enhanced photodestruction of non-compact PAH molecules or clusters, i.e. those with bay regions. This concept is well-supported theoretically. If the carrier of the emission is large PAHs, it has long been established that irregular PAH molecules are less stable than their compact brethren \citep{1985ApJ...293L..45C, 2011ApJ...729...94R}, on the other hand if the carrier is clusters of PAHs then the van der Waals bonds which hold them together are certaintly weaker than the bonds within a PAH molecule.

The results of \citet{P15} are largely based on the interpretation of the spectral maps of NGC~2023 north, in which it is clear that the PAH emission occurs in different regions to the continuum emission, and that some of the putative PAH bands actually follow the continuum rather than the other PAH bands. In our sources the continuum and the PAH emission is largely co-incident, probably due to the much greater distances to the \HII\ regions. However, NGC7023 is a second source in support of these assignments as the continuum and PAH emission peak spatially distinct (Figure~\ref{fig:N7023maps}). It's worth noting that in this source G7.8 is more similar to the continuum emission while the G8.2 component exhibits weaker emission at the location of the PAH emission. This suggests that our decomposition needs further fine-tuning in order to single out this larger-sized component accurately. Higher spectral-resolution spectral maps (as e.g. provided by JWST) will be essential to further disentangle the different sub-populations to the 7-9~\micron\ PAH/cluster/VSG emission. 

Similar arguments can be invoked to explain the relative weakening of the 5-10~\micron\ plateau (as defined in the typical `Spline method' fashion, see Paper~I) in such environments. Many studies have concluded that the plateau features are likely due to very small grains, which are thought to be carbonaceous dust grains with sizes of order N$_C$ $>$ 500 (e.g., \citealt{2005A&A...429..193R, 2012A&A...542A..69P, P15}). Such dust grains are also thought to be susceptible to destruction in harsh environments, where it has been suggested that they evaporate (e.g., \citealt{2012A&A...542A..69P}).

\subsection{The 7.8~\micron\ Band Peak Position} \label{sec:disc78}

We found that the average 7.8~\micron\ band peak position moves to shorter wavelengths in the \HII\ region data as compared to the RNe by around 0.05~\micron\ (or approximately three resolution elements; Table~\ref{tables:params}). We investigate this phenomenon by looking for spatial correlations between the 7.8~\micron\ band peak position in the constrained fits and other derived spectral properties. The main conclusion is that the 7.8~\micron\ band blue-shift coincides with increased 7.8~\micron\ emission. 

\begin{figure}
  \centering
    \includegraphics[width=8cm]{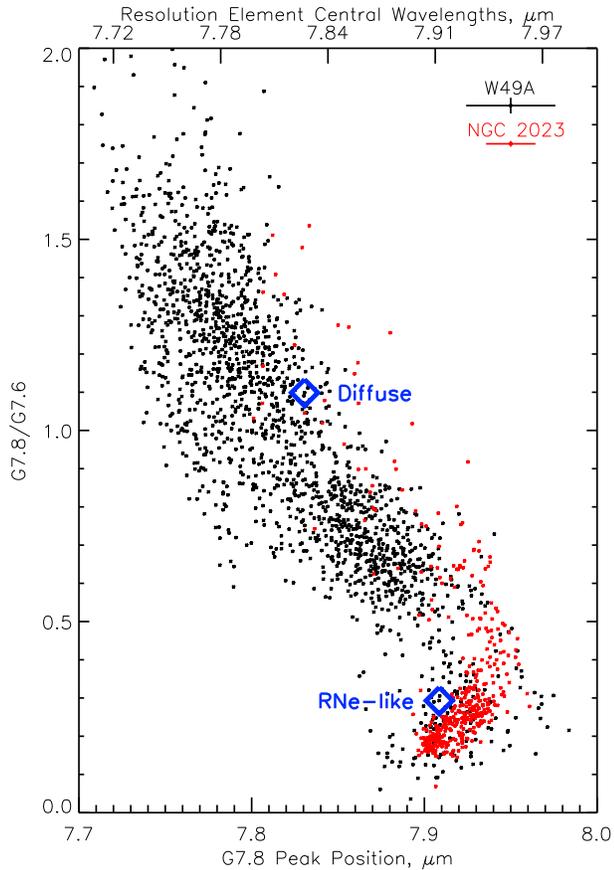}
    \caption{Comparison of the ratio of the G7.8 and G7.6 component fluxes and the G7.8 peak position wavelength for the constrained fits (i.e. those with peak positions and FWHM allowed to vary within a small window). Mean uncertainties for both datasets are indicated in the top right corner. We have indicated the locations of the `Diffuse' and `RNe-like' spectra presented in Figure~\ref{fig:spectra_w49a} with blue diamonds and labels. The `CC Core' spectrum is not included as it was not measured along with the other spectra due to difficulties of fitting consistent continua in regions of strong continuum emission (see Section~\ref{sec:data}).}
    \label{fig:bandshift}
\end{figure}

We show this relationship in Figure~\ref{fig:bandshift}, where we have used the constrained fits (i.e. those where the peak positions and FWHM are allowed to vary) and plotted the G7.8 peak position versus the ratio of the intensities of the G7.8 and G7.6 components. For this discussion we focus on the data for W49A and NGC~2023 as they seem to show the patterns described shortly most clearly, however the data are representative of the other sources. It is clear from inspection of Figure~\ref{fig:bandshift} that the blueward shift of the G7.8 peak position corresponds with the G7.8 band increasing its share of the 7.7~\micron\ PAH complex compared to the G7.6 band. In the W49A data there is a very large spread of points at high ($>$0.5) G7.8/G7.6 values, while for NGC~2023 the concentration at around [7.93,0.2] seems to dominate in terms of the number of points. The differences seen in G7.8/G7.6 ratio in Figure~\ref{fig:bandshift} are also clearly visible in the spectra, with the `Diffuse' spectrum possessing much stronger emission at around 7.8~\micron\ than the other two spectra, both of which have a pronounced `knee' at around 7.9~\micron\ (e.g. Figure~\ref{fig:spectra_w49a}). It is also important to note that the shifting peak position corresponds to FWHM changes in both the G7.6 and G7.8~\micron\ components, and that this drives much of the variation in flux ratio (0.1--$\sim$2) seen in Figure~\ref{fig:bandshift}, which is around double that seen for the `fixed' fits. 

\begin{figure}
  \centering
    \includegraphics[width=8cm]{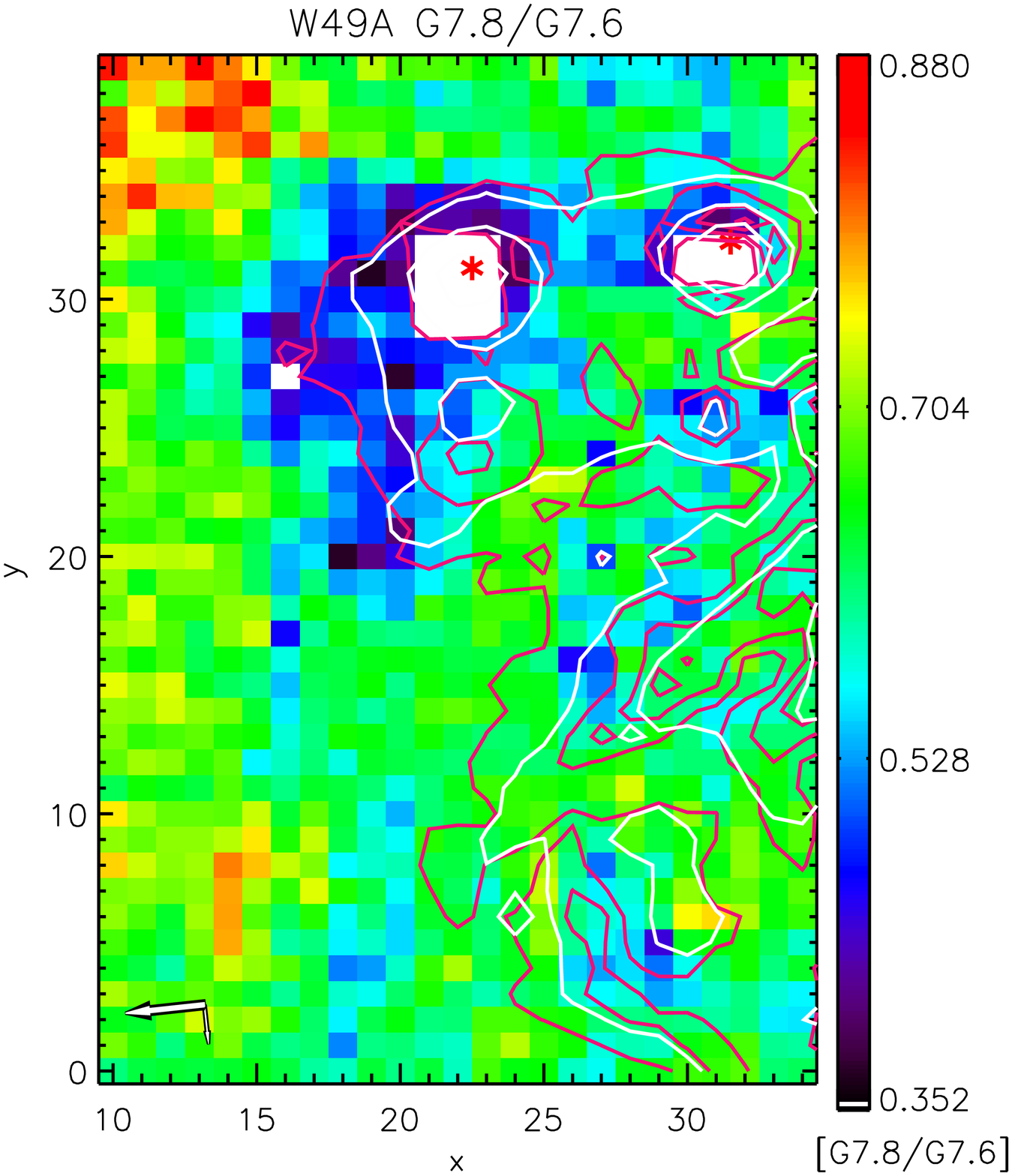}
    \includegraphics[width=8cm]{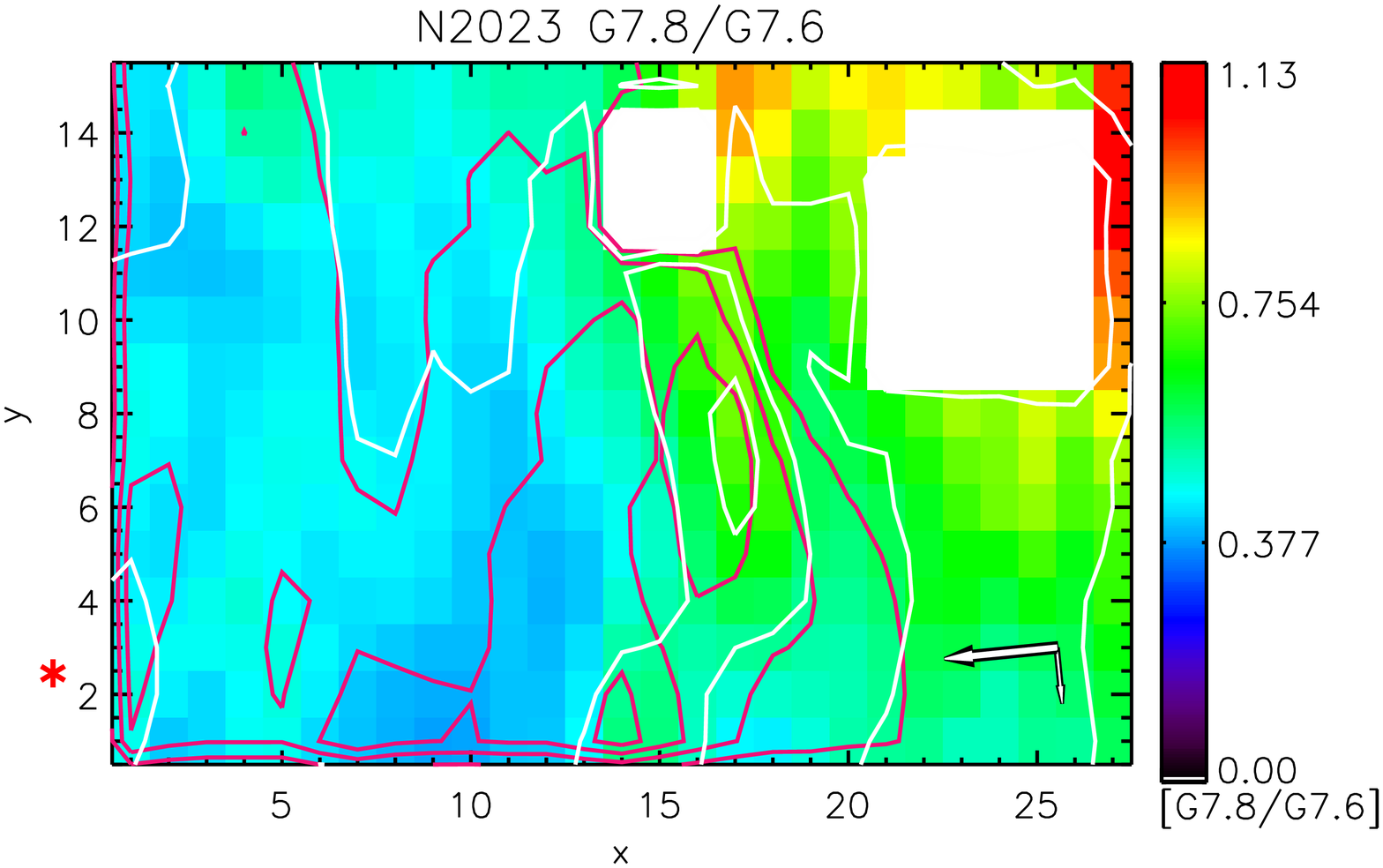}
    \includegraphics[width=8cm]{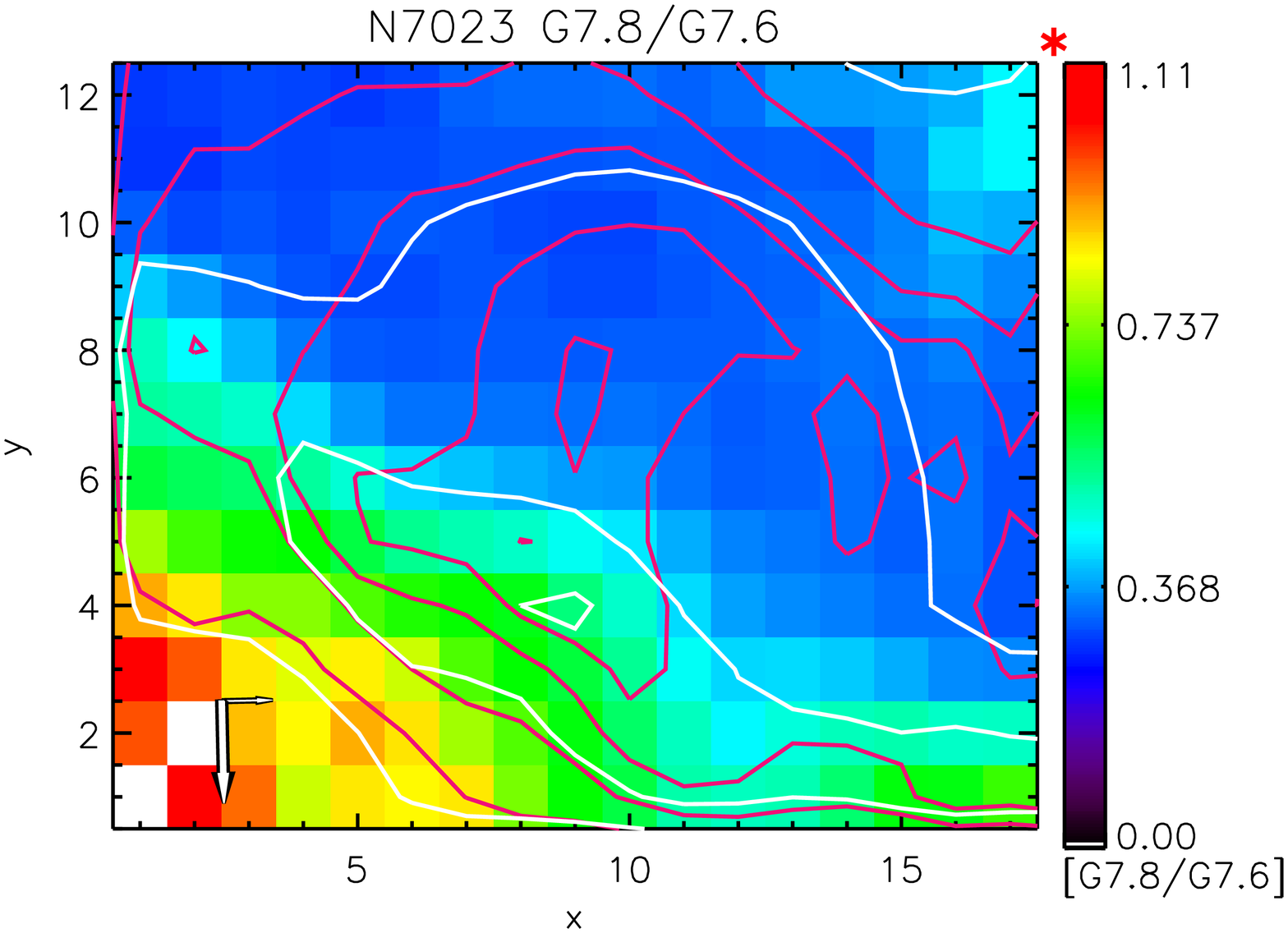}
    \caption{Spatial maps of the ratio of the G7.8 and G7.6 components, showing low values near sources of UV radiation (the locations of the dominant exciting stars are marked with red stars) and high values elsewhere. In these maps the white regions have been masked for three distinct regions, a) in W49A we have masked the regions around the red stars as these suffer from poor continuum subtraction in regions of very high UV flux; b) for W49A pixel [17,27] and NGC~7023 pixels [1,1] and [2,2] there is insufficient SNR; and c) for NGC~2023 there are two separate regions with bright protostars which have been removed. North and east are indicated in each panel by the long and short arrows respectively.}
    \label{fig:7876maps}
\end{figure}

As the move to shorter wavelength peak positions coincides with the strengthening of the G7.8 band, we interpret this phenomenon as being the result of the constrained fitting process. In terms of the spectra, the change between the `RNe-like' and `Diffuse' spectra can be summarized as follows: the peak of the `diffuse' spectrum becomes more rounded  as the flux between about 7.7 and 7.9~\micron\ increases, meanwhile the red wing of the 7.7~\micron\ PAH complex becomes much steeper but does not seem to move in wavelength space\footnote{Incidentally, we find agreement with \citet{2014ApJ...795..110B} as the increased G7.8/G7.6 is accompanied by a slight rise in the red wing of the 11.2~\micron\ PAH emission complex. }.  In terms of our fit, the steepening of the red side of the 7.7~\micron\ emission leads to a slight shift of the G7.8 component towards the blue and a decrease in FWHM in order to capture the shape of the peak. As such, the blueward shift in peak wavelength seems only to be indicative of the relative strengthening of the G7.8 band. As the average G7.8 peak position presented in Table~\ref{tables:params} is somewhere between the `red' and `blue' extremes, our `fixed' fits still recover acceptable fits even in regions of strong or weak G7.8 emission. 

Furthermore, the relationship between the G7.8/G7.6 ratio and the physical conditions persists in our `fixed' fits, which we refer to for the remainder of the paper. We show this in Figure~\ref{fig:7876maps} where we present maps of the G7.8/G7.6 ratio for W49A, NGC~2023 and NGC~7023 using the fits with fixed peak position and FWHM. It is obvious from these maps that low ratios seem to accompany regions near sources of UV radiation (marked with red stars in Figure~\ref{fig:7876maps}). For W49A this means that the areas around UC-\HII\ regions CC and DD have very low ratios, as well as some scattered pixels towards the arc region described by \citet{2014ApJ...791...99S}. However for NGC~2023 and 7023, we see spatially resolved examples of this behavior, where there is a smooth gradient from low ratios to high ratios, corresponding very well with increasing physical distance from the exciting star. \citet{2014ApJ...795..110B} shows a similar map, albeit inverted (i.e. G7.6/G7.8) and derived using direct integration, which superficially agrees with the map created using our Gaussian components. However, the \citet{2014ApJ...795..110B} map of G7.6/G7.8 does not capture the strong ratio increase in the corner of the map furthest from the exciting star. We attribute this to their applied extraction of the strength of the 7.7~\micron\ PAH complex (and hence the 7.6 and 7.8~\micron\ PAH components) which includes emission underneath the 8.6~\micron\ PAH band (which in our decomposition method is incorporated in the G8.2 and G8.6 PAH components). In the NGC 2023 map we also see a slight jump in the ratio at the location of the PDR front labeled `South Ridge' by \citet{P15}.

\subsection{Band Intercorrelations} 
Our observation that the G7.8/G7.6 ratio seems to trace physical conditions helps explain the appearance of the correlation plots presented in Figure~\ref{fig:plots}, where good band inter-correlations were found for all of the Gaussian components in the \HII\ regions sample while for the RNe the `inner' and `outer' sets of components seemed to correlate. This appears to be an observational effect, in the sense that the \HII\ region maps systematically sample much larger physical areas, thus including large areas which are not directly associated with the strong UV fields dominating the named objects in each map. The upshot of this is that the \HII\ regions sample in the correlation plots are dominated by spectra of the outskirts of star forming regions (i.e. the diffuse spectrum in Figure~\ref{fig:spectra_w49a}), while the RNe samples are dominated by spectra representing irradiated PDRs (the RNe-like spectrum in Figure~\ref{fig:spectra_w49a}). Thus the \HII\ regions sample spans a much larger parameter space in the correlation plots as it contains spectra from weakly- and strongly-illuminated regions, while the RNe spectra are exclusively from strongly irradiated regions.

While this difference does impact the appearance of the 7.7~\micron\ PAH emission complex in that the 7.8~\micron\ component is sensitive to the ambient radiation field, we conclude that there is no intrinsic spectral difference between the PAH emission from strongly irradiated RNe PDRs and from strongly irradiated \HII\ region PDRs in terms of their 7.7~\micron\ PAH complex emission. However, there is a strong difference between the PAH spectra of the highly-irradiated regions and the diffuse outskirts which dominate the \HII\ regions data sample. For the weakly irradiated regions each of the components of our Gaussian decomposition intercorrelate, perhaps representing the increased diversity of emitting PAH species in environments with low PAH destruction rates.

\subsection{Physical Conditions and PAH Profile Classes} 

Given that we have found a qualitative link between the G7.8/G7.6 ratio and the presence of UV radiation sources, we now seek to quantify this relationship. In Paper~1  (Table~5) we gave estimates or literature values of G$_0$ for the sources with simple geometries, i.e. those with a resolved\footnote{Not necessarily resolved in the MIR, in most cases the radius measurements come from continuum observations at radio wavelengths, see Paper I.} ~distance between the star and the PDR front. We summarize this data in Table~\ref{tables:physconds} and extend it with entries relevant for the RNe at the PDR fronts. In addition, we select spectra from each source from the region of the PDR front, so as to ensure that the PAH spectra are from regions with G$_0$ close to that quoted in Table~\ref{tables:physconds}. Uncertainties for the PAH ratio were calculated by combining the SNR measurements for each band except for the Horsehead PDR, where due to low SNR measurements we took the average of a large section of the PDR front and quote the standard deviation of that set of ratios as the uncertainty. For the $G_0$ measurements we have adopted an undertainty of 10\%. 

\begin{table}
\caption{\label{tables:physconds} UV-Field Intensities and PAH Band Ratios at PDR Fronts}
\begin{center}
\begin{tabular}{c c c c c l}
\hline\hline
Object            & \multicolumn{2}{c}{Pixel}	& 7.8G/7.6G	& log(G$_0$) & Ref. \\
		  & $x$		& $y$\\
\hline
\\
W49A/CC           & 20 		& 28 		&  0.35 $\pm$ 0.01	& 4.67	& 1 \\
W49A/DD           & 33 		& 33 		&  0.45 $\pm$ 0.01	& 4.67	& 1 \\
G37.55-0.11       &  8 		&  4 		&  0.51 $\pm$ 0.01 	& 4.38	& 1 \\ 
IR~12063-6259 	  &  8 		&  3 		&  0.48 $\pm$ 0.01 	& 4.31	& 1 \\
HH PDR     	  & 15:18$^a$	& 8:10$^a$	&  1.07 $\pm$ 0.11 	& 2.00	& 2 \\
NGC~2023          & 17		&  7 		&  0.71 $\pm$ 0.01 	& 3.60	& 3 \\
NGC~7023          &  8 		&  4 		&  0.65 $\pm$ 0.01 	& 3.41	& 4 \\
\\
\hline\hline
\end{tabular}
\end{center}
$^a$: The lower signal to noise found in the HH PDR cube neccessitated the PAH ratio being measured as an average along the PDR front in a box defined by the given coordinate ranges.

\tablerefs{(1) \citet{S15}; (2) \citet{2015AA...576A...2O}; (3) \citet{1998PASA...15..194B}; (4) \citet{1988ApJ...334..803C}}
\end{table}

\begin{figure}
  \centering
    \includegraphics[width=8cm]{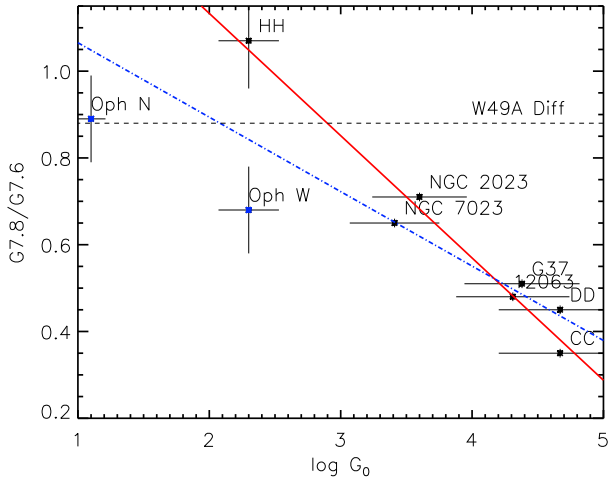}
    \caption{Response of the G7.8/G7.6 intensity ratio to increasing radiation field intensity (G$_0$; \citealt{1968BAN....19..421H}). Black points represent points in our sample of maps for which we have a literature prediction for G$_0$, blue points represent ISO diffuse ISM spectra from \citet{boul_ism}. The red line (the parameters of which are given in Equation~\ref{eq:linreg}), represents a fit to the black points, while the blue dashed line represents a fit to the black \textit{and} blue points. In addition we have shown the maximum G7.8/G7.6 intensity ratio found in the outskirts of the W49A map as a horizontal dashed line (denoted `W49A Diff').}
    \label{fig:7876g0}
\end{figure}

In Figure~\ref{fig:7876g0} we show the response of the G7.8/G7.6 band intensity ratio to the strength of the incident radiation field, G$_0$ (measured in Habings; \citealt{1968BAN....19..421H}). We find a strong negative correlation between the intensity ratio and the radiation field strength which follows the relationship:

\begin{equation}\label{eq:linreg}
\textrm{I}_{7.8}/\textrm{I}_{7.6} = (1.70 \pm 0.13) - (0.28 \pm 0.03) \textrm{ log  G}_0. 
\end{equation}

We conclude from this that the flux ratio of our 7.8 and 7.6~\micron\ Gaussian components can be used to trace the strength of the radiation field G$_0$ in the range of 2 $<$ log G$_0$ $<$ 5. We note at this point that this relationship has been found for only ISM type sources, that is sources with \citet{2002A&A...390.1089P} PAH Class A\footnote{\citet{2002A&A...390.1089P} classified PAH emission in a sample of ISO-SWS spectra on the basis of the 7.7~\micron\ PAH complex peak position, with Class A being the bluest, and almost exclusively comprised of ISM sources, and Classes B and C indicating differing degrees of redshift and usually representing stars with circumstellar environments (e.g. post AGB stars, PNe, HAeBe stars). }.

Previous studies have found similar relationships involving the 7.7~\micron\ PAH complex and the physical conditions. In particular \citet{2005ApJ...621..831B} found that the peak position of the 7.7~\micron\ PAH complex was systematically lower for higher values of G$_0$ within RNe, which we have shown here can be extended between objects with a simple Gaussian fit. Other studies, including those by \citet{2008ApJ...679..310G, 2012A&A...542A..69P} and \citet{2015ApJ...806..121B}, have drawn links between physical conditions and PAH ratios, usually using a simple PAH flux ratio like 6.2 / 11.2 and the ionization parameter $\gamma$ (where $\gamma$ is typically defined as G$_0$ / n$_e$ $\times$ T$^{0.5}$). As it is challenging to precisely isolate G$_0$ from the results of these studies such that they are directly comparable to the results presented here, we do not discuss them further. 

\paragraph{Low-G$_0$ constraints} We have well sampled the high end of the correlation with our set of ultra-compact \HII\ regions, in fact, we have argued that the strong radiation fields found near these regions seem to start breaking down the carriers of the 7.8~\micron\ component and changing the appearance of the 7.7~\micron\ PAH complex beyond the `normal' ISM profile (e.g. Section~\ref{sec:disc8}). Moreover it is the low-G$_0$ end of the correlation which is most interesting as these conditions are prevalent in the diffuse ISM. We can gain some understanding of how much we should trust an extrapolation of this curve to low-G$_0$ by investigating published spectra of diffuse ISM regions. This is especially important as we only have one low-G$_0$ source in our sample, the Horsehead PDR, and so we do not know whether this is a good prototype for the appearance of the PAH spectra at low-G$_0$. Although, we note here that if we exclude the Horsehead PDR point from the fit, the resulting relationship barely changes and it still consistent with the Horsehead PDR point. \citet{boul_ism} provide a set of ISOCAM spectra of such regions with accompanying G$_0$ estimates for North and West pointings of the Ophiuchus diffuse cloud (henceforth Oph-N and Oph-W). We have fitted these spectra in the same manner as the Spitzer-SL sample and find that for Oph-N (log G$_0$ $\sim$ 2.3) the G7.8/G7.6 ratio is 0.68, while for Oph-W (log G$_0$ $\sim$ 1.1) the G7.8/G7.6 ratio is 0.89. Qualitatively these values agree with our trend, i.e. higher G7.8/G7.6 corresponds to lower G$_0$, however it must be pointed out that these two points lie below the trend found in Equation~\ref{eq:linreg}. If we include these two points in our fit, we arrive at a much shallower line with the following fit  parameters:

\begin{equation}\label{eq:linreg2}
\textrm{I}_{7.8}/\textrm{I}_{7.6} = (1.24 \pm 0.12) - (0.17 \pm 0.03) \textrm{ log  G}_0. 
\end{equation}

We have argued here, and in \citet{2014ApJ...791...99S}, that the outskirts of the W49A spectral map are more representative of the PAH spectra seen for diffuse ISM sightlines than the \HII\ regions and PDRs which dominate the emission from the map. As one might expect, at the periphery of the W49A cube we see higher ratios of G7.8/G7.6 of up to 0.88 (Figure~\ref{fig:7876maps}, top panel), which is expected given that we have argued that the peripheral regions represent low-UV field conditions. If we take the maximum G7.8/G7.6 value for W49 and use Equation~\ref{eq:linreg} to derive a corresponding log(G$_0$) value, we obtain a log(G$_0$) of 2.9, while if we use Equation~\ref{eq:linreg2}, we find a log(G$_0$) of 2.1. Clearly the latter value is closer to typical values for diffuse ISM material (e.g. G$_0$ $\sim$ 100) than the former, implying that the shallower correlation including the diffuse cloud observations may be more appropriate for diffuse objects. NGC~2023 and NGC~7023 display similar trends, with parts of the maps furthest from the ionizing sources displaying high G7.8/G7.6 ratios of around 1.1 (Figure~\ref{fig:7876maps}, middle and bottom panels). In cuts through both nebulae \citet{2012A&A...542A..69P} estimated that the G$_0$ experienced by these regions was of order 10$^3$, which would put both regions somewhat above the linear regression presented in Equation~\ref{eq:linreg}, and well above that given by Equation~\ref{eq:linreg2}. These extra data points suggest that there is likely a dependence on other variables, with the obvious candidates being the density or temperature of the PDR. The fact that the NGC~2023 and 7023 PDRs possess regions with moderate G$_0$ and higher G7.8/G7.6 ratios than Equation~\ref{eq:linreg} predicts, whilst the diffuse ISM regions with low G$_0$ have lower G7.8/G7.6 ratios than Equation~\ref{eq:linreg} predicts, suggests that the missing variable might be density, as both the diffuse ISM regions have low densities compared to the PDRs of the RNe. Further study, either in the form of adding additional sources or making spatially resolved measurements of the PDR densities and radiation fields for our sources, would be required to fully resolve this dilemma.

\paragraph{Profile Variations} At the beginning of this section we remarked that the dataset we used to uncover this correlation consists entirely of `ISM' type sources, where an `ISM' source corresponds to Class A using the \citet{2002A&A...390.1089P} classification scheme for PAH spectral profile appearance. Objects with circumstellar material (CSM; e.g. P-AGB, PPNe, PNe, Herbig AeBe, etc), mostly display redshifted PAH band profiles which are classified as Classes B and C when the peak of the 7.7~\micron\ PAH complex shifts into the range between 7.6 and 8.2 \micron, and to $\sim$8.2~\micron\ respectively. Therefore we have investigated applying our decomposition routines developed here to the sample of ISO spectra of CSM sources from \citet{2002A&A...390.1089P} as well as the sample of Herbig AeBe stars used by \citet{2008ApJ...684..411K}. We find that although good fits can be made to these data\footnote{Good fits for the class B sources were arrived at by allowing the central wavelengths of the bands to redshift slightly whilst maintaining fixed widths.}, the results do not seem to be compatible with the earlier discussion. 

For the Herbig AeBe stars, we found that the relationship between the stellar effective temperature and the central wavelength of the 7.7~\micron\ PAH complex (found by \citealt{2007ApJ...664.1144S,2008ApJ...684..411K}) could be recovered using the G7.8/G7.6 ratio in place of the 7.7~\micron\ PAH complex central wavelength, however all but one of the ratios found for the \citet{2008ApJ...684..411K} sample were also well above that seen for the ISM sources. Our evolved star results largely corroborated that picture, because while the sample spans a range of G$_0$ values, the G7.8/G7.6 ratios found were all significantly higher than either of the correlations presented earlier, with G7.8/G7.6 generally falling between one and four. It should be noted though that the evolved star sample is dominated by PNe which were found by \citet{2007ApJ...664.1144S} not to fit the T$_{eff}$ versus central wavelength correlation. This could be further evidence that the density, which may generally be higher for the CSM sources compared to the ISM sources, could also be playing a role here. Alternatively it could be that the PAH populations present in the CSM sources is dramatically different from those seen in the ISM sources, resulting in the incompatibility of these results.

\paragraph{Outcomes} The main conclusion we draw from this Section is that there appears to be a correlation between the ratio of two of the Gaussian components (G7.6 and G7.8), and the intensity of the incident UV radiation field G$_0$. While we have discussed at length the possible problems with the relationships we have derived it seems inescapable that there is a linear relationship between our band ratio and log G$_0$ over three orders of magnitude from G$_0$ = 10$^3$ to G$_0$ = 10$^5$, at least for dense, strongly irradiated PDRs such as the RNe and the \HII\ regions. At the low-G$_0$ end of this correlation there appears to be large scatter in the data, however further work is necessary to determine whether this is the effect of some variable we are not considering (e.g. density) or whether there is intrinsically scatter in the observations. From our limited investigation of the CSM sources it seems that they have systematically higher G7.8/G7.6 ratios, and presumably have systematically higher PDR densities, which leads us to speculate that there may be parallel relationships to that shown in Figure~\ref{fig:7876g0} representing constant densities at different G$_0$ levels. This is consistent with our assignments as it would imply that the G7.8 carriers were better shielded from destruction in high density environments.

\section{Conclusions}

We have shown that a fixed decomposition of the 7.7~\micron\ PAH emission complex into four Gaussian components with fixed positions and widths can provide a good fit to more than 7000 MIR spectra from a set of nine objects of differing physical conditions and object types (\HII\ regions and RNe). Initially we allowed the Gaussian components some freedom in terms of their peak positions and widths (of order 0.2 \micron) however it was found that the fits were very stable across our sample. As such we created a fixed decomposition with four components located at 7.55, 7.87, 8.25 and 8.59~\micron\ and with FWHM of 0.44, 0.40, 0.29 and 0.35~\micron\ respectively. These parameters provide a good fit across all of our objects, despite slight systematic differences between object types. We then studied the inter-correlations of the fluxes in our four bands, and found some striking changes with object type. For the RNe the outer (G7.6 and G8.6) and inner (G7.8 and G8.2) sets of bands were seen to only correlate with each other, i.e. G7.6 correlates with G8.6, while for the \HII\ regions each of the bands was observed to correlate with its neighbors. The reason for this dichotomy was found to be that the RNe maps are dominated by spectra which possess strong 7.6 and 8.6~\micron\ bands and weak 7.8 and 8.2~\micron\ bands, while for the \HII\ regions only a subset of our spectra displayed this pattern (mainly areas associated with high UV fluxes), but the dominant spectral component in terms of the number of pixels possessed a much stronger 7.8~\micron\ band and is associated with the more diffuse outer regions of our spectral maps which we identify as being representative of the diffuse ISM.

We conclude that the correlations seen are reflective of changes in the overall PAH population in different environments, i.e. irradiated PDRs versus more quiescent outskirts. In addition, we established that the 7.8~\micron\ Gaussian component is negatively correlated with the strength of the UV field. We have quantified this relationship by showing that the ratio of the 7.8 to 7.6~\micron\ components is negatively correlated with G$_0$, the radiation field strength. This relationship between the physical conditions and the Gaussian components was observed again for the 8.2~\micron\ component, which behaves strangely in the core of some of the \HII\ regions, almost disappearing completely in some cases. We attribute this effect to the ease of photodestruction for the most common carriers of emission at around 8.2 \micron: PAH molecules with ragged edge structure including bays or clusters of such PAH molecules. It is not clear if such an assignment can explain our observation that the 7.8~\micron\ band is also weaker in irradiated PDRs as there are fewer specific predictions for the types of molecules contributing to such emission. However, these data add additional support to the results of \citet{P15} in that the observed 7.8~\micron\ emission is consistent with its origins being in larger species such as clusters and possibly very small grains.

\section*{Acknowledgments}

DJS and EP acknowledge support from an NSERC Discovery Grant and an NSERC Discovery Accelerator Grant. 

This work is based on observations made with the \textit{Spitzer Space Telescope}, which is operated by the Jet Propulsion Laboratory, California Institute of Technology under a contract with NASA.

This research has made use of NASA's Astrophysics Data System Bibliographic Services; the SIMBAD database, operated at CDS, Strasbourg, France; and the IDL Astronomy Library \citep{1993ASPC...52..246L}.

\bibliographystyle{apj}

\appendix
\section{Maps of Other Sources}
\begin{figure*}
  \centering
    \includegraphics[width=8cm]{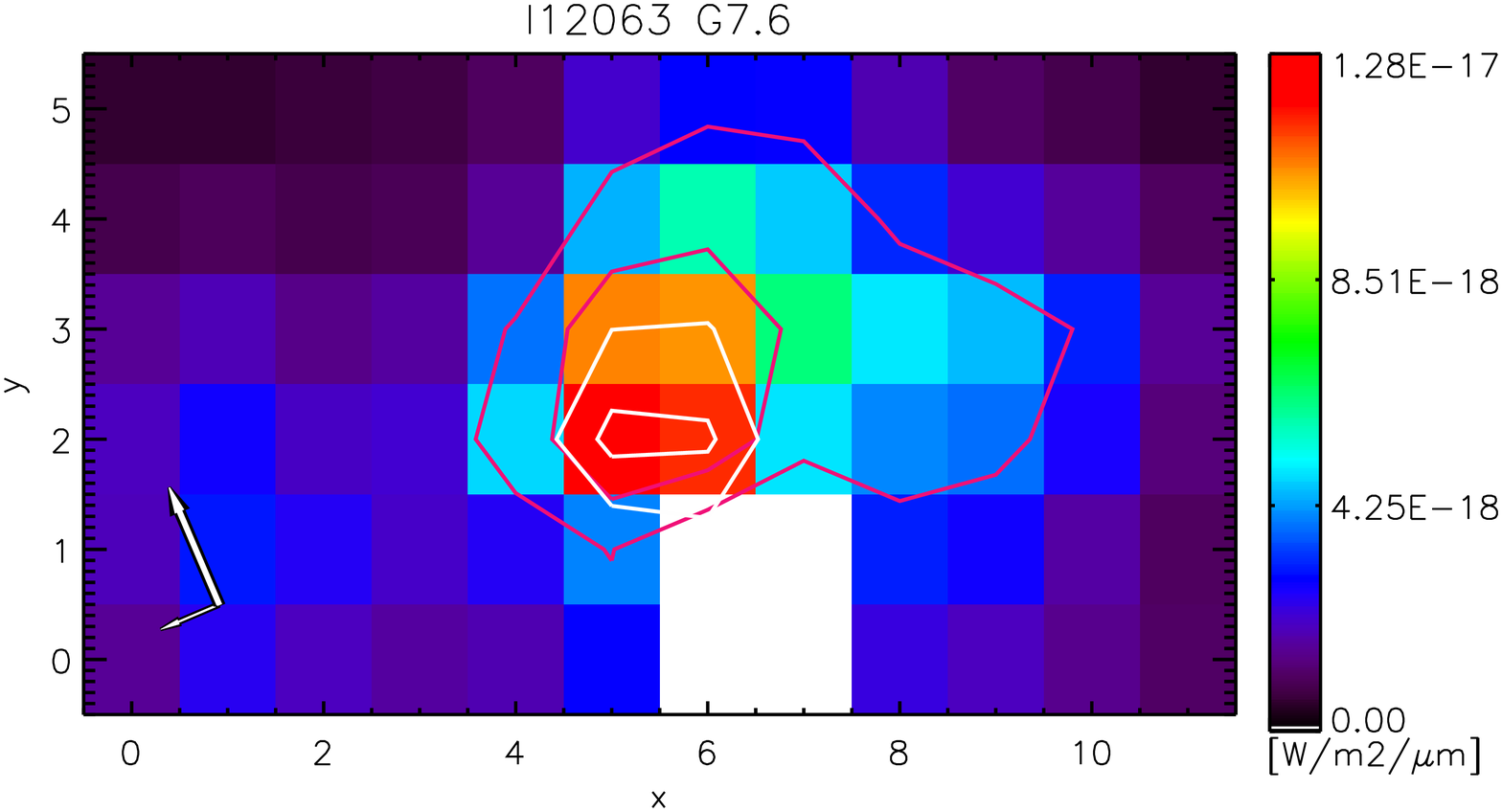}
    \includegraphics[width=8cm]{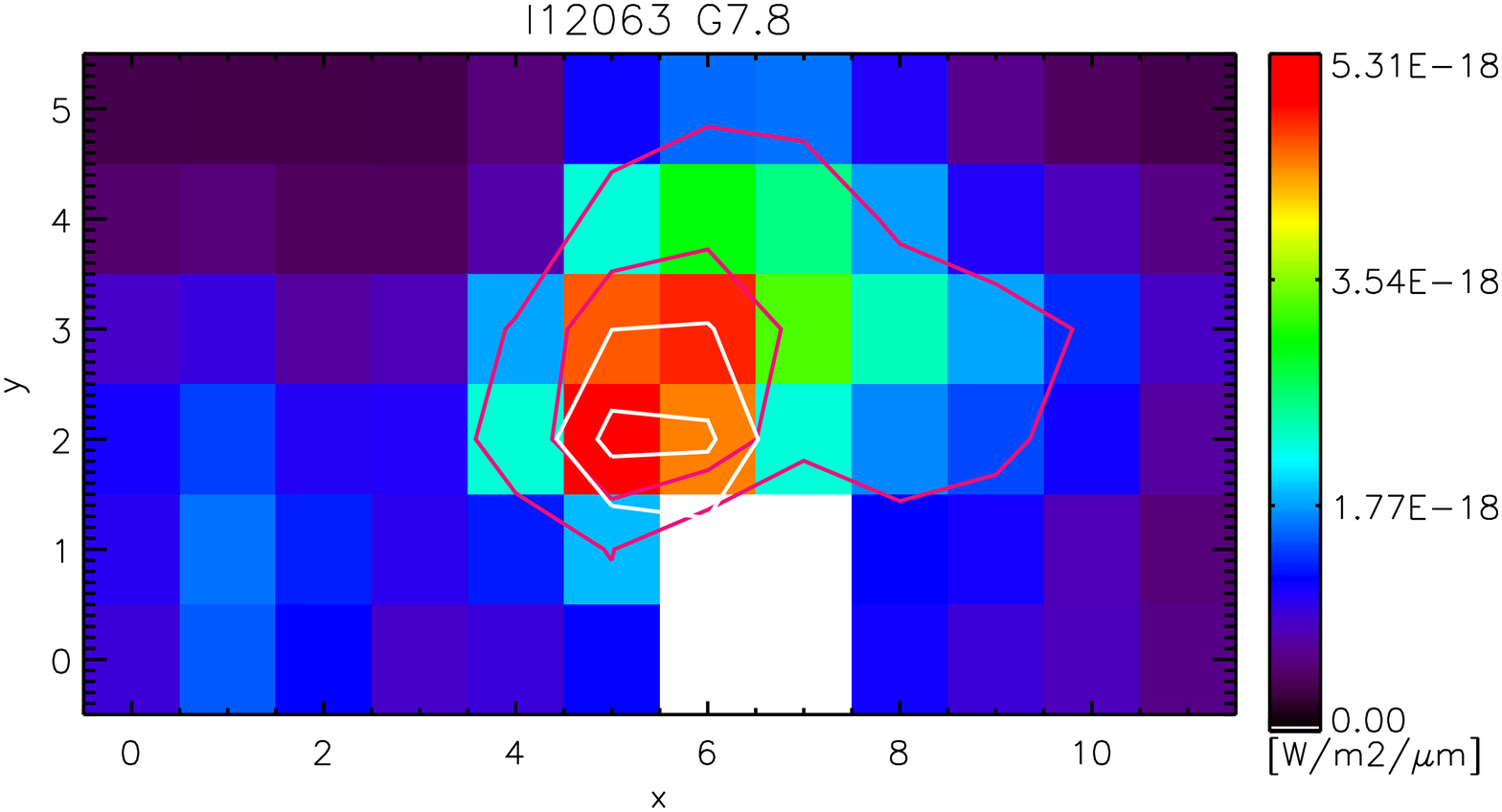}
    \includegraphics[width=8cm]{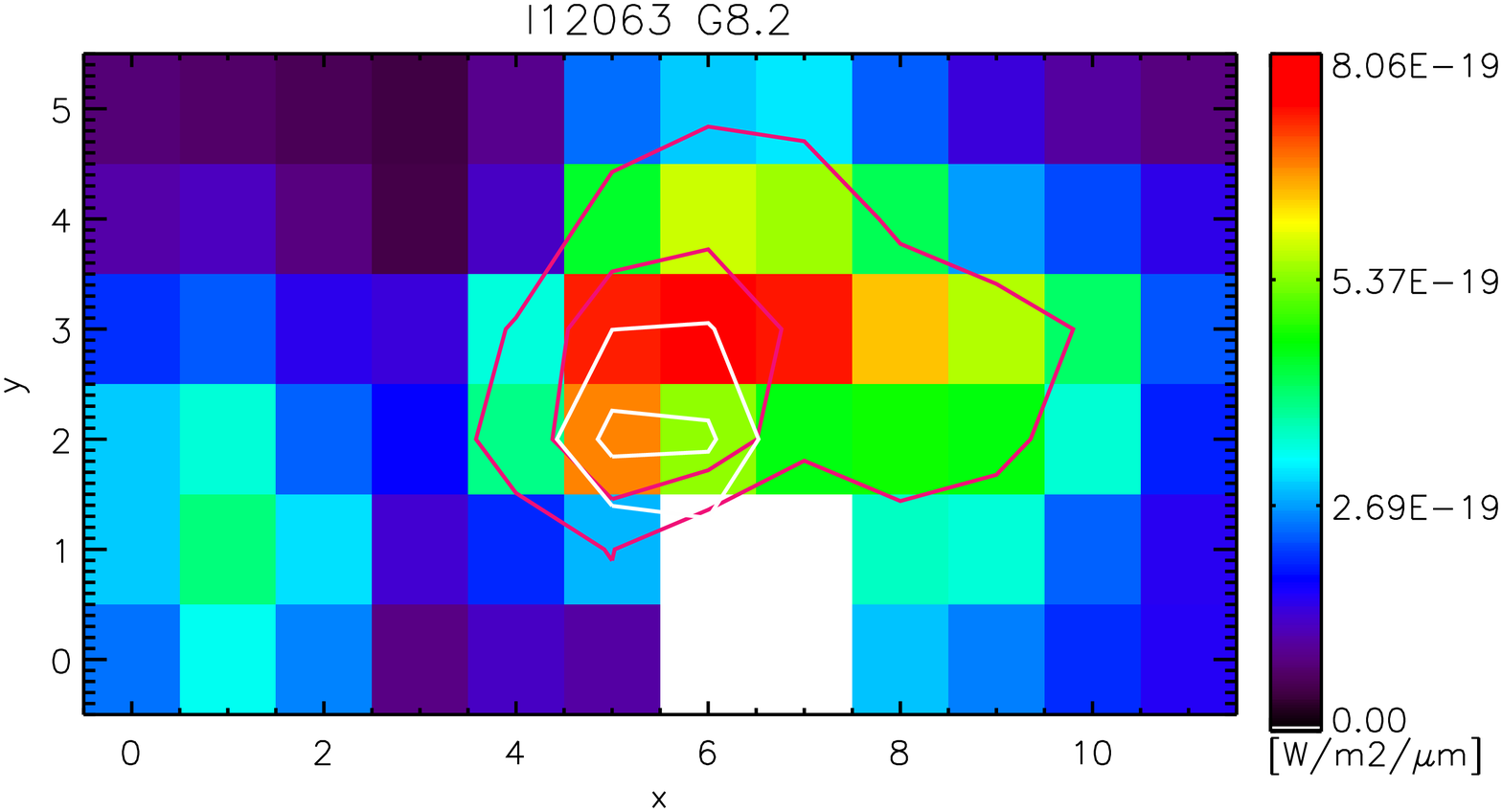}
    \includegraphics[width=8cm]{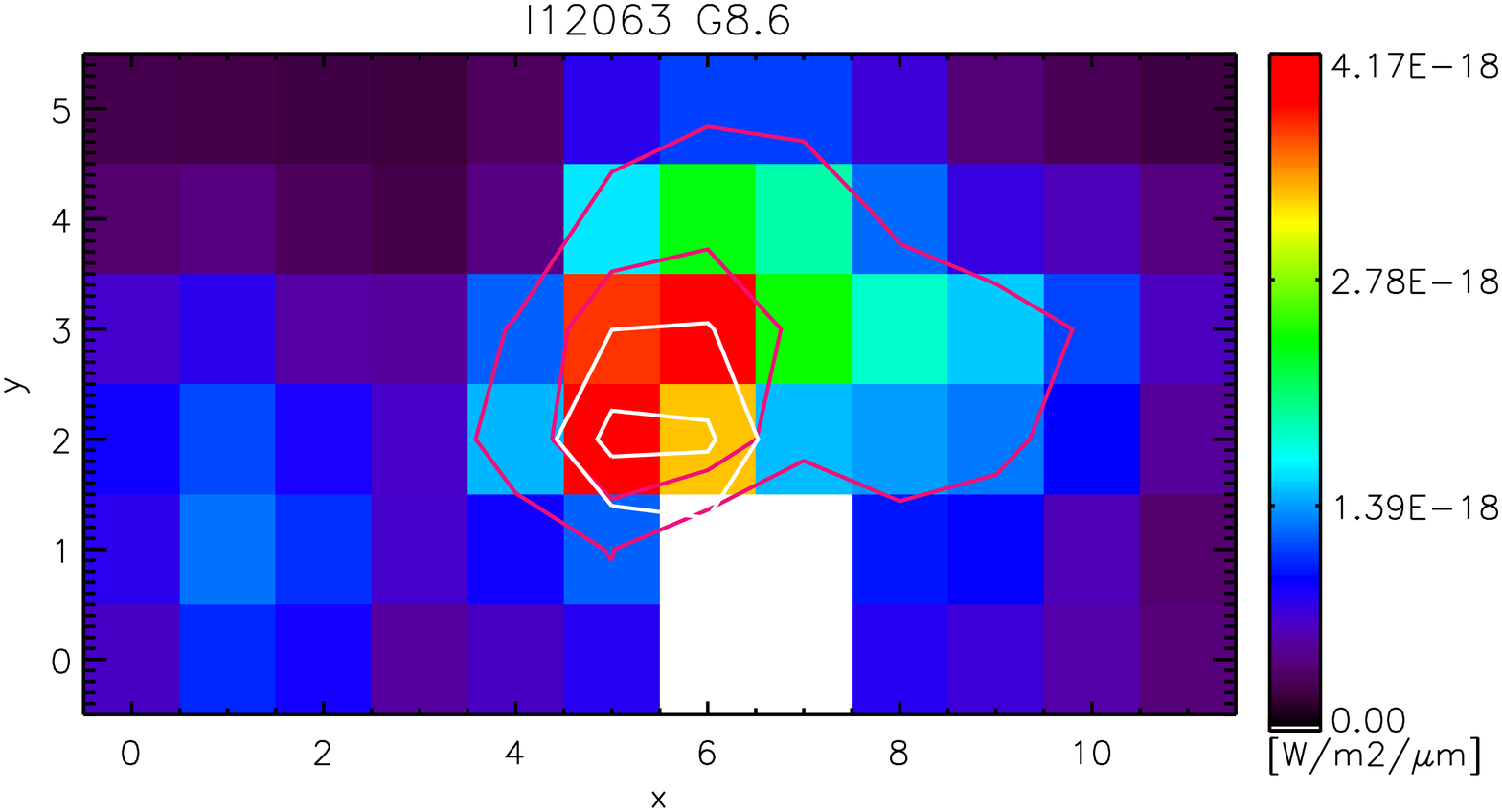}
    \includegraphics[width=8cm]{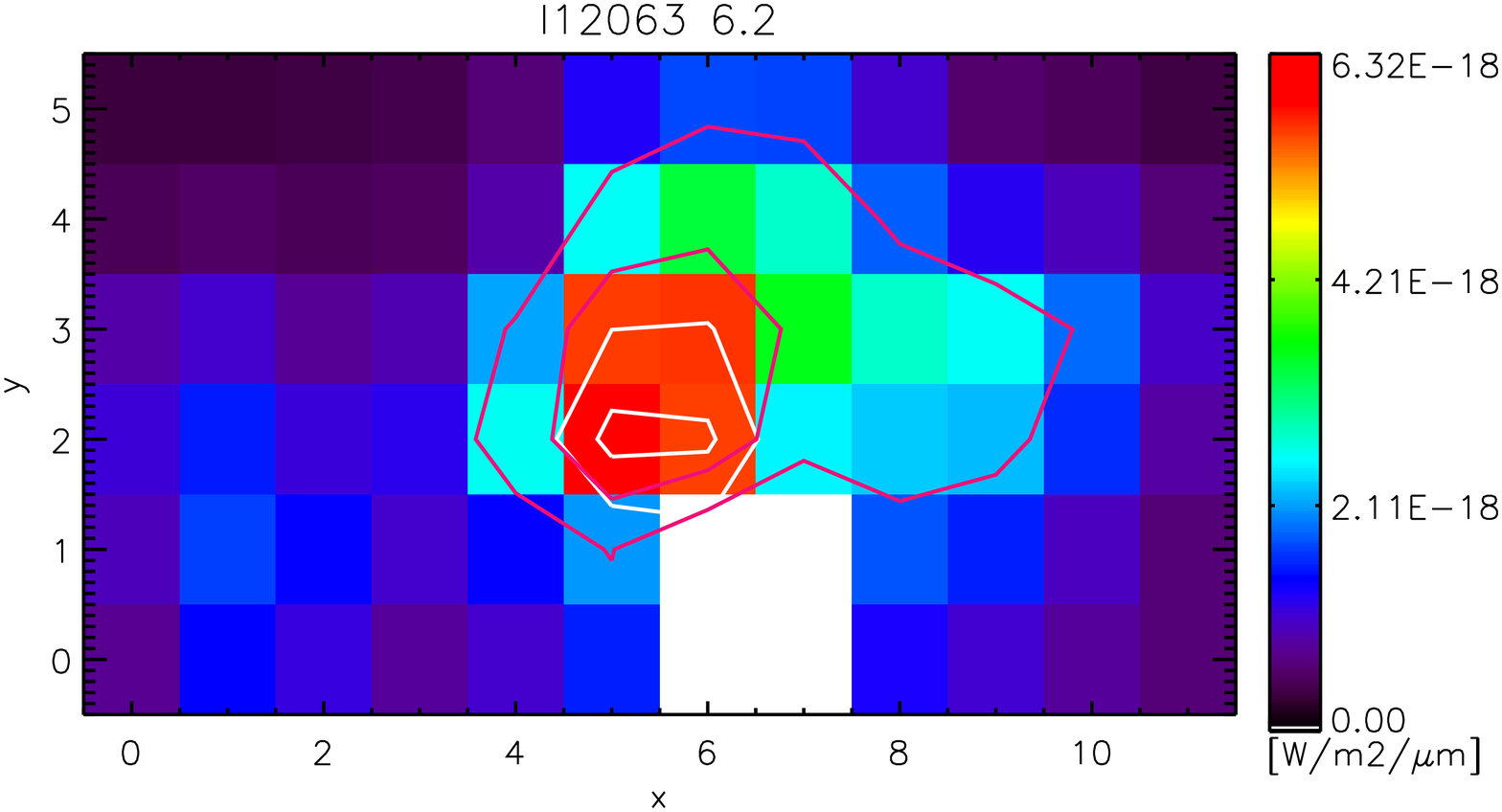}
    \includegraphics[width=8cm]{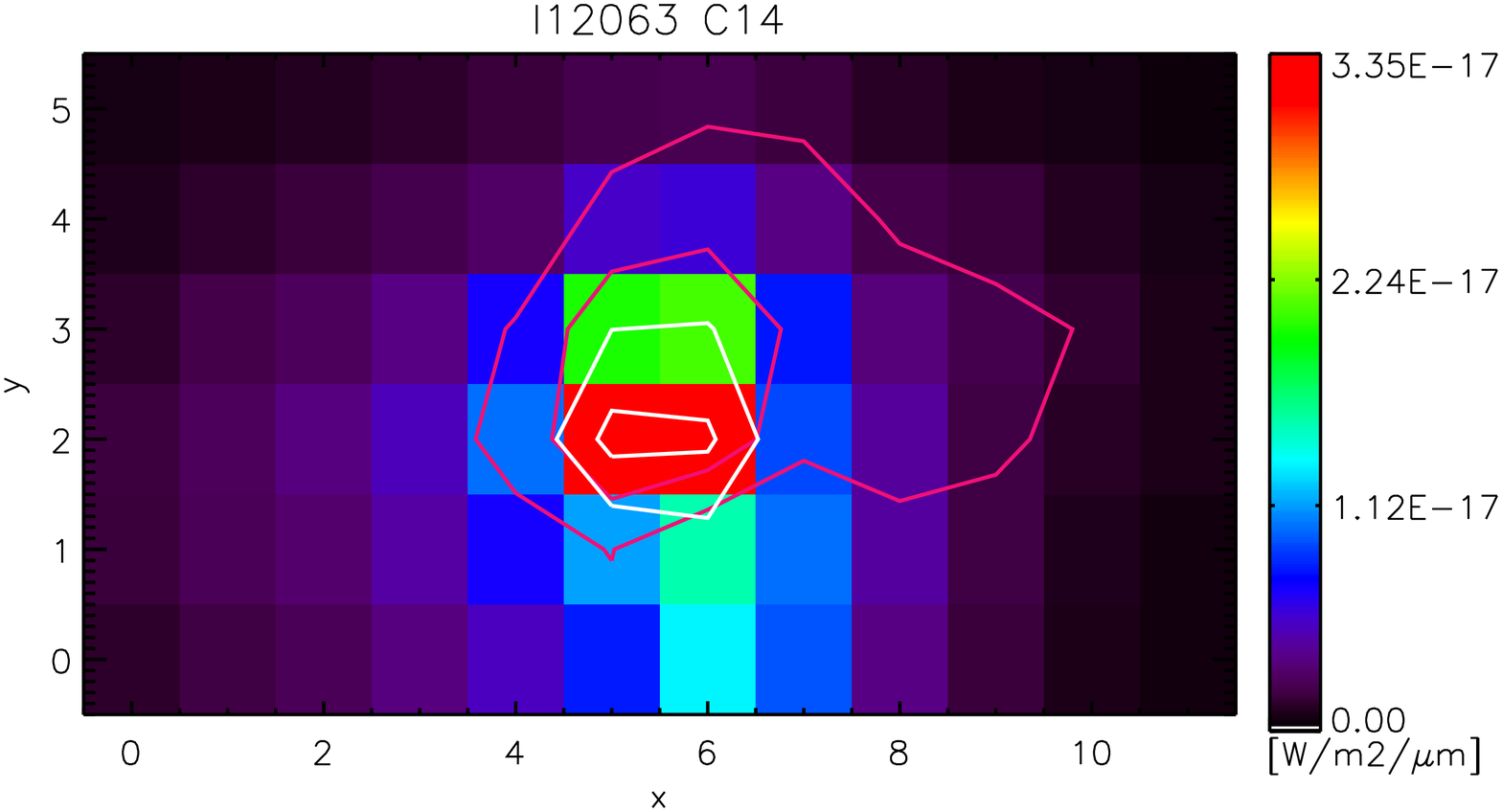}
    \caption{IRAS 12063-6259 spatial emission maps: (top two rows) the four components making up the 7.7~\micron\ emission complex; (bottom row) maps of the `pure' PAH 6.2~\micron\ PAH band and the continuum strength at 14 \micron. The component maps and the 6.2~\micron\ PAH map are in units of W m$^{-2}$ per pixel and the continuum is in units of W m$^{-2}$ \micron$^{-1}$ per pixel. The square of white pixels represents a region where our continuum subtraction method breaks down as described in Section~\ref{sec:data}. The white and red contours represent the distribution of 6.2~\micron\ PAH emission and 14~\micron\ continuum emission respectively. North and east are indicated in the top left panel by the long and short arrows respectively.}    
    \label{fig:I12063maps}
\end{figure*}

\begin{figure*}
  \centering
    \includegraphics[width=5cm]{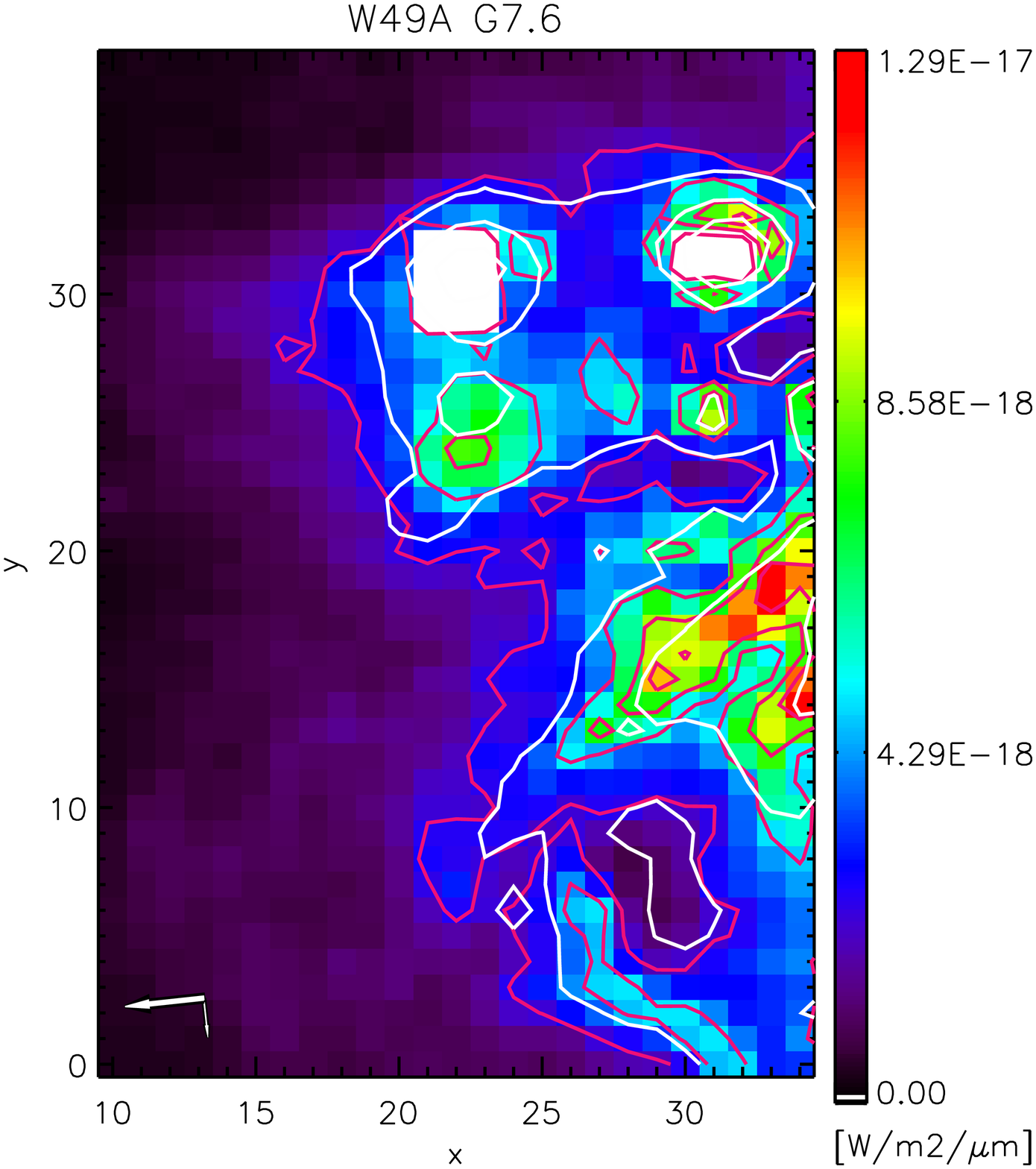}
    \includegraphics[width=5cm]{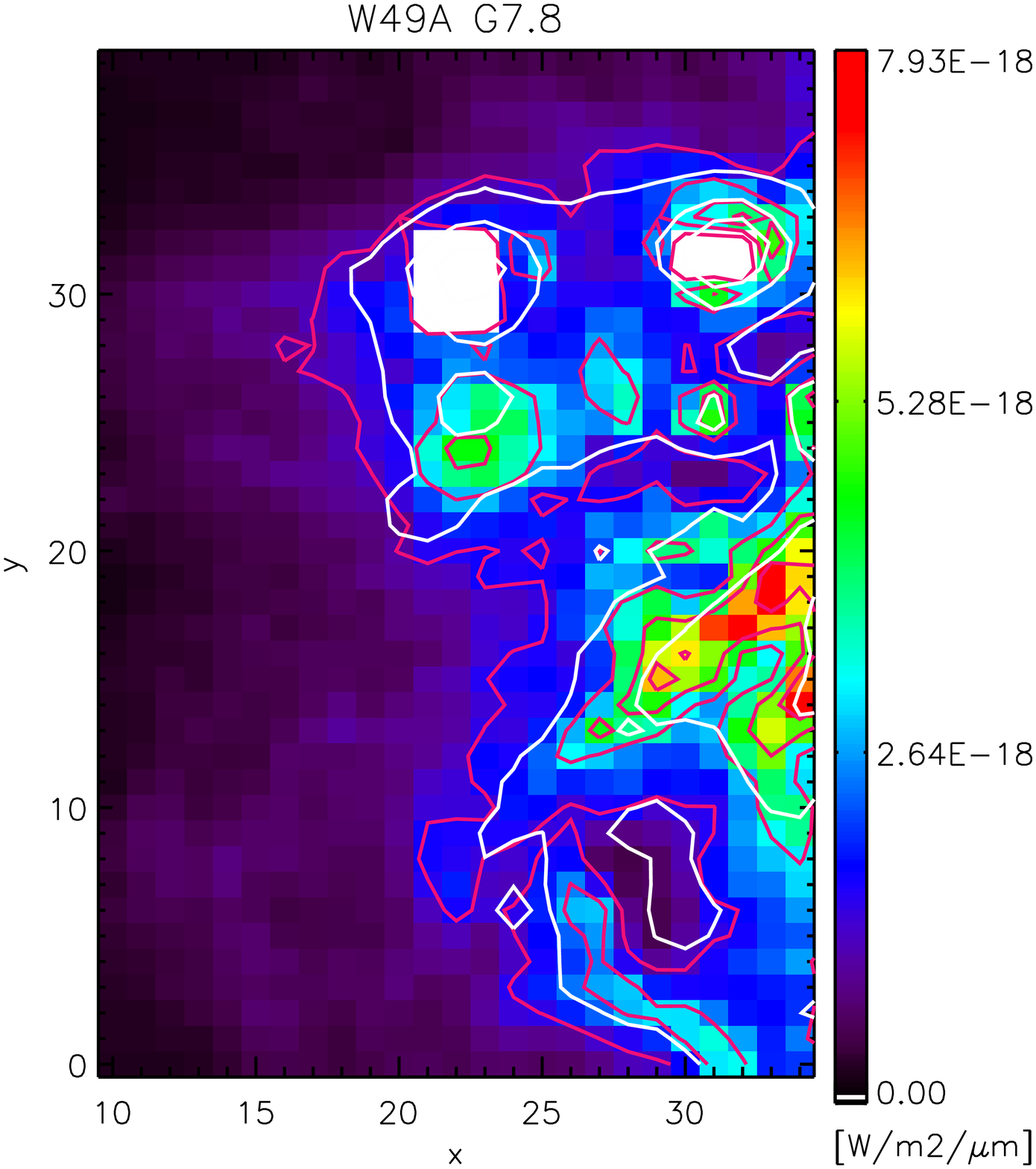}
    \includegraphics[width=5cm]{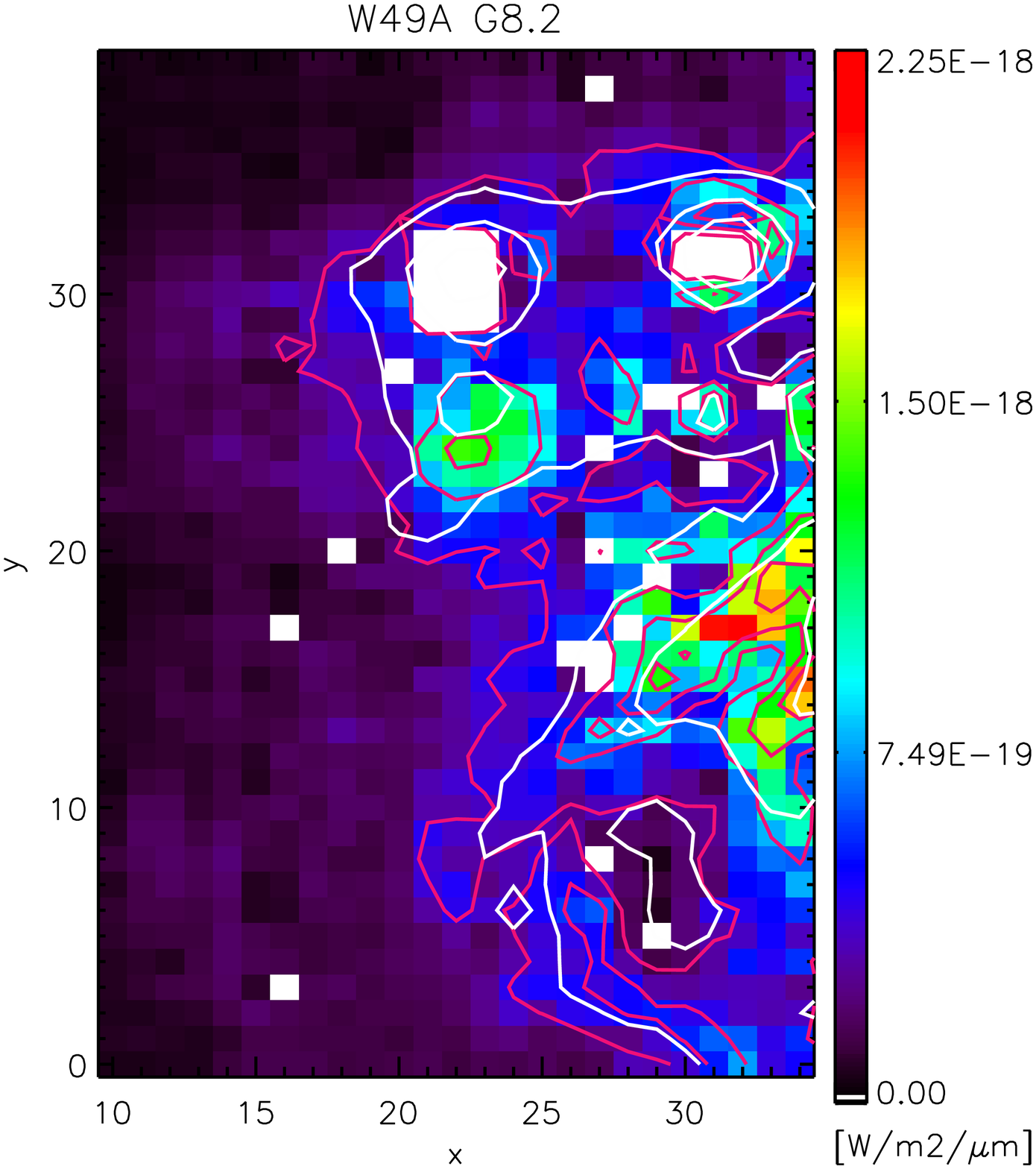}
    \includegraphics[width=5cm]{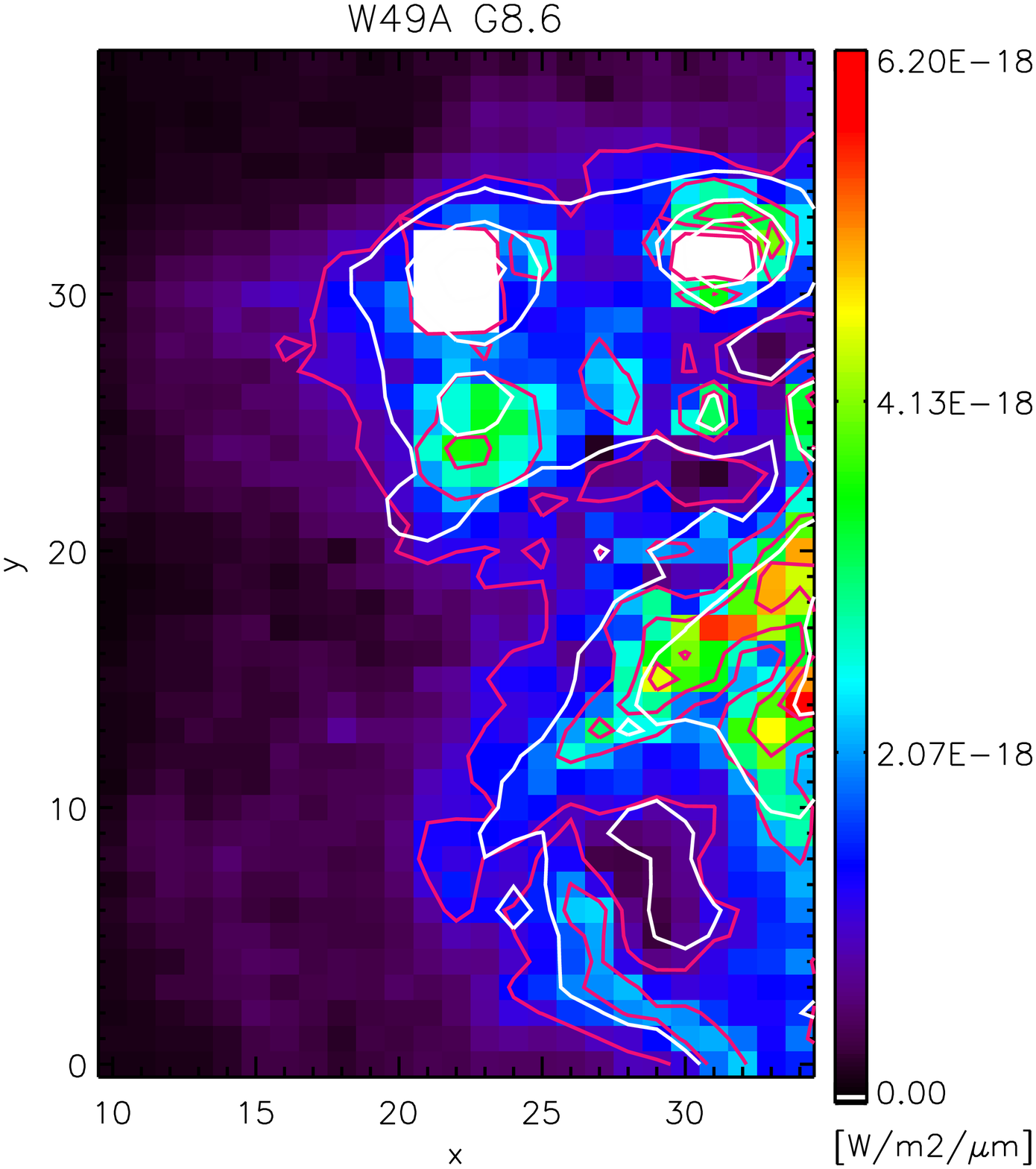}
    \includegraphics[width=5cm]{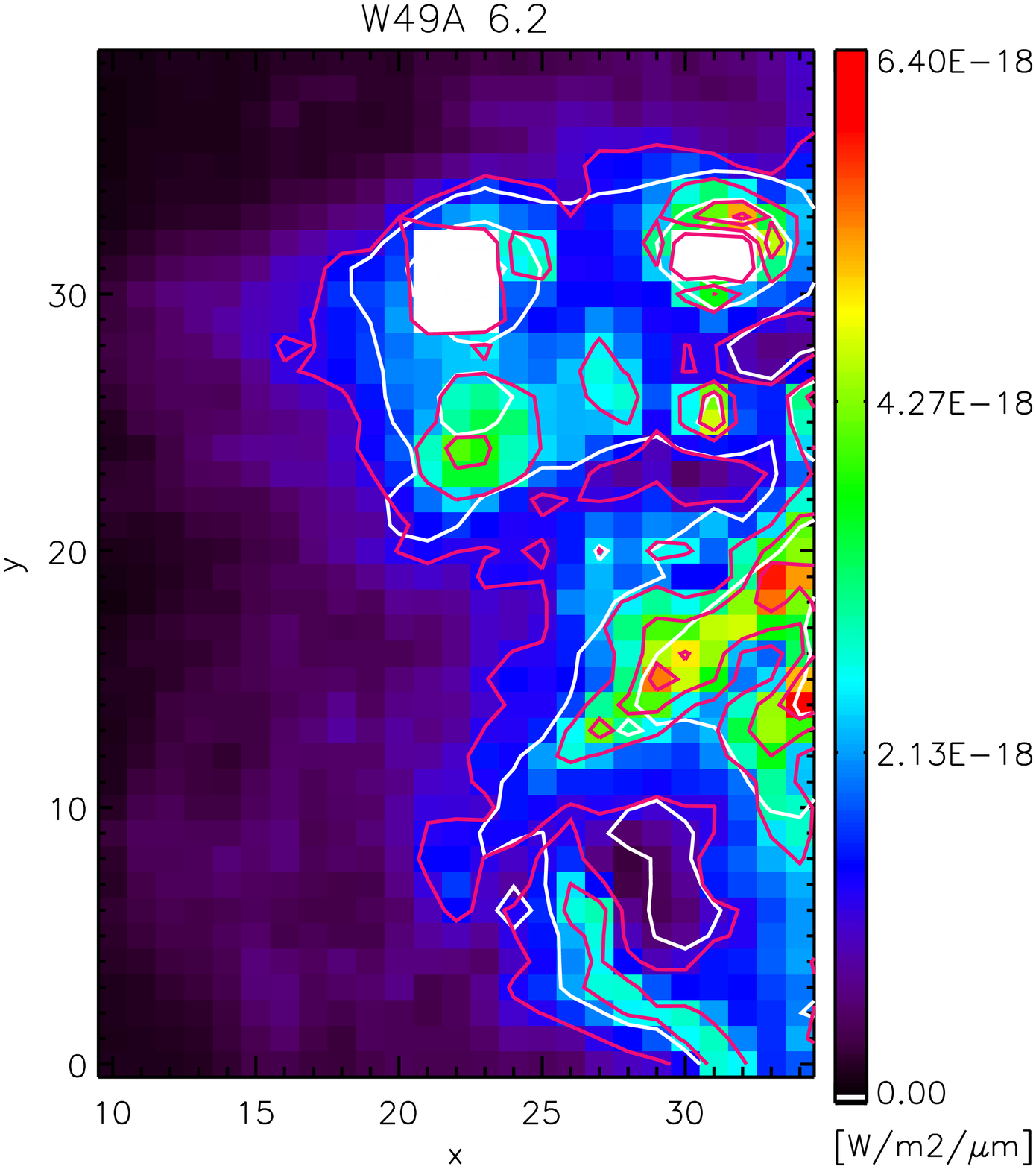}
    \includegraphics[width=5cm]{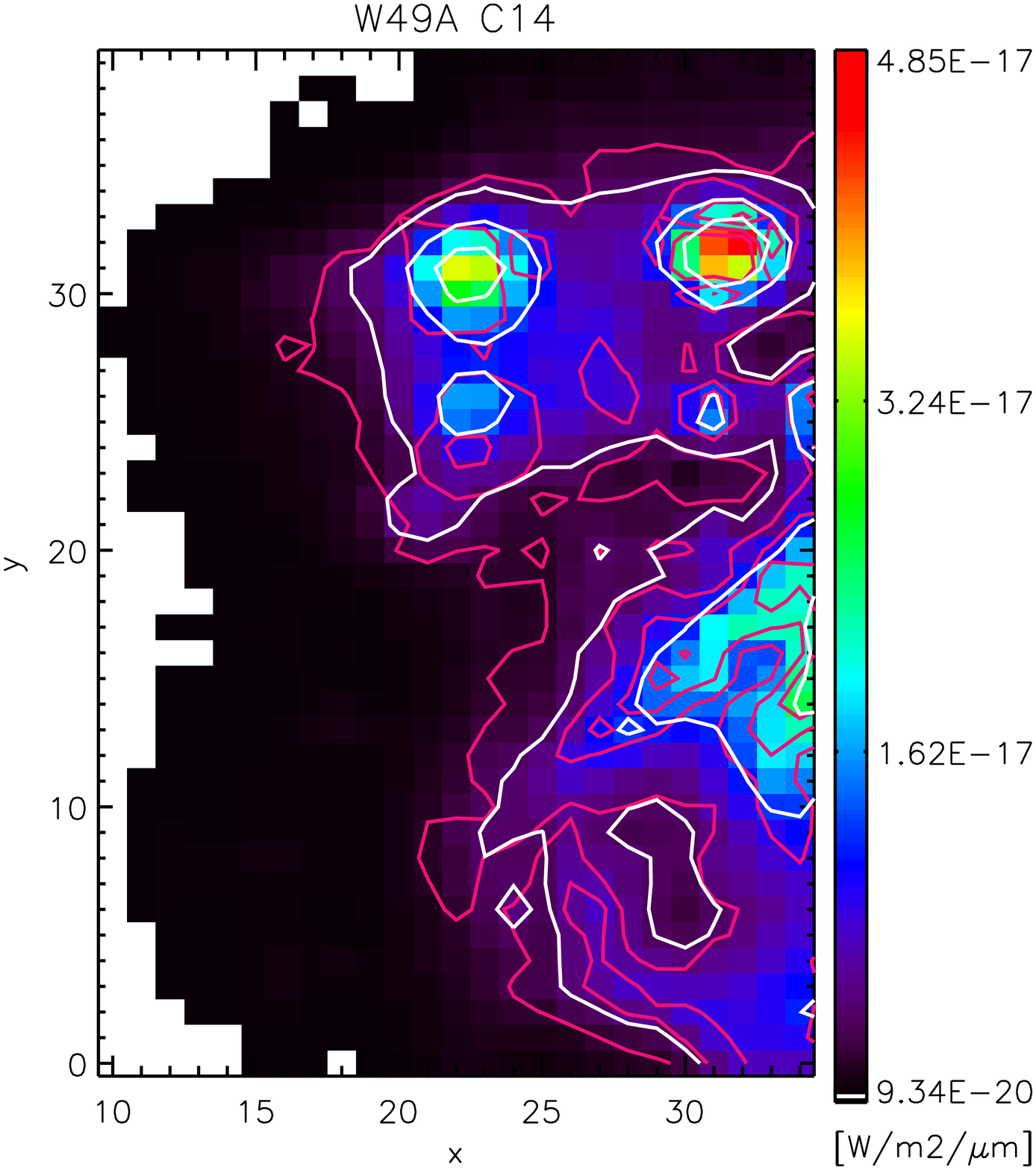}
    \caption{W49A spatial emission maps: (top two rows) the four components making up the 7.7~\micron\ emission complex; (bottom row) maps of the `pure' PAH 6.2~\micron\ PAH band and the continuum strength at 14 \micron. The component maps and the 6.2~\micron\ PAH map are in units of W m$^{-2}$ per pixel and the continuum is in units of W m$^{-2}$ \micron$^{-1}$ per pixel. The prominent squares of white pixels represent the sources W49A-CC and W49A-DD which possess very strong continuum emission leading to poor continuum subtraction, as described in Section~\ref{sec:data}. The other white pixels represent regions of low SNR.  The white and red contours represent the distribution of 6.2~\micron\ PAH emission and 14~\micron\ continuum emission respectively. North and east are indicated in the top left panel by the long and short arrows respectively.}   
    \label{fig:W49Amaps}
\end{figure*}

\begin{figure*}
  \centering
    \includegraphics[width=8cm]{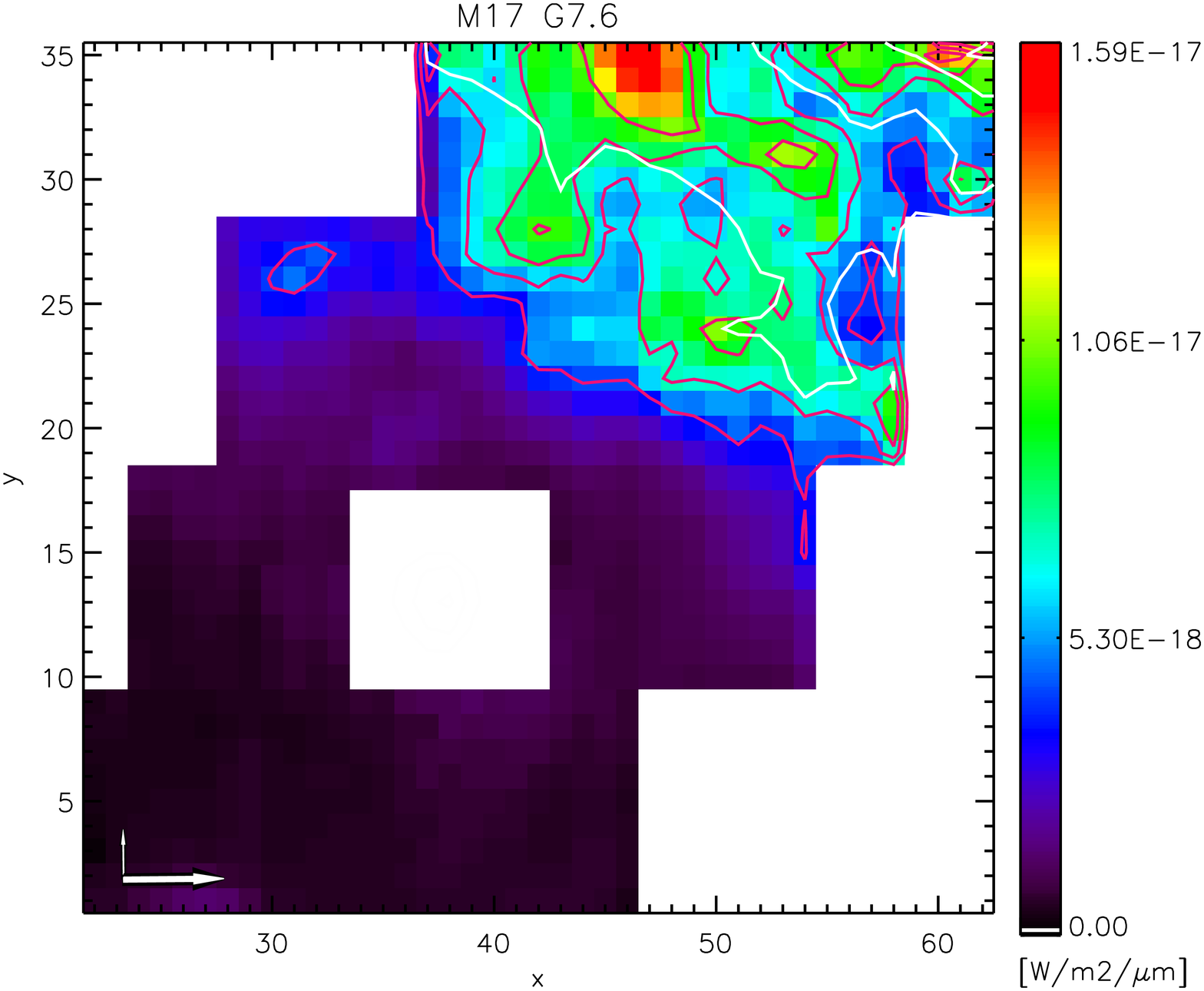}
    \includegraphics[width=8cm]{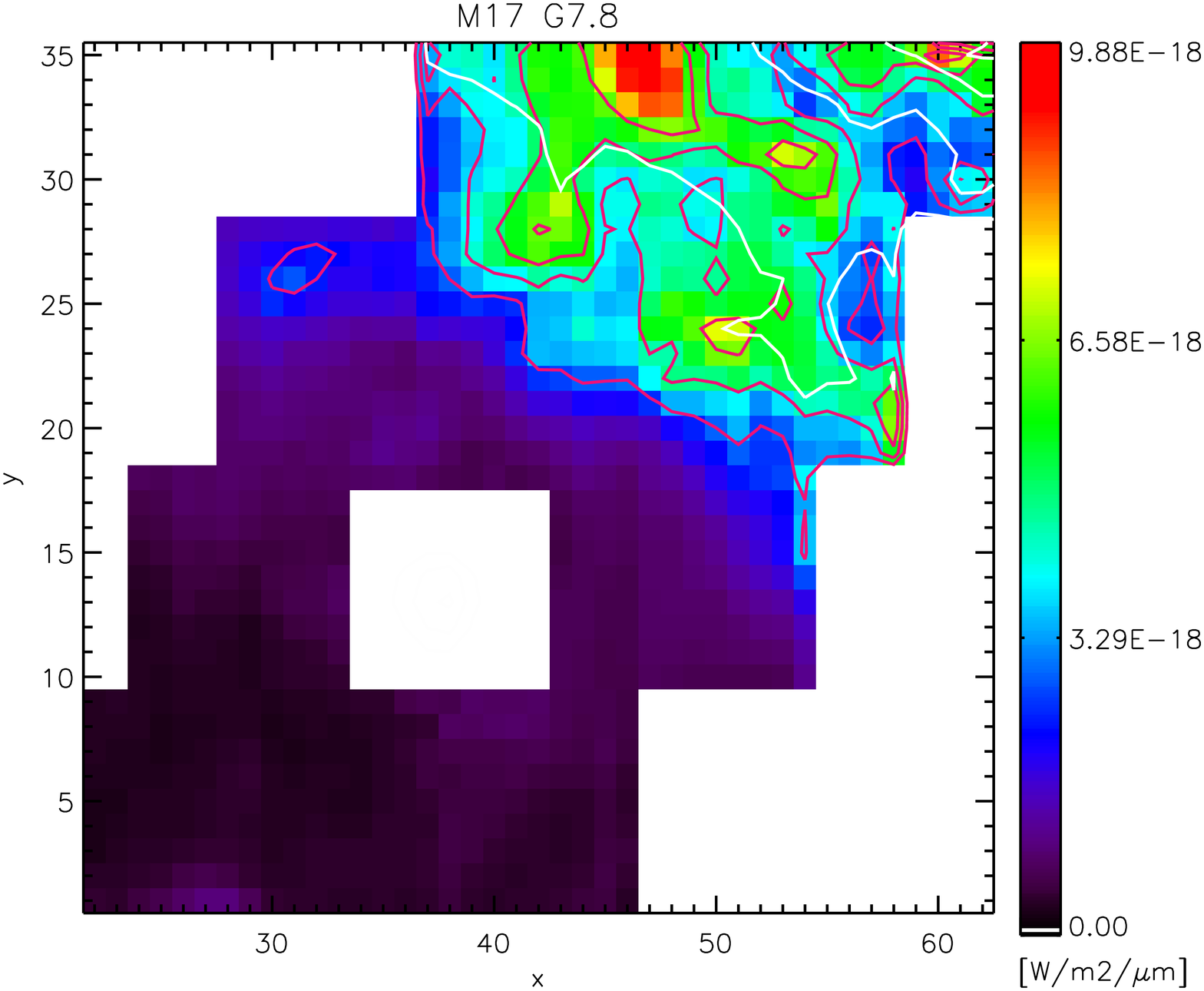}
    \includegraphics[width=8cm]{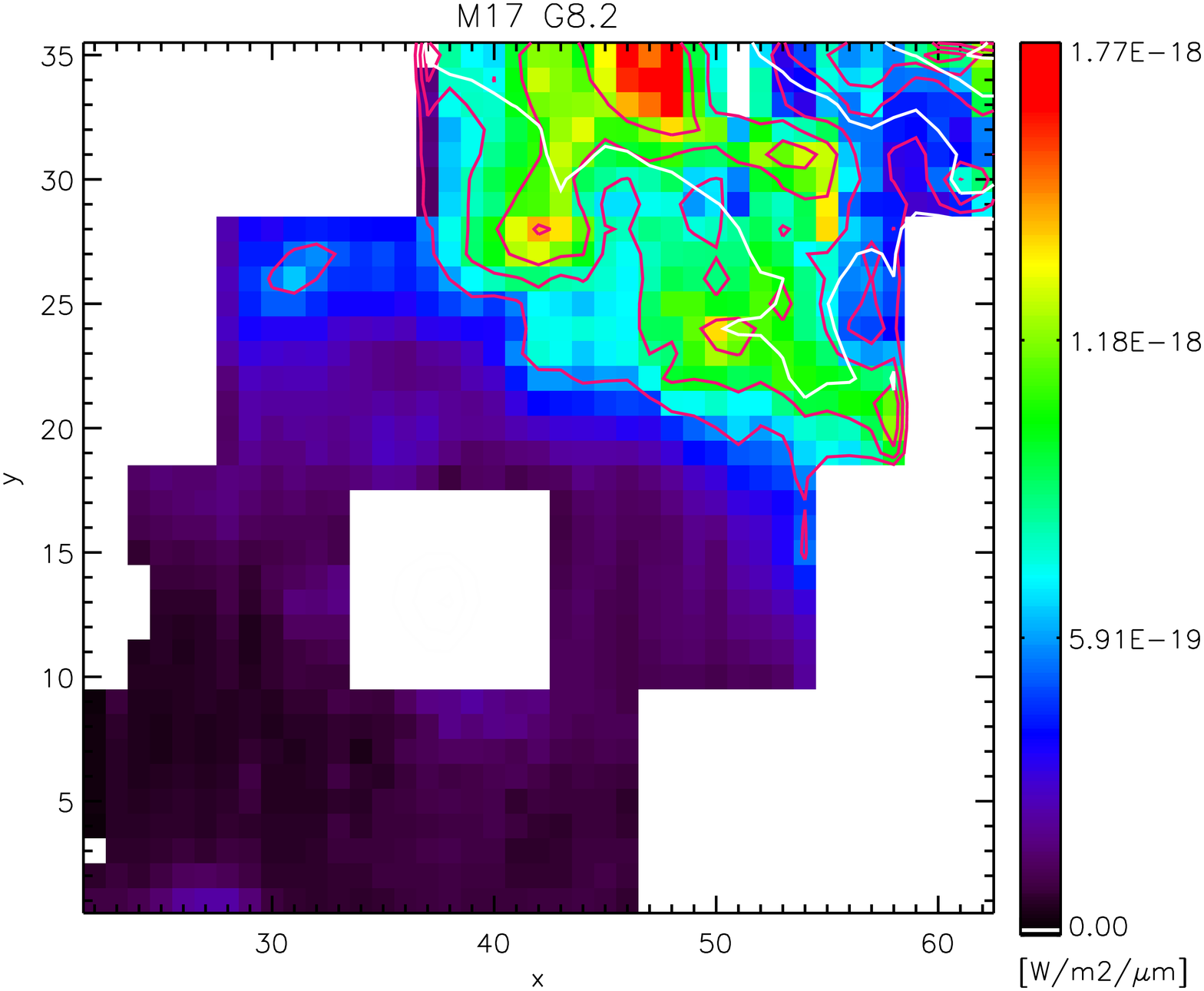}
    \includegraphics[width=8cm]{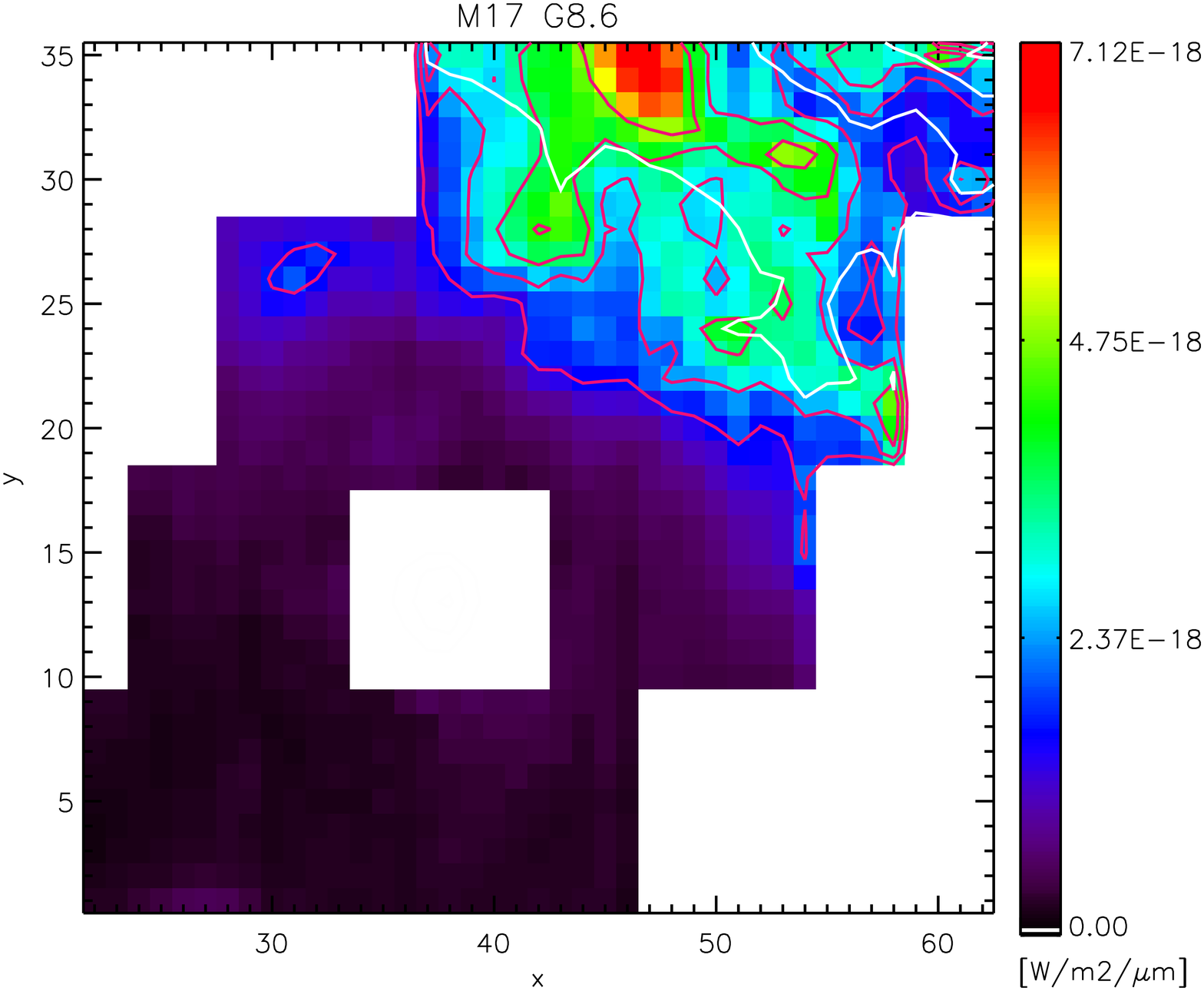}
    \includegraphics[width=8cm]{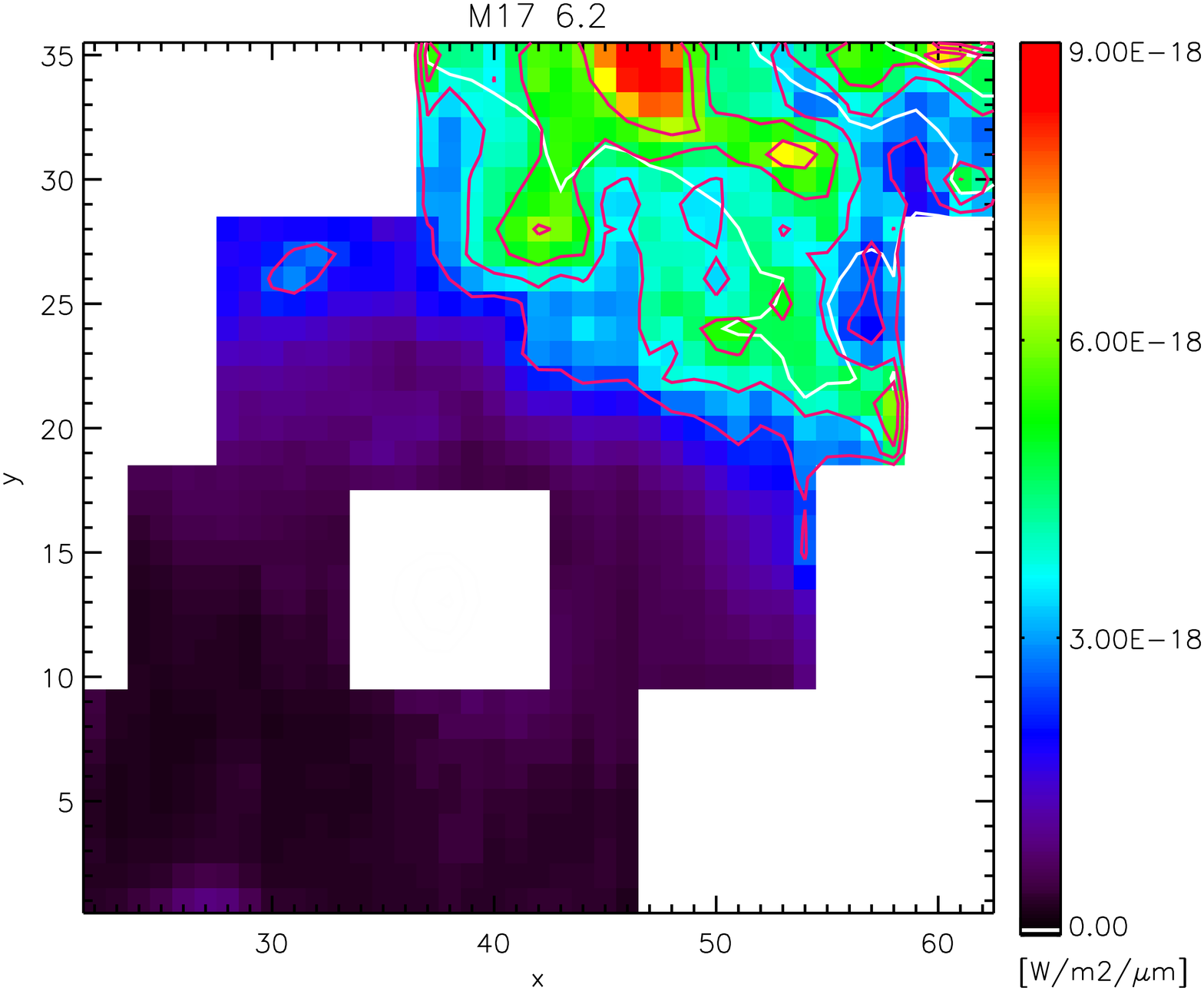}
    \includegraphics[width=8cm]{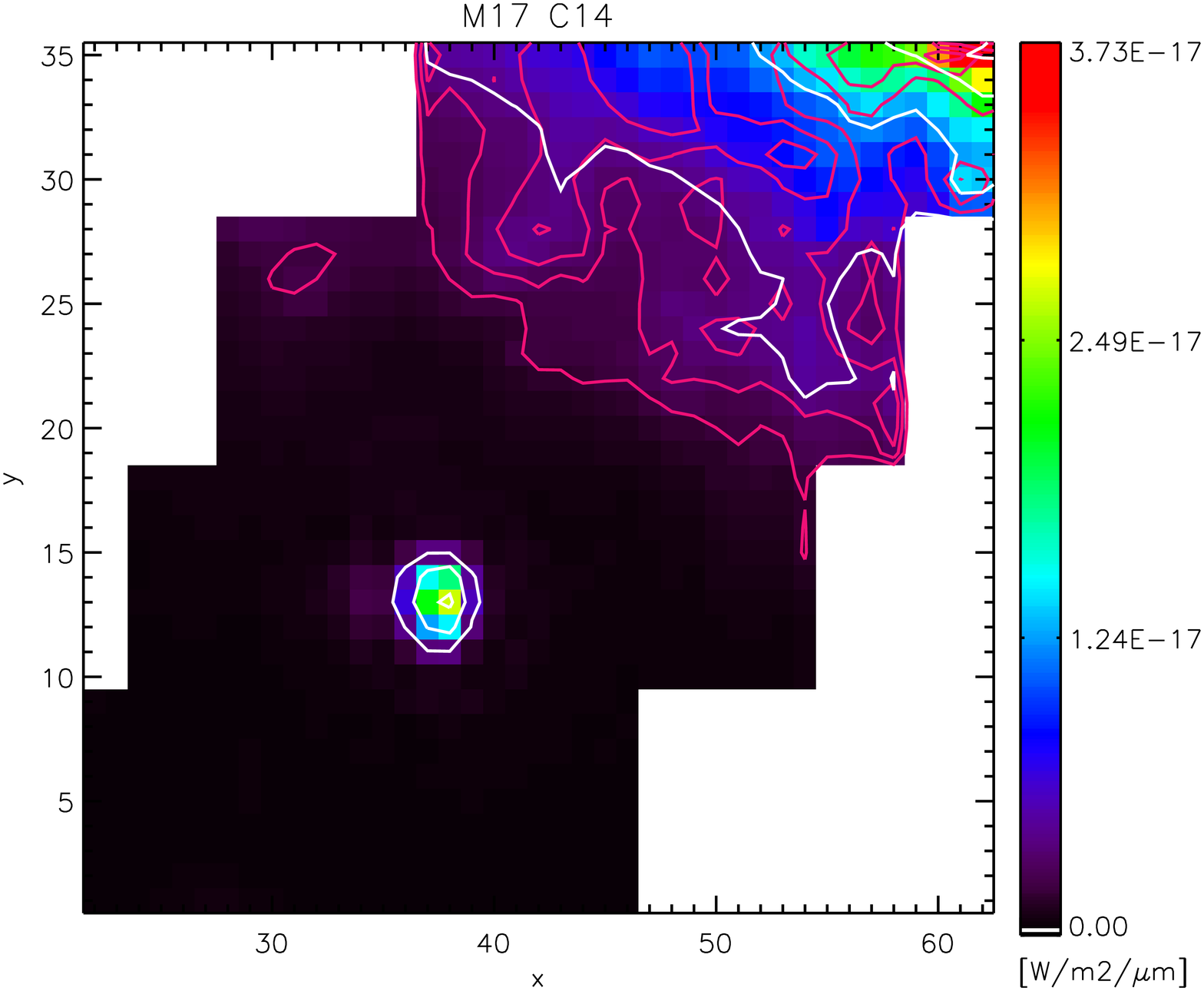}
    \caption{M 17 spatial emission maps: (top two rows) the four components making up the 7.7~\micron\ emission complex; (bottom row) maps of the `pure' PAH 6.2~\micron\ PAH band and the continuum strength at 14 \micron. The component maps and the 6.2~\micron\ PAH map are in units of W m$^{-2}$ per pixel and the continuum is in units of W m$^{-2}$ \micron$^{-1}$ per pixel. The large white square in the center of the map has been removed as the emission in those pixels is dominated by a bright protostar -- as evident in the continuum map. The white and red contours represent the distribution of 6.2~\micron\ PAH emission and 14~\micron\ continuum emission respectively. North and east are indicated in the top left panel by the long and short arrows respectively.}    
    \label{fig:M17maps}
\end{figure*}

\begin{figure*}
  \centering

    \includegraphics[width=5cm]{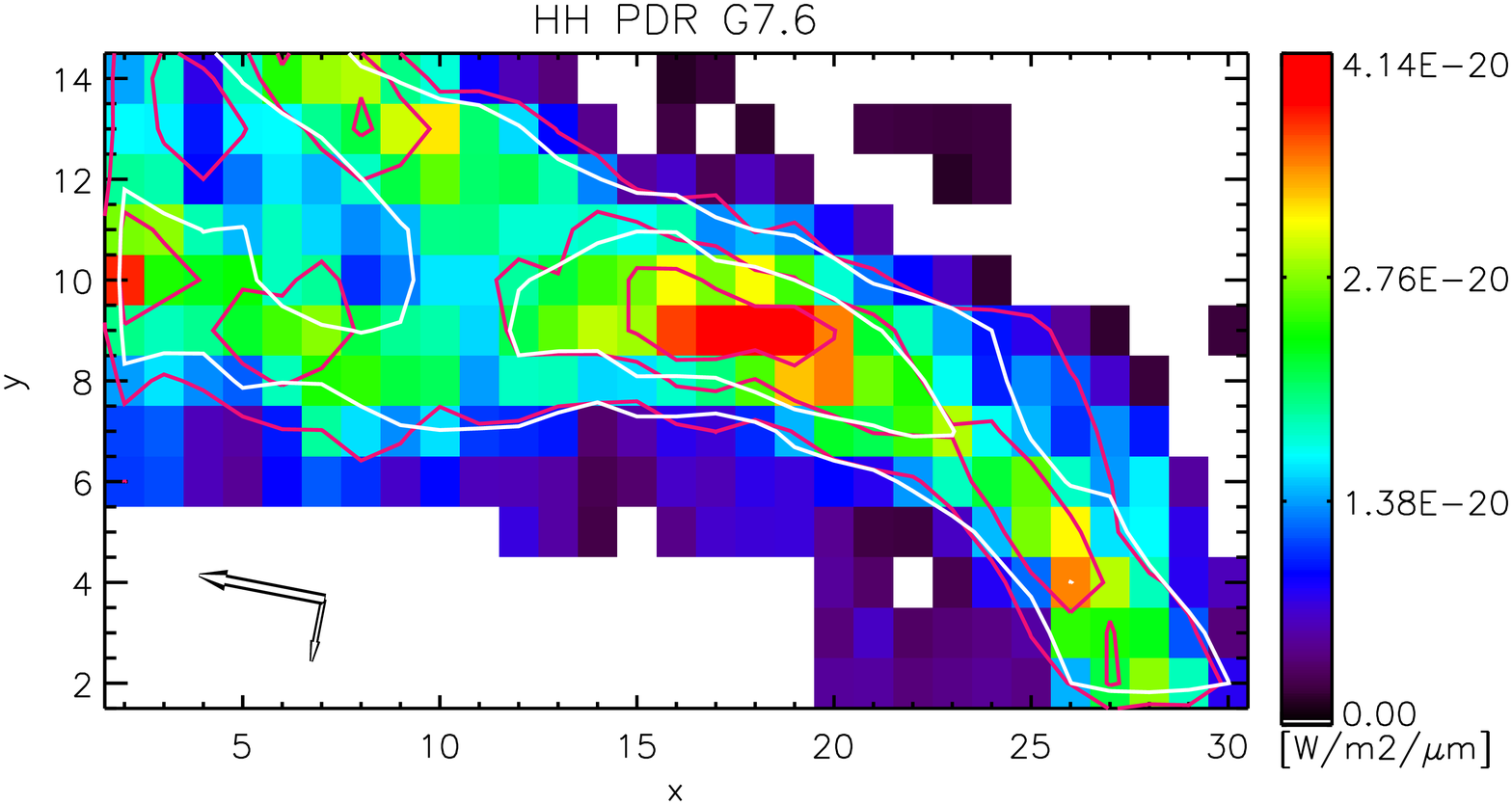}
    \includegraphics[width=5cm]{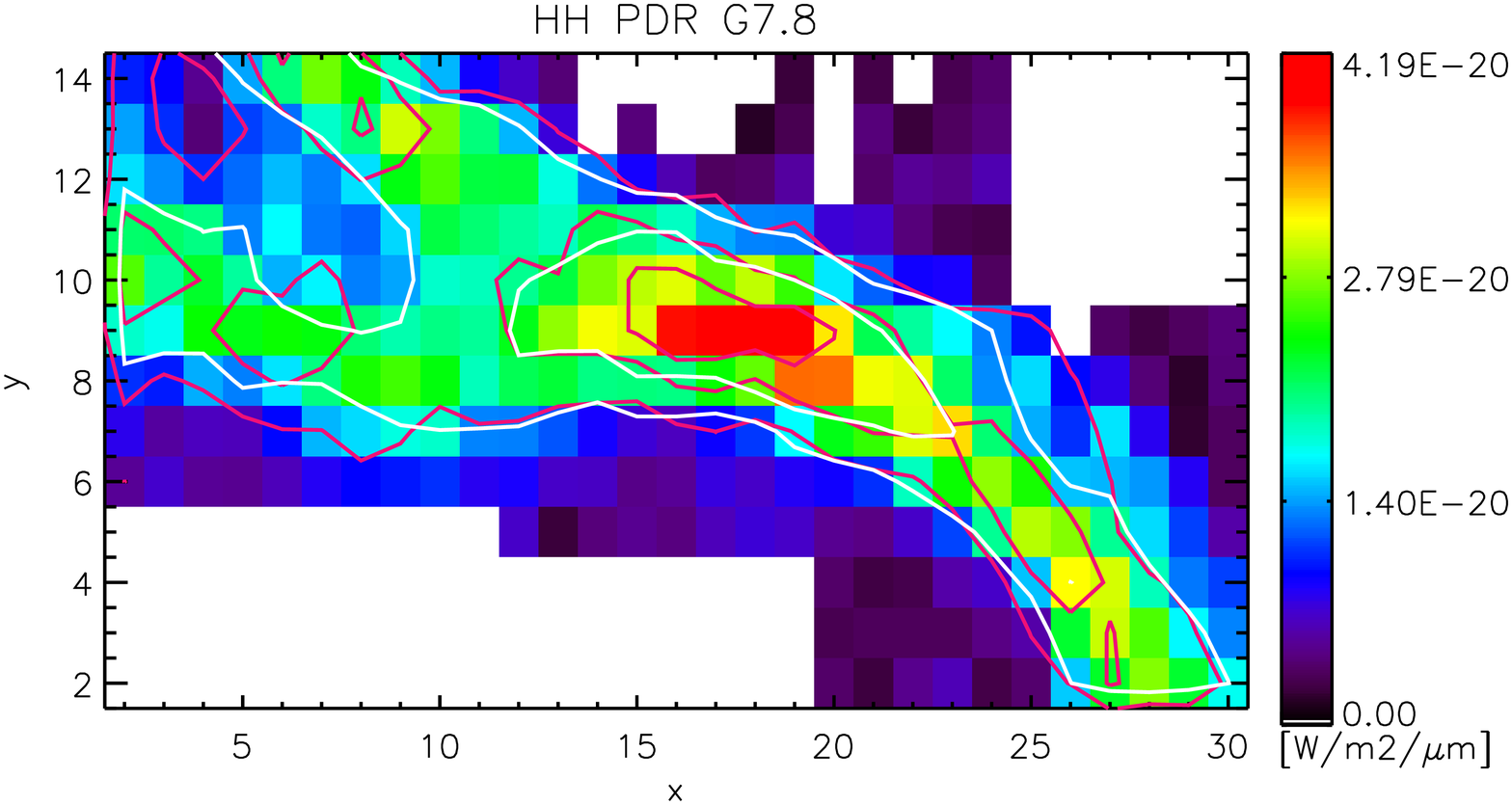}
    \includegraphics[width=5cm]{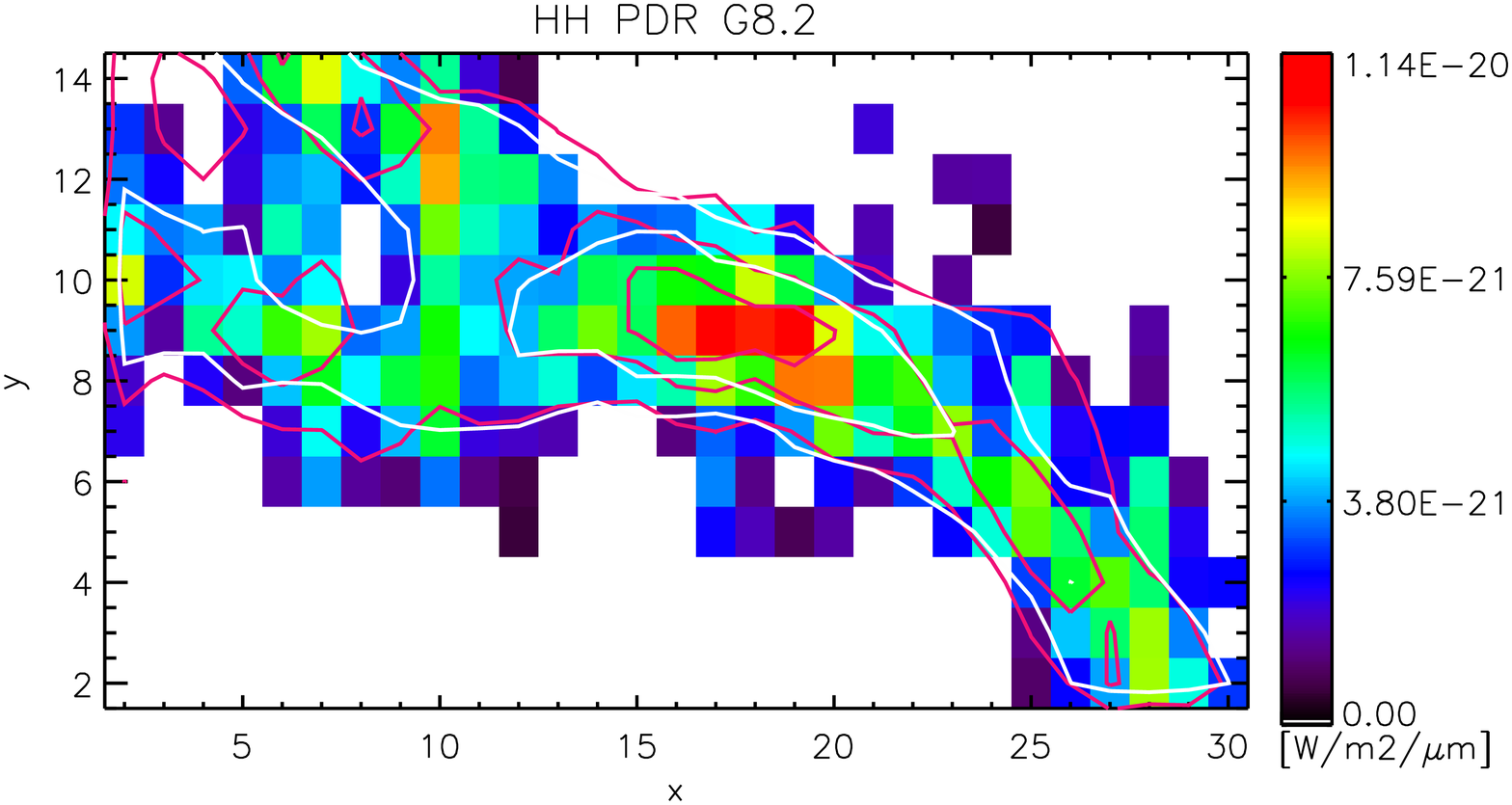}
    \includegraphics[width=5cm]{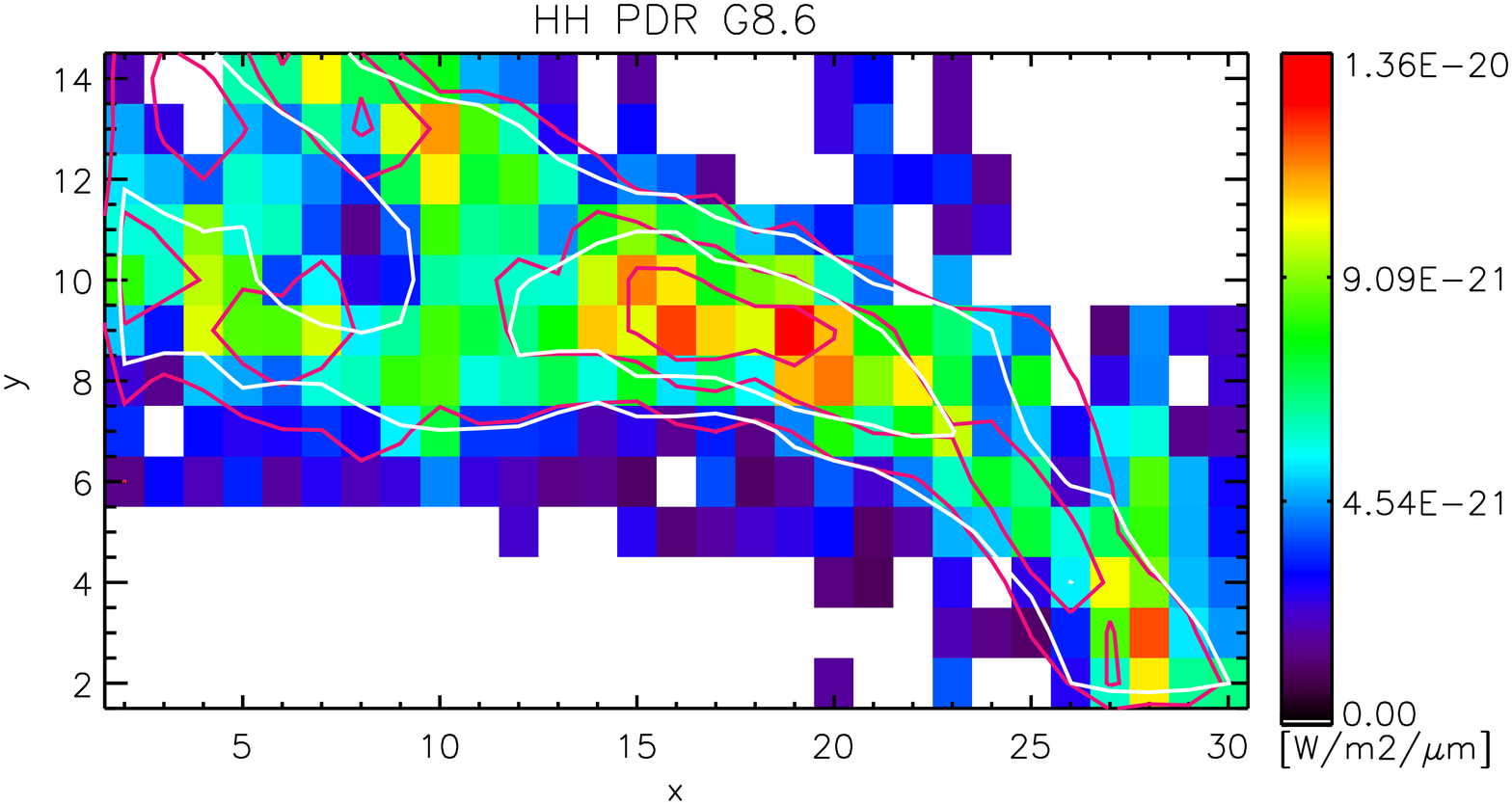}
    \includegraphics[width=5cm]{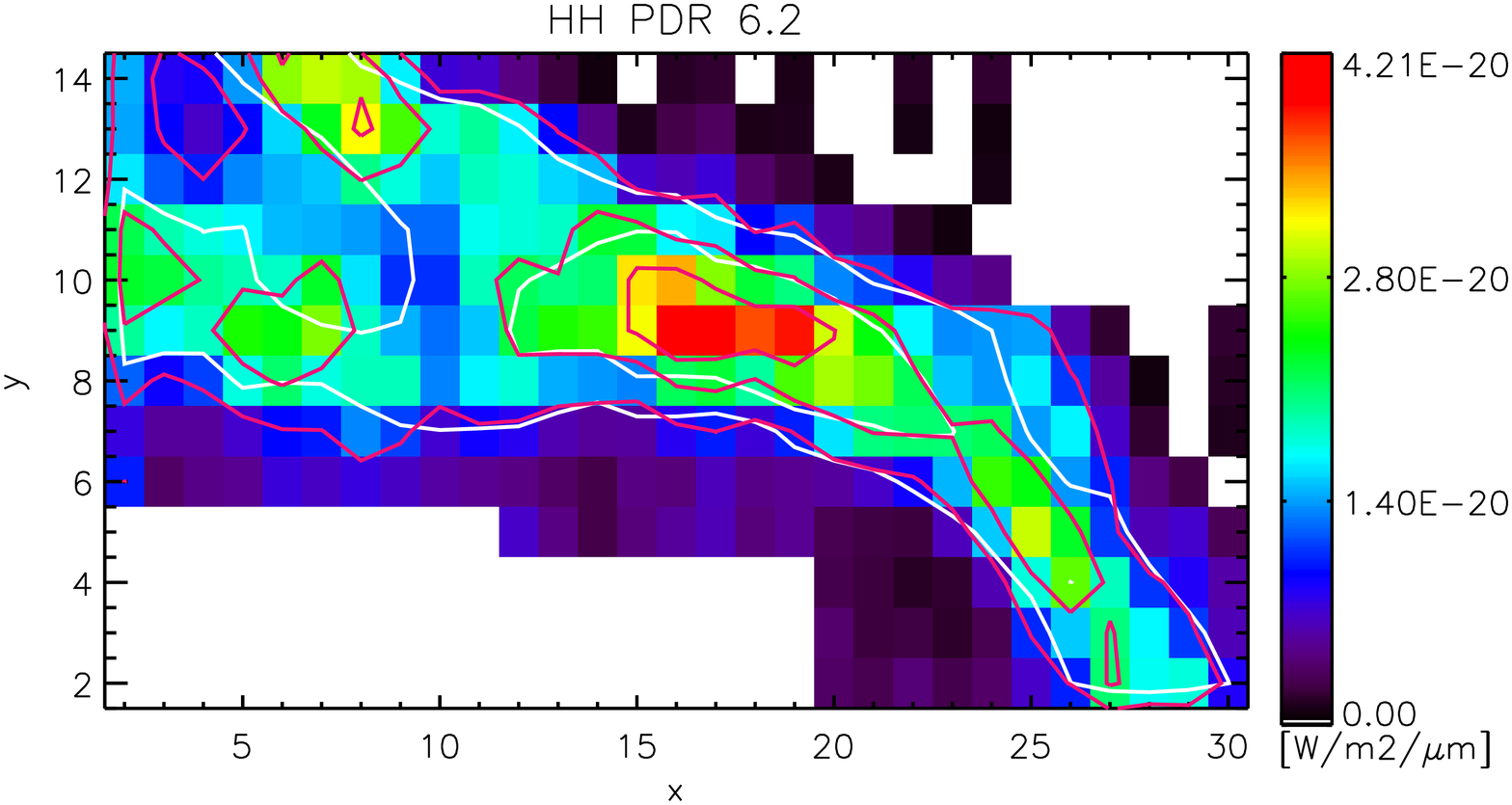}
    \includegraphics[width=5cm]{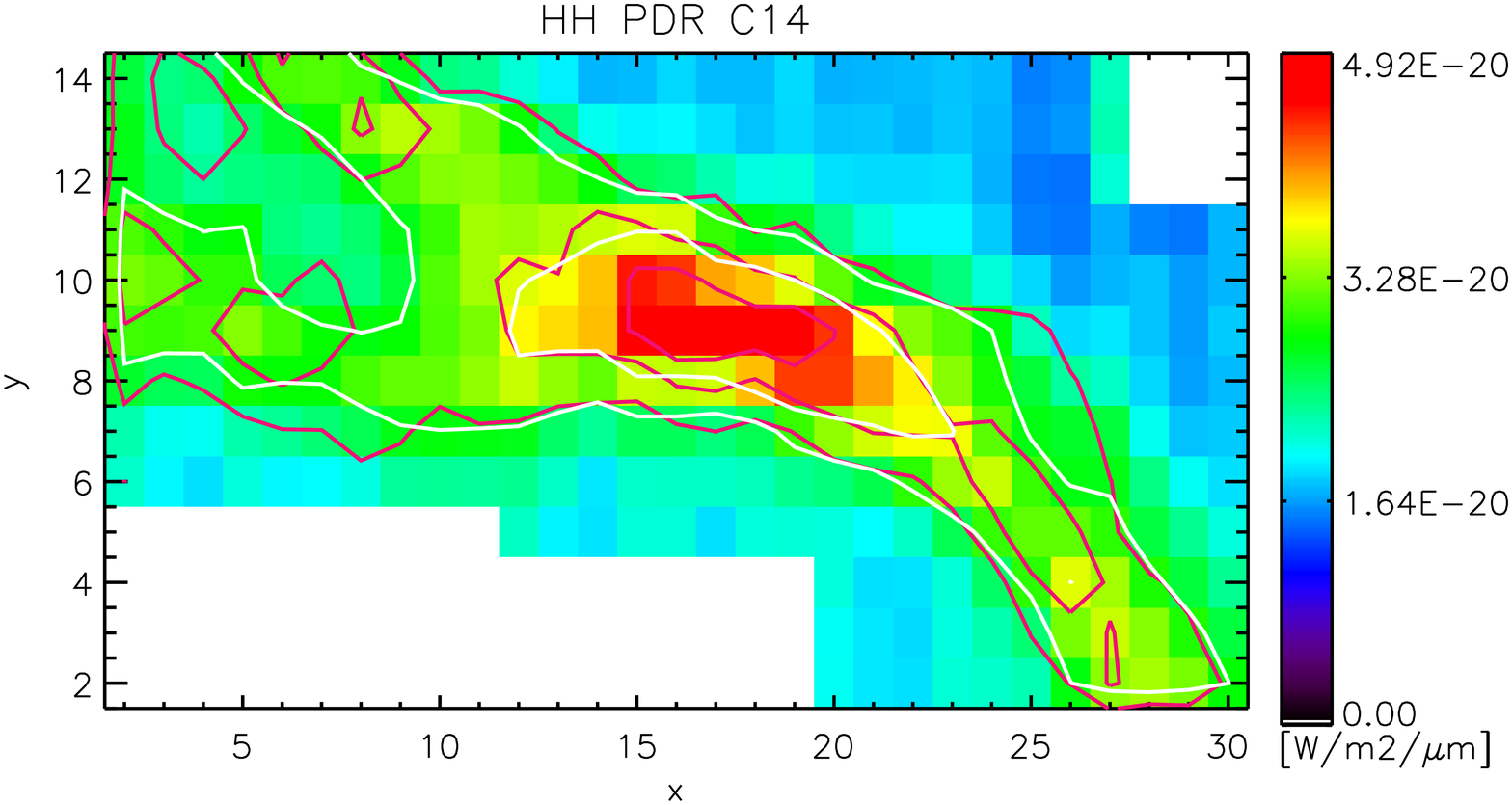}
    \caption{Horsehead nebula spatial emission maps: (top two rows) the four components making up the 7.7~\micron\ emission complex; (bottom row) maps of the `pure' PAH 6.2~\micron\ PAH band and the continuum strength at 14 \micron. The component maps and the 6.2~\micron\ PAH map are in units of W m$^{-2}$ per pixel and the continuum is in units of W m$^{-2}$ \micron$^{-1}$ per pixel. Regions of low SNR have been shown in white. The white and red contours represent the distribution of 6.2~\micron\ PAH emission and 14~\micron\ continuum emission respectively. North and east are indicated in the top left panel by the long and short arrows respectively.}   
    \label{fig:IC434maps}
\end{figure*}

\begin{figure*}
  \centering
    \includegraphics[width=5cm]{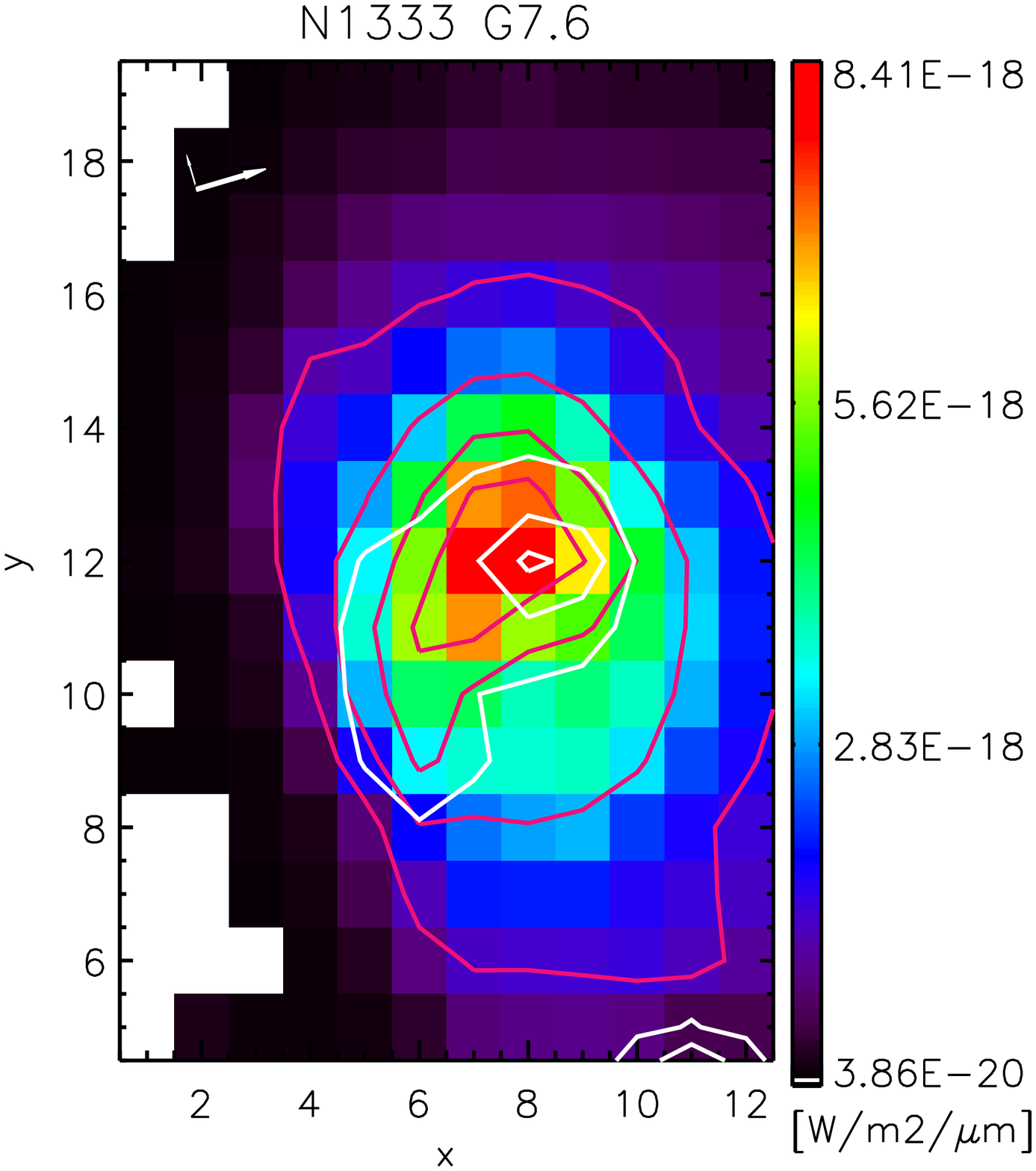}
    \includegraphics[width=5cm]{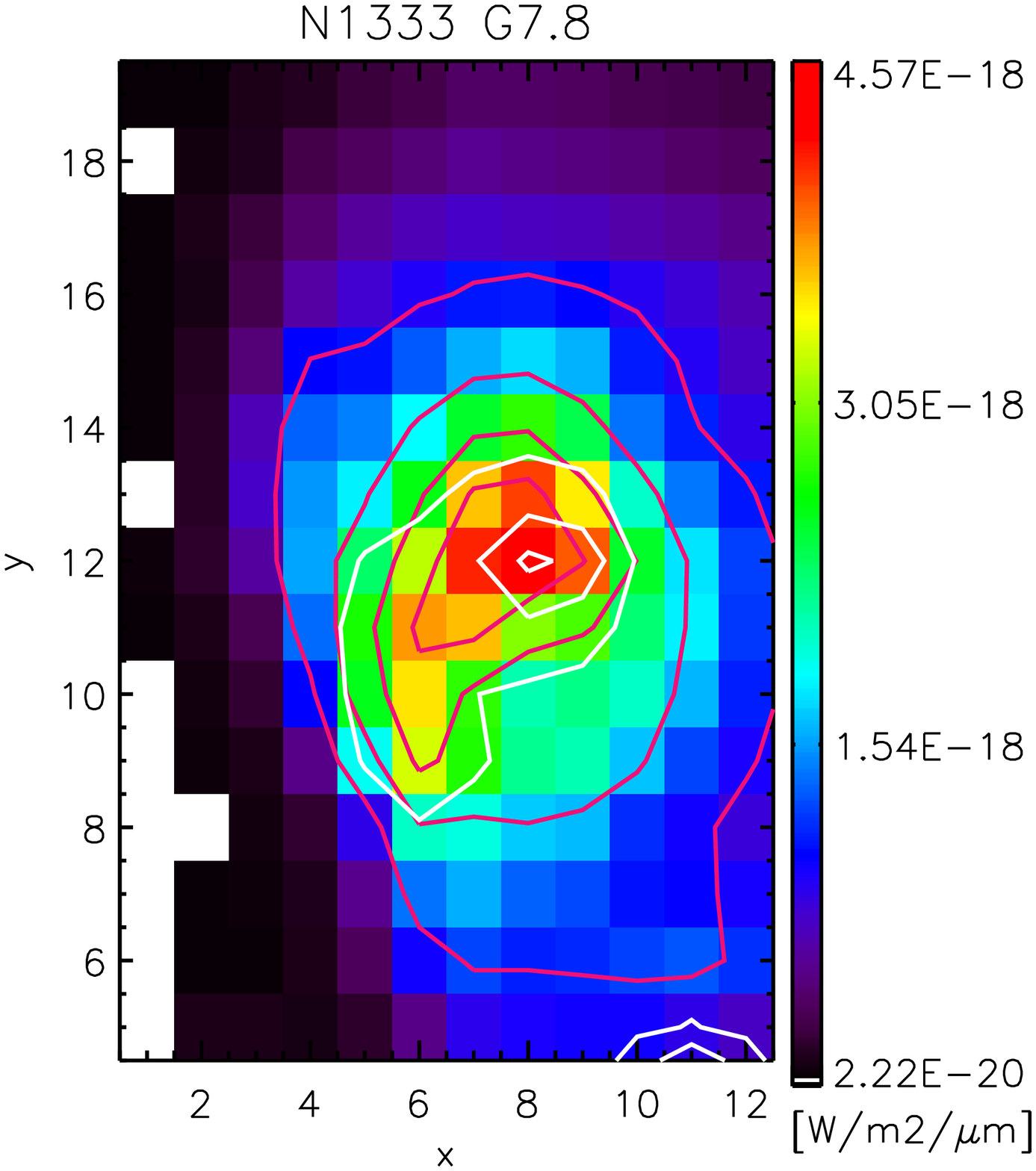}
    \includegraphics[width=5cm]{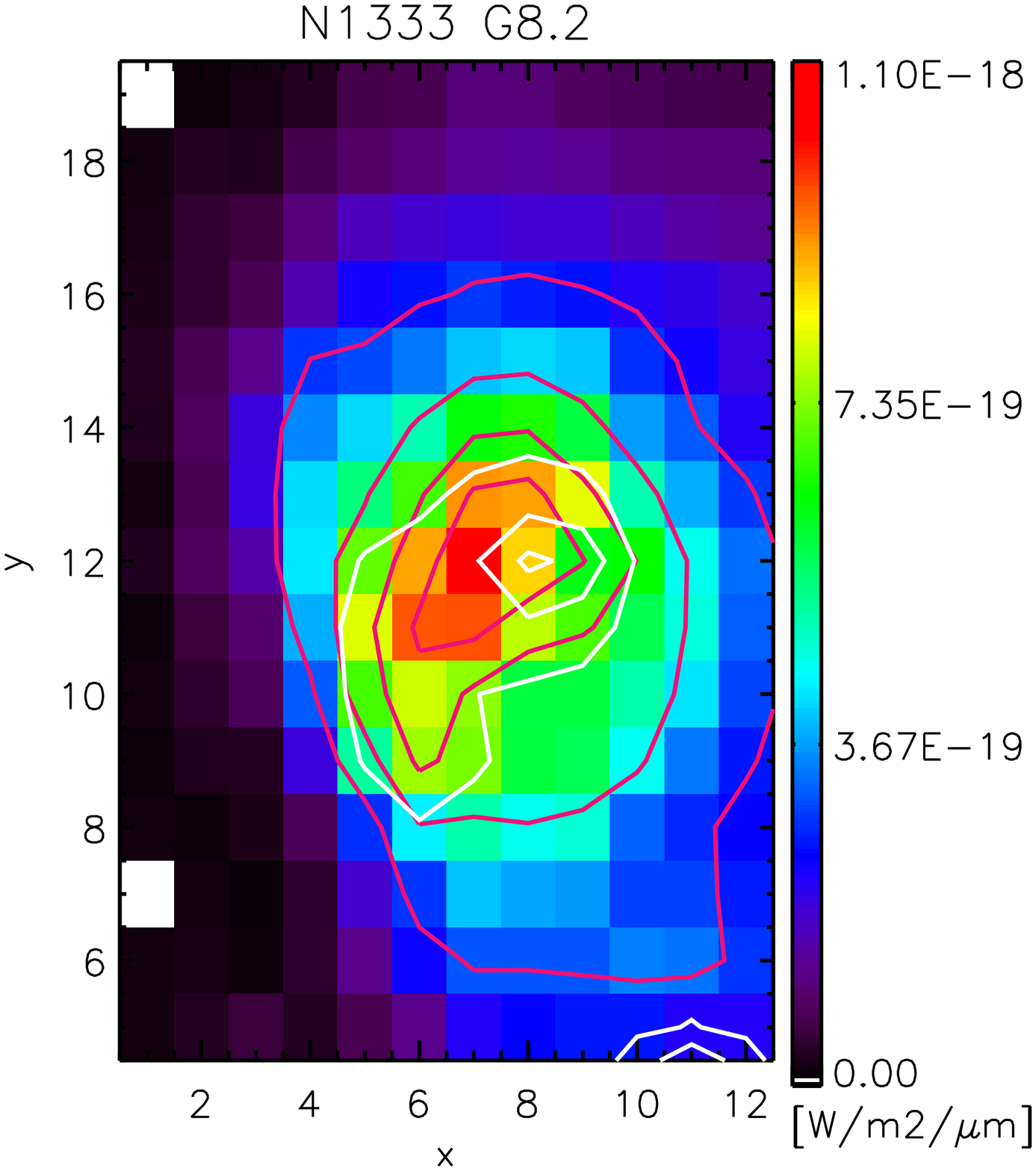}
    \includegraphics[width=5cm]{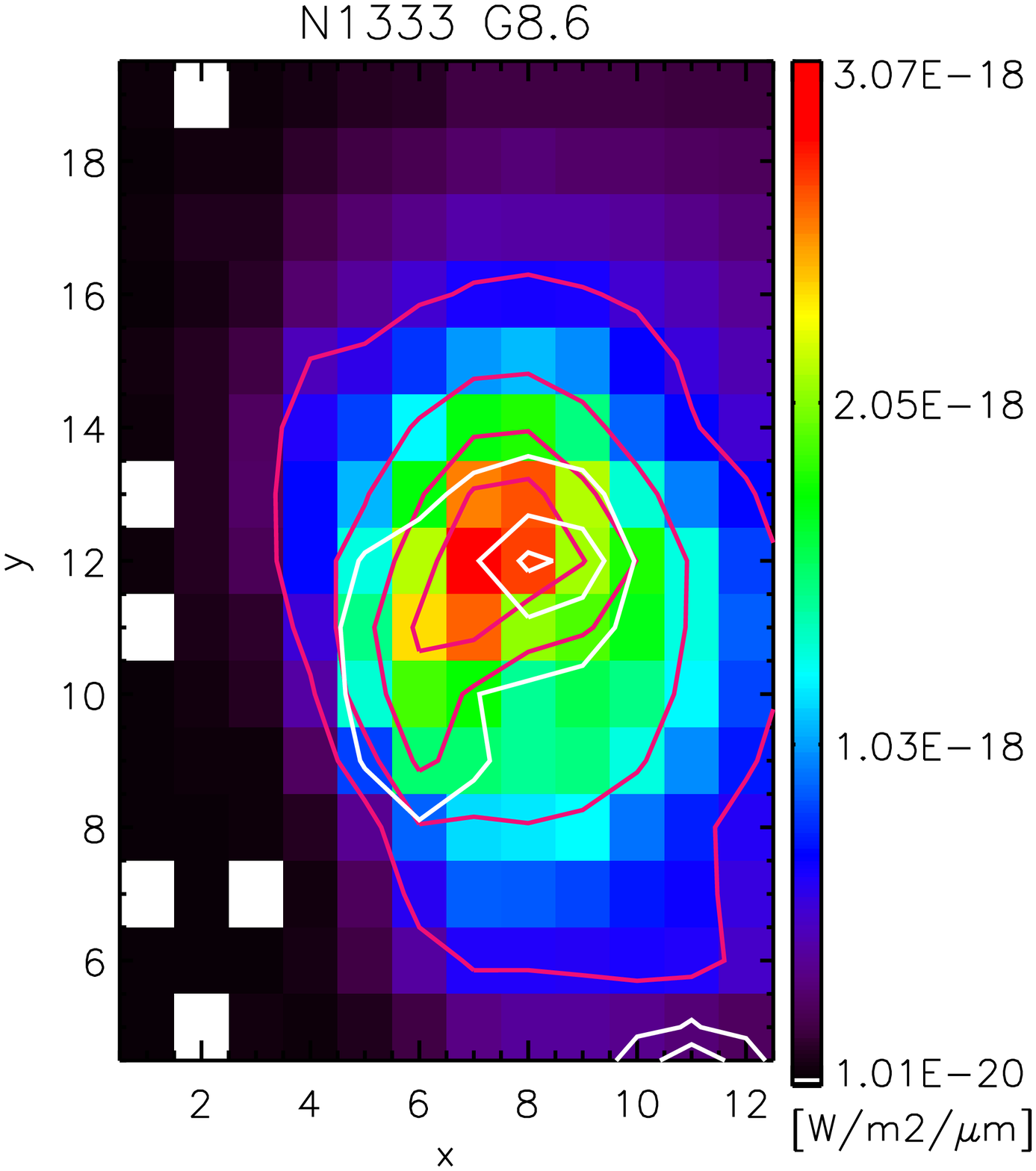}
    \includegraphics[width=5cm]{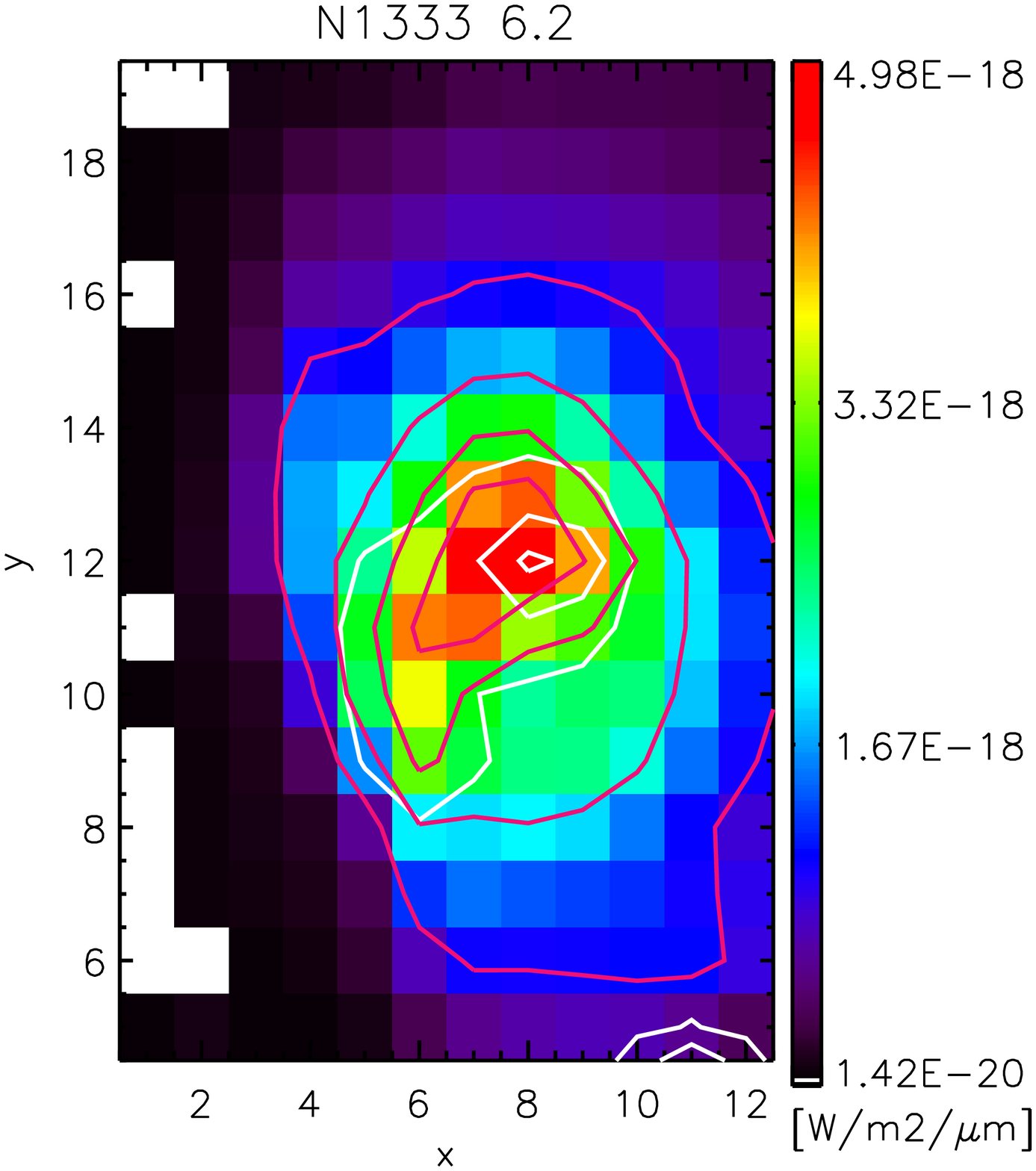}
    \includegraphics[width=5cm]{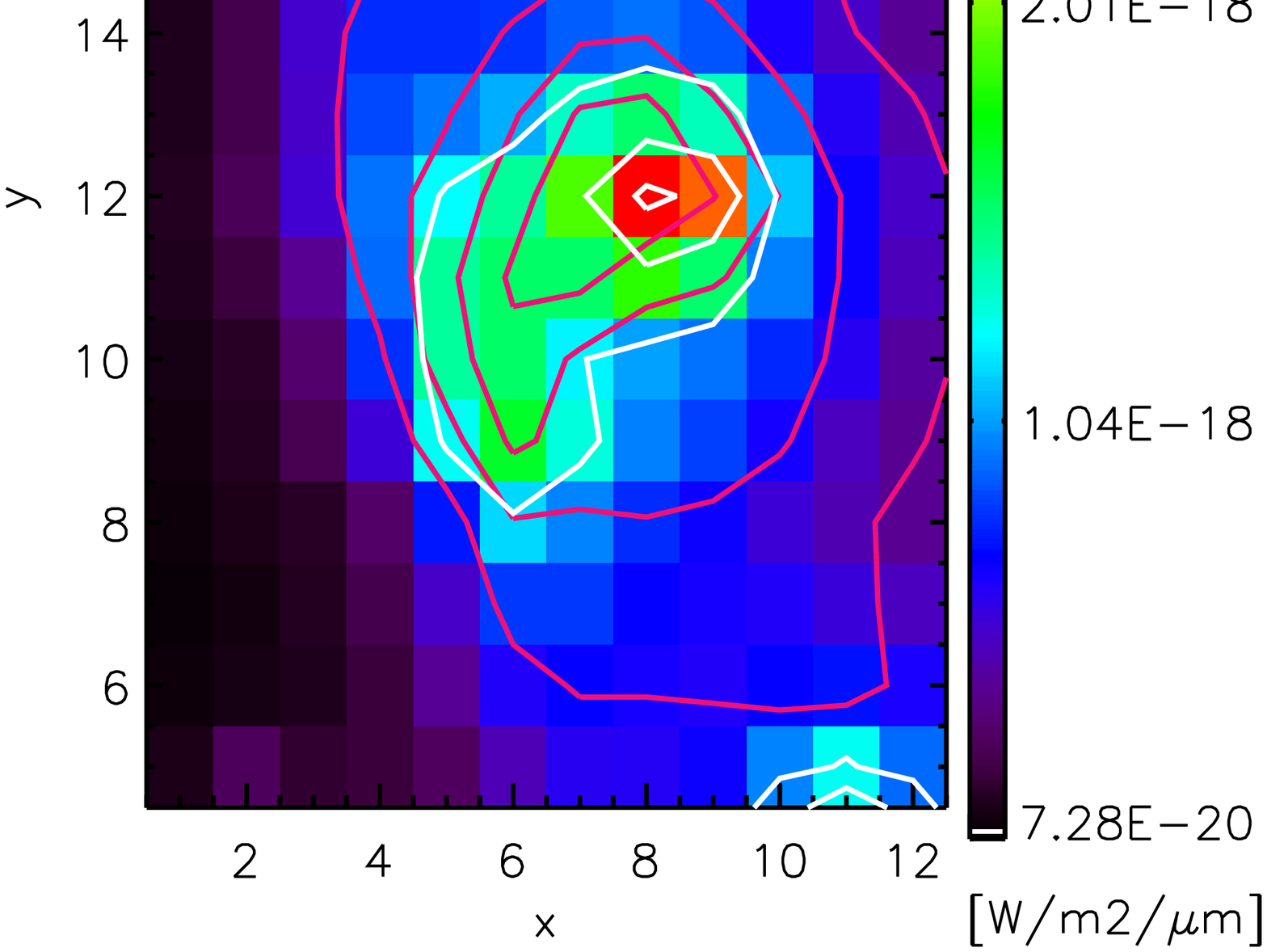}
    \caption{NGC~1333 spatial emission maps: (top two rows) the four components making up the 7.7~\micron\ emission complex; (bottom row) maps of the `pure' PAH 6.2~\micron\ PAH band and the continuum strength at 14 \micron. The component maps and the 6.2~\micron\ PAH map are in units of W m$^{-2}$ per pixel and the continuum is in units of W m$^{-2}$ \micron$^{-1}$ per pixel. Regions of low SNR have been shown in white. The white and red contours represent the distribution of 6.2~\micron\ PAH emission and 14~\micron\ continuum emission respectively. North and east are indicated in the top left panel by the long and short arrows respectively.}   
    \label{fig:N1333maps}
\end{figure*}

\begin{figure*}
  \centering

    \includegraphics[width=5cm]{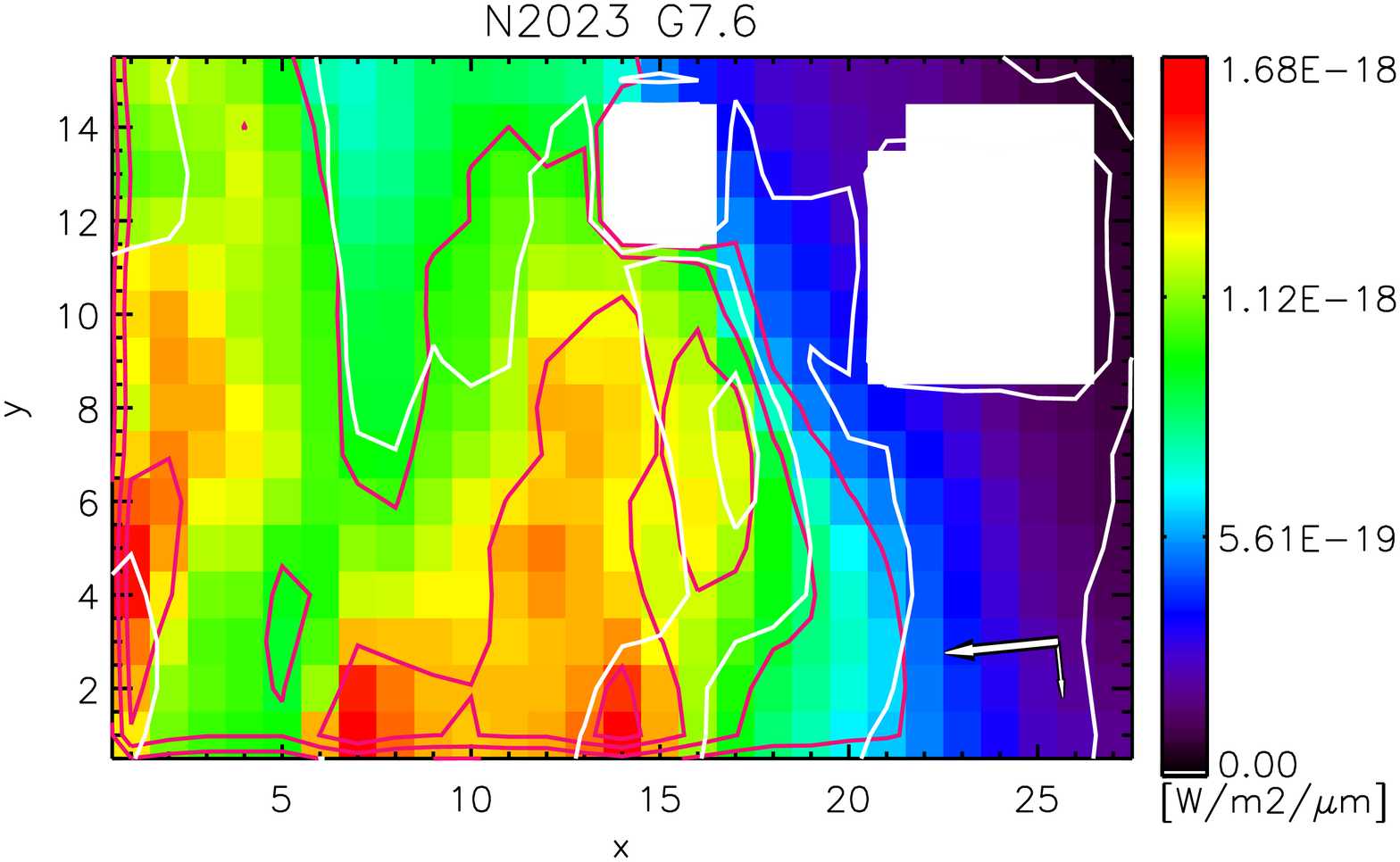}
    \includegraphics[width=5cm]{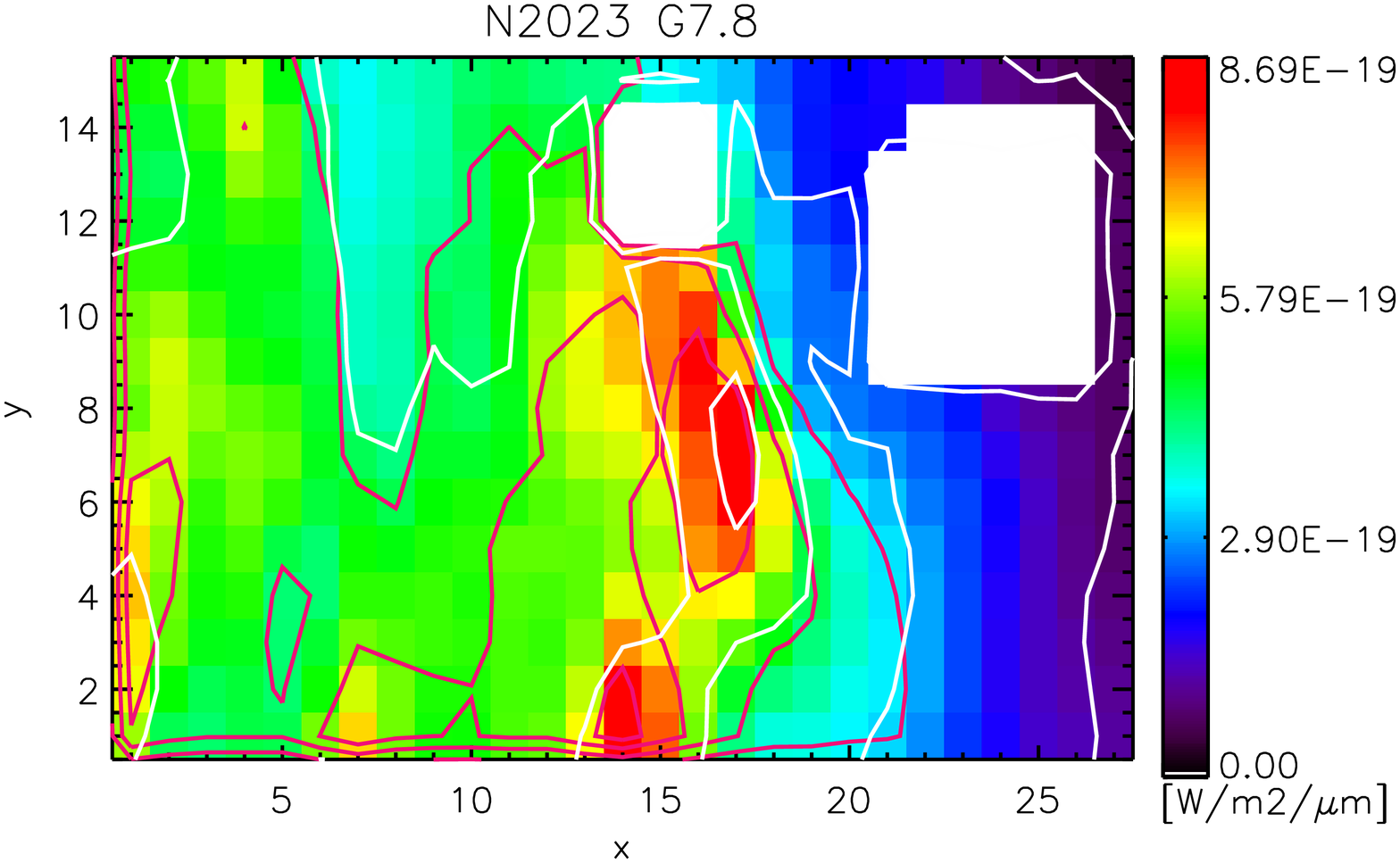}
    \includegraphics[width=5cm]{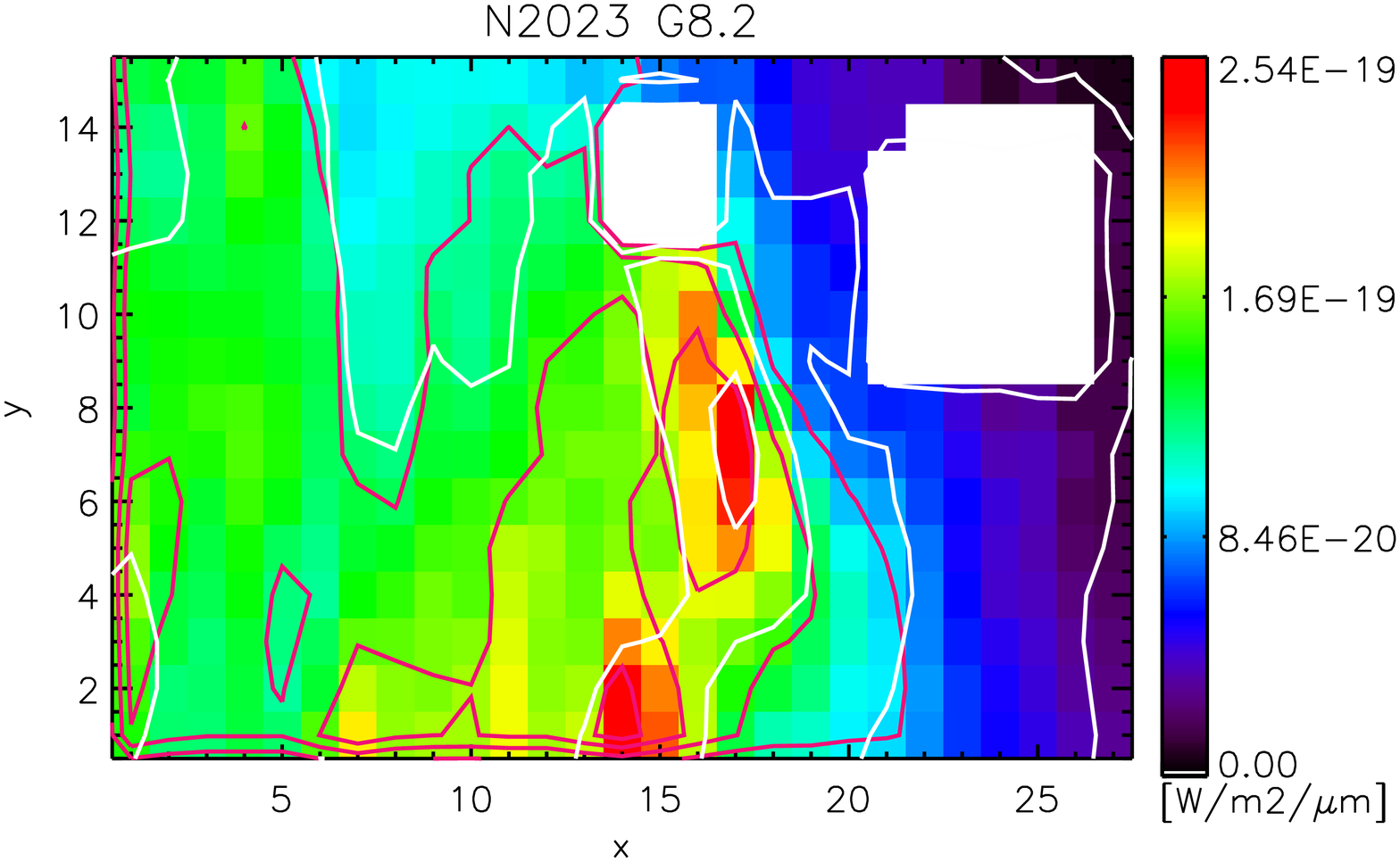}
    \includegraphics[width=5cm]{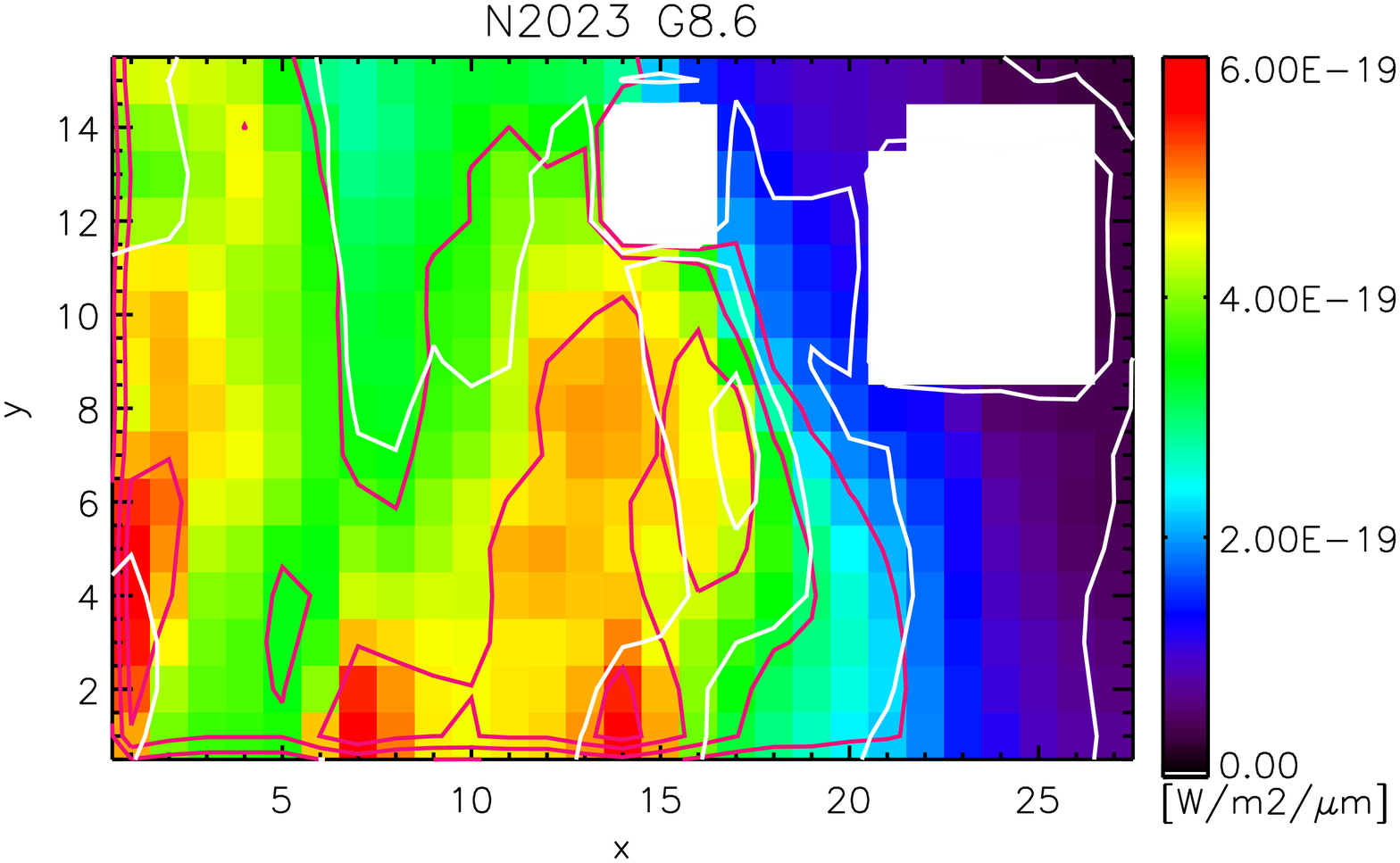}
    \includegraphics[width=5cm]{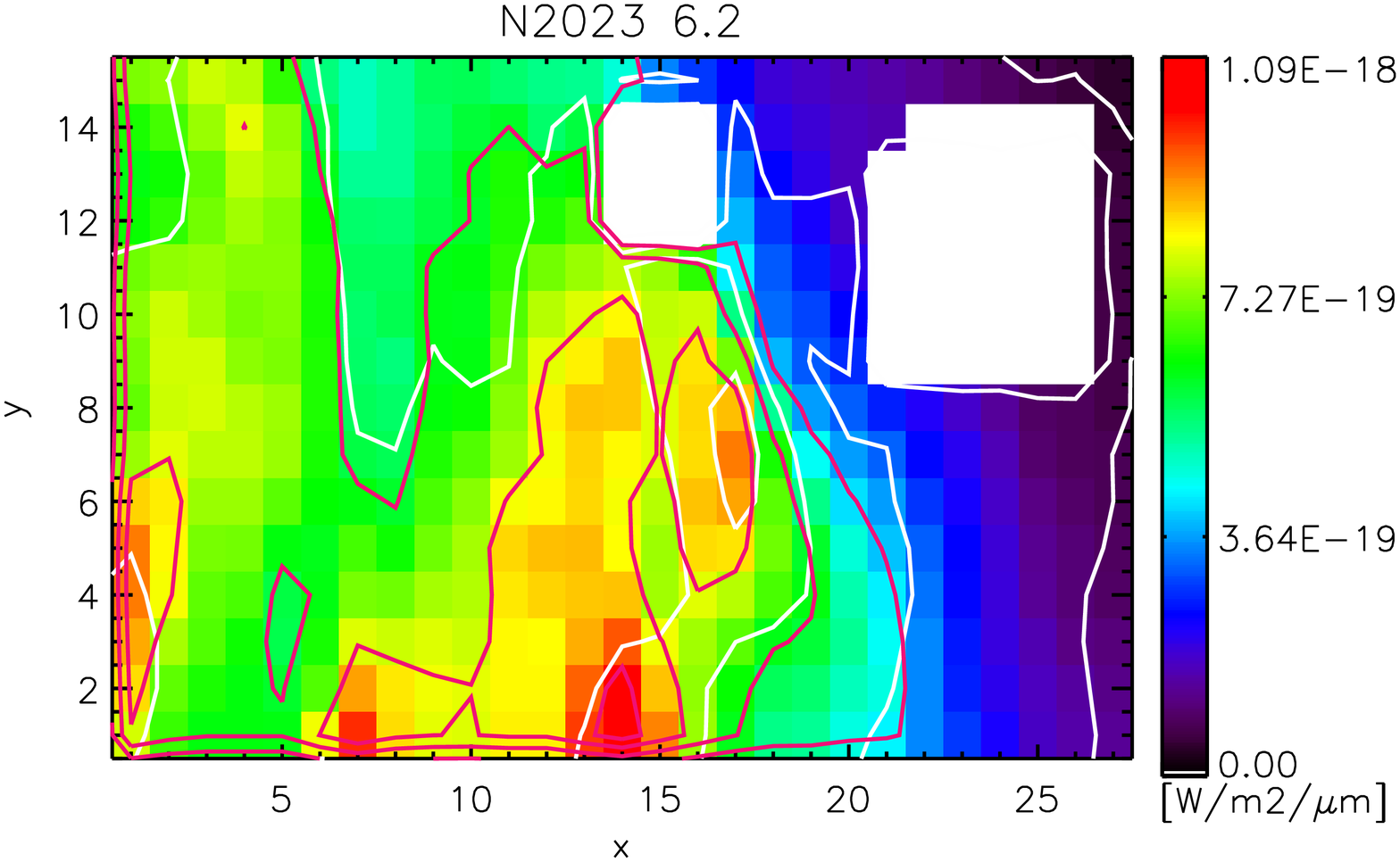}
    \includegraphics[width=5cm]{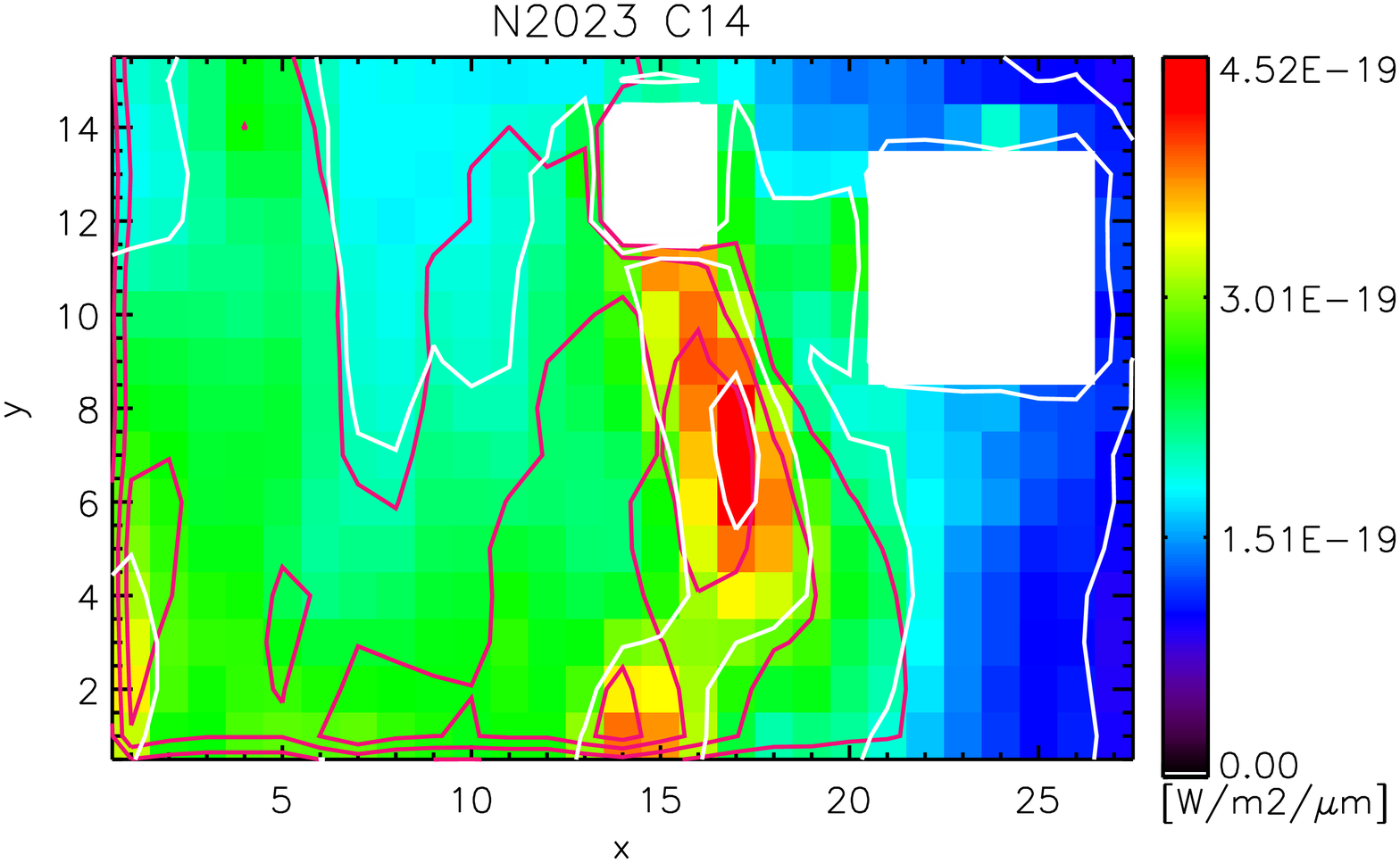}
    \caption{NGC~2023 spatial emission maps: (top two rows) the four components making up the 7.7~\micron\ emission complex; (bottom row) maps of the `pure' PAH 6.2~\micron\ PAH band and the continuum strength at 14 \micron. The component maps and the 6.2~\micron\ PAH map are in units of W m$^{-2}$ per pixel and the continuum is in units of W m$^{-2}$ \micron$^{-1}$ per pixel. The two white boxes represent regions where we the spectra are dominated by bright protostars. The white and red contours represent the distribution of 6.2~\micron\ PAH emission and 14~\micron\ continuum emission respectively. North and east are indicated in the top left panel by the long and short arrows respectively.}   
    \label{fig:N2023maps}
\end{figure*}

\begin{figure*}
  \centering

    \includegraphics[width=5cm]{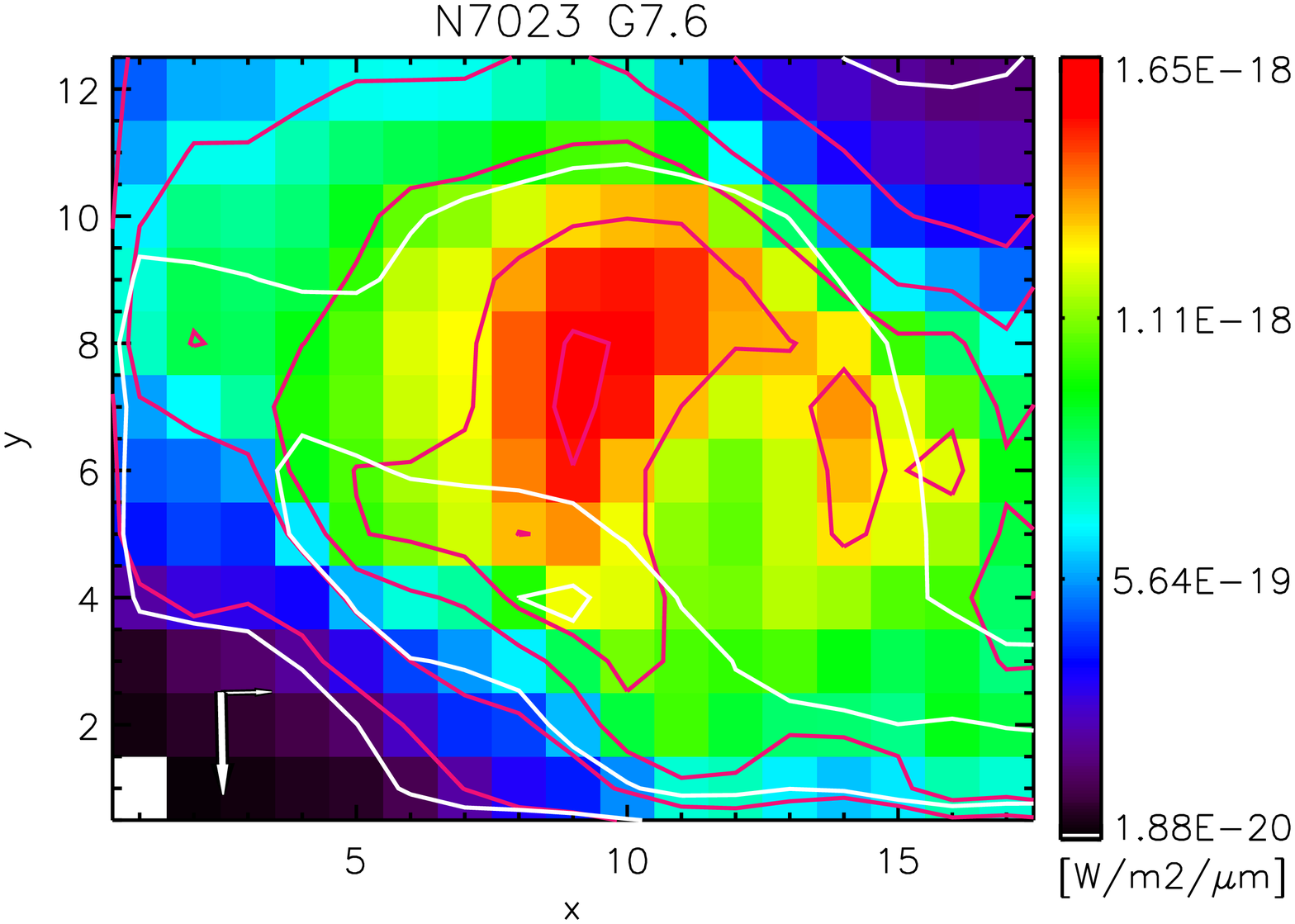}
    \includegraphics[width=5cm]{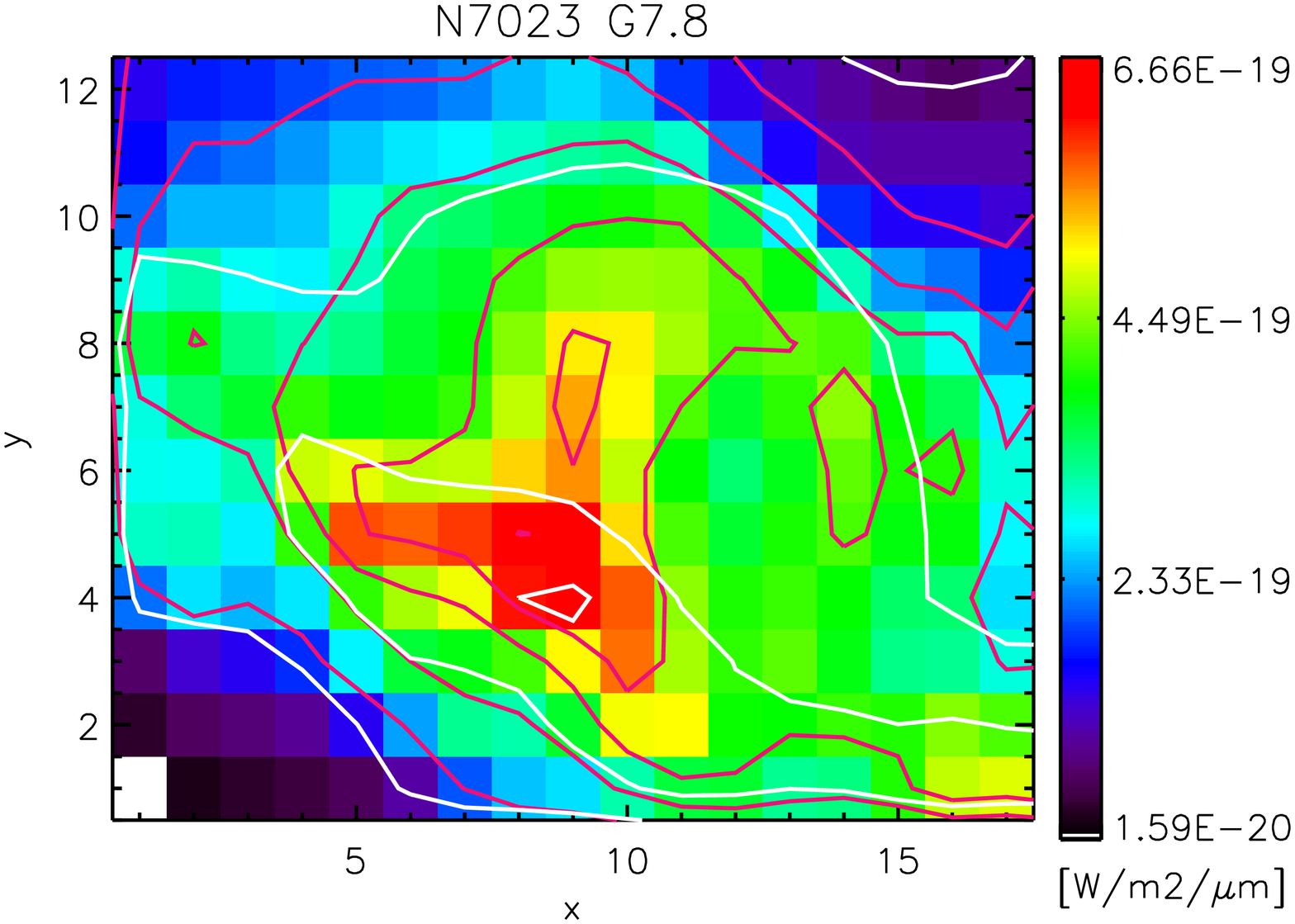}
    \includegraphics[width=5cm]{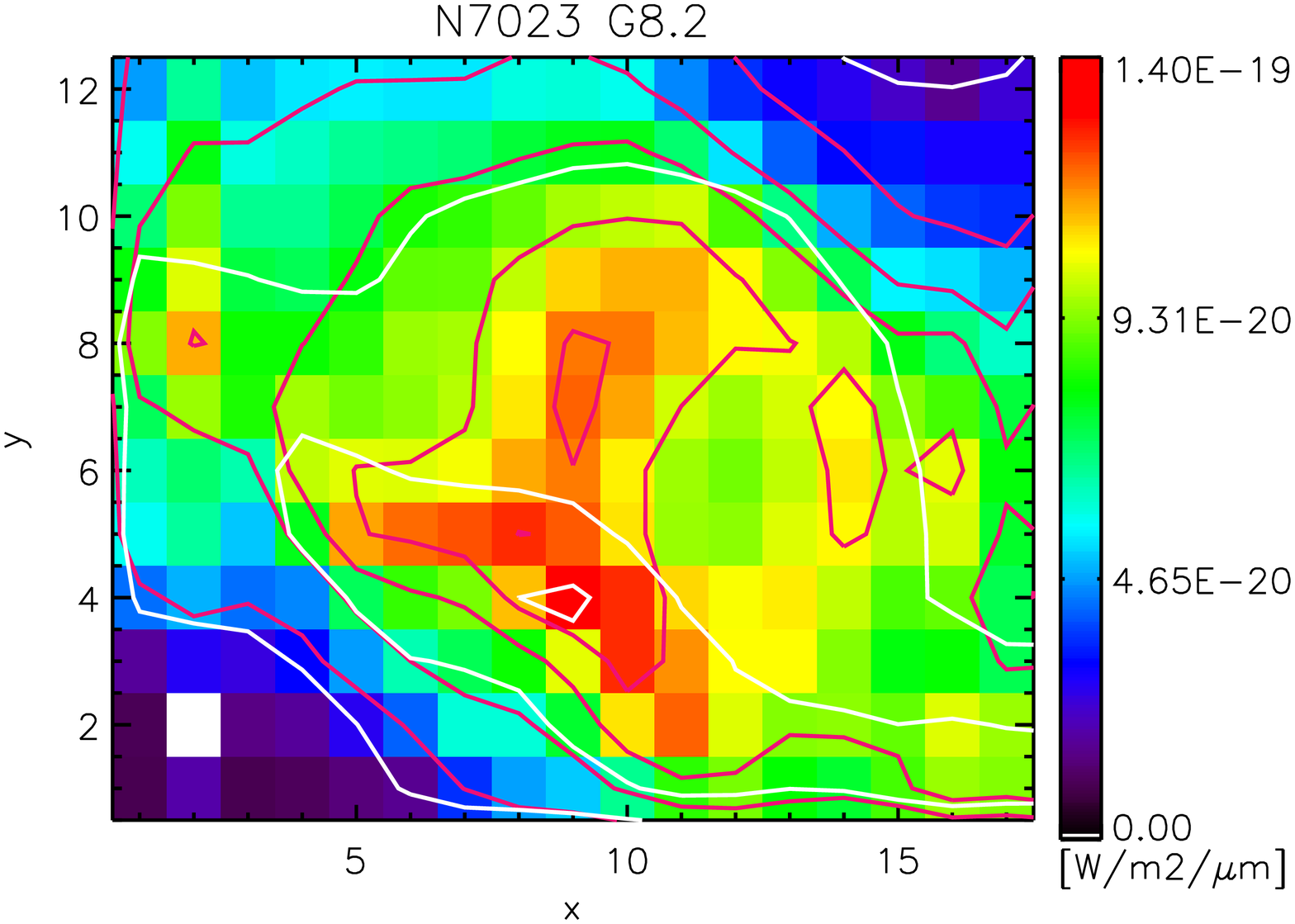}
    \includegraphics[width=5cm]{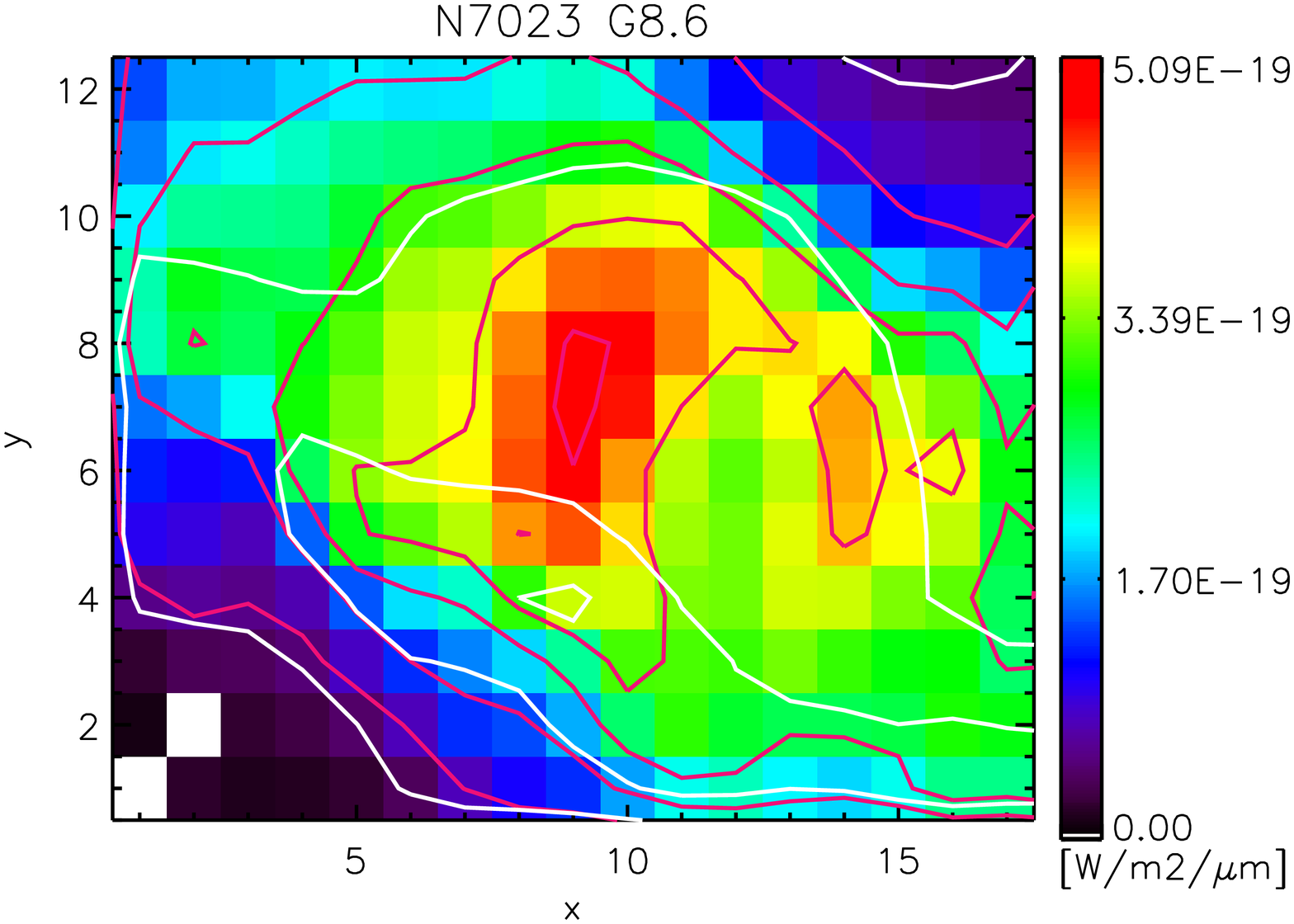}
    \includegraphics[width=5cm]{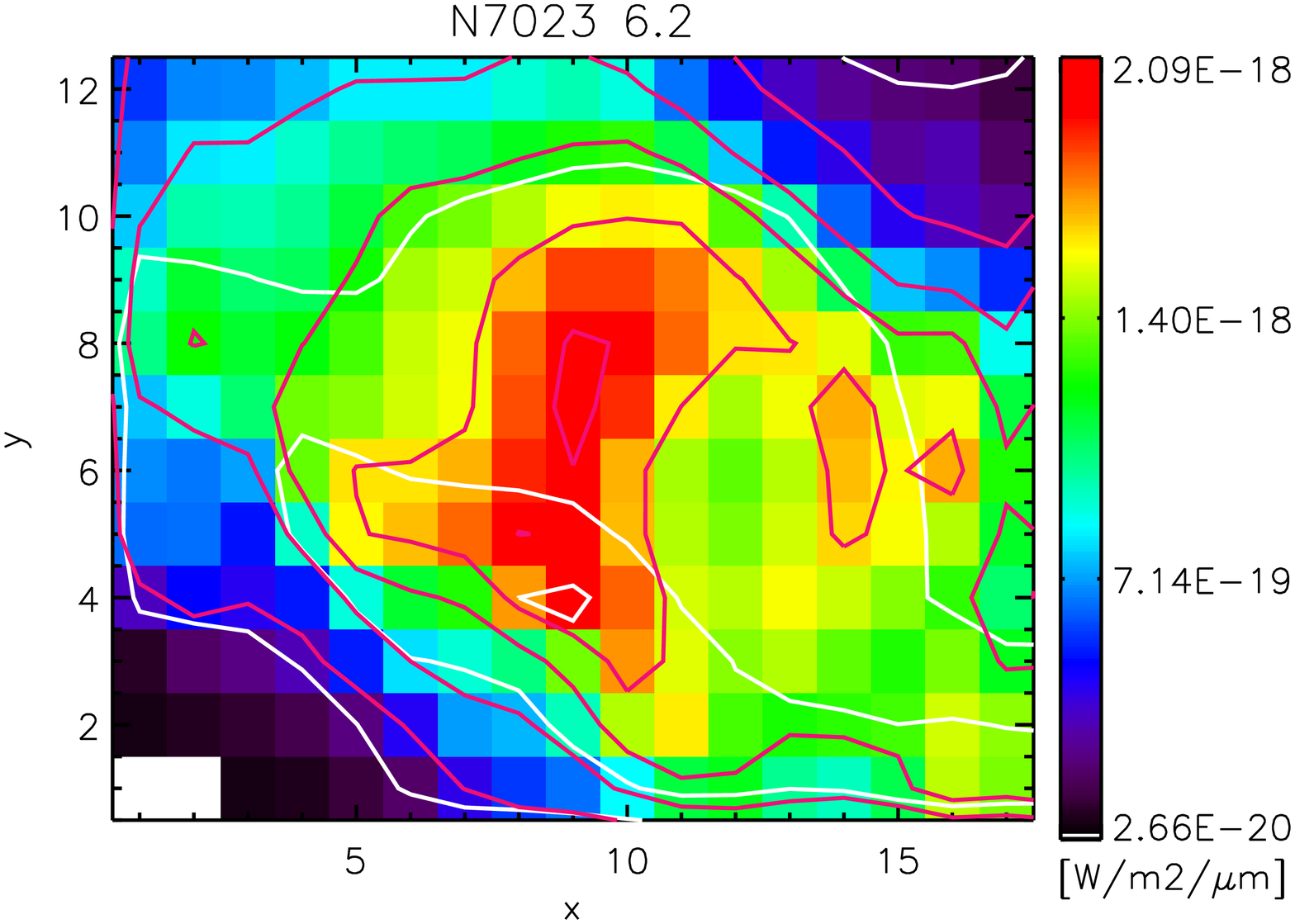}
    \includegraphics[width=5cm]{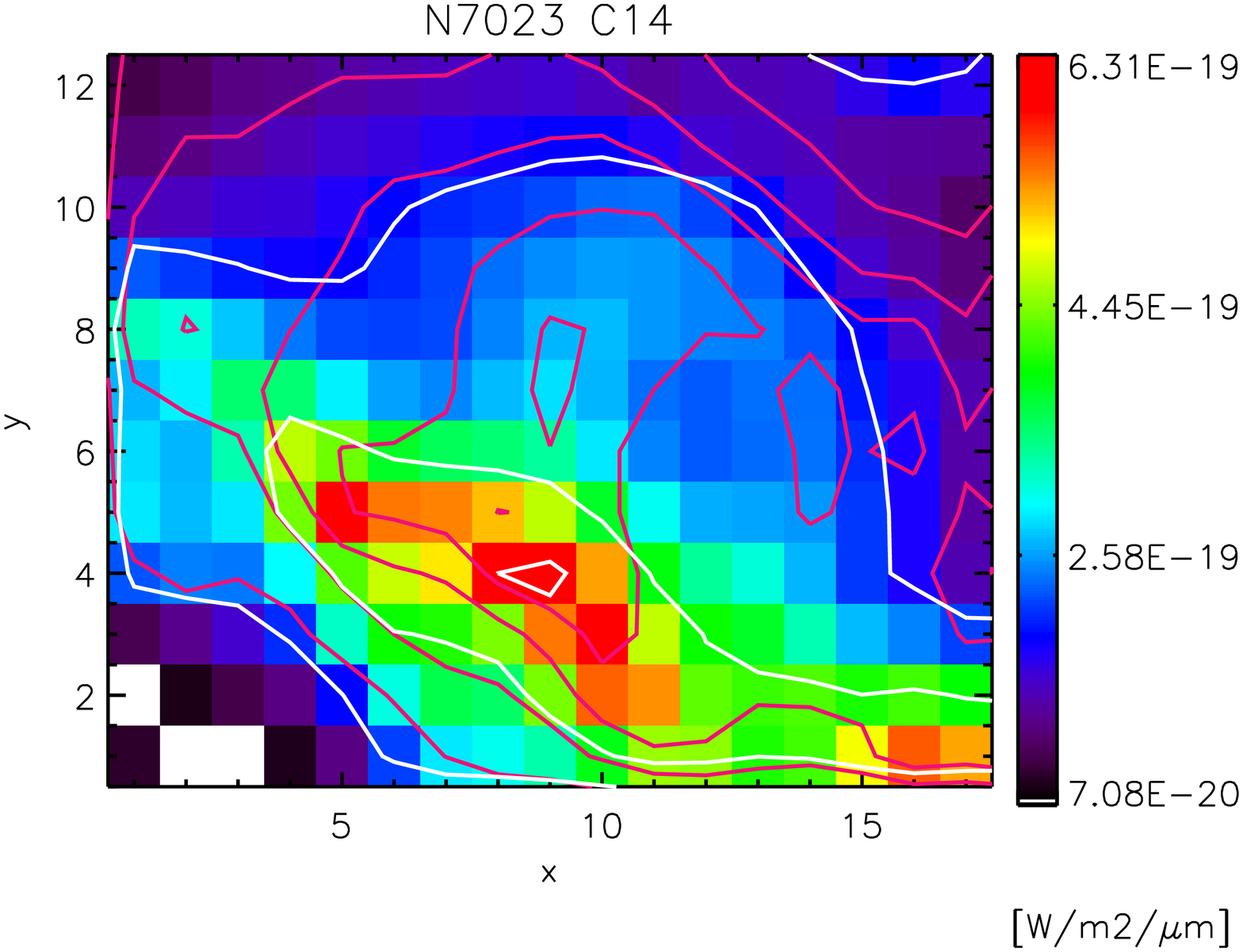}
    \caption{NGC~7023 spatial emission maps: (top two rows) the four components making up the 7.7~\micron\ emission complex; (bottom row) maps of the `pure' PAH 6.2~\micron\ PAH band and the continuum strength at 14 \micron. The component maps and the 6.2~\micron\ PAH map are in units of W m$^{-2}$ per pixel and the continuum is in units of W m$^{-2}$ \micron$^{-1}$ per pixel. Regions of low SNR have been shown in white.  The white and red contours represent the distribution of 6.2~\micron\ PAH emission and 14~\micron\ continuum emission respectively. North and east are indicated in the top left panel by the long and short arrows respectively.}  
    \label{fig:N7023maps}
\end{figure*}

Figures~\ref{fig:I12063maps},~\ref{fig:W49Amaps},~\ref{fig:M17maps},~\ref{fig:IC434maps},~\ref{fig:N1333maps},~\ref{fig:N2023maps} and~\ref{fig:N7023maps} show the intensity maps for IRAS~12063-6259, W49A, M17, the Horsehead PDR, NGC 1333, NGC 2023 and NGC 7023 respectively in the same format as for G~37 which is included in the main text (Figure~\ref{fig:g37maps}).

\end{document}